\documentclass[aps,pre,onecolumn,superscriptaddress,showpacs]{revtex4}

\usepackage{amssymb}
\usepackage{epsfig}
\usepackage{graphicx}
\usepackage{float}
\usepackage{amsmath}
\usepackage{times}
\usepackage{multirow}
\usepackage{caption}
\usepackage{graphicx,epstopdf}
\usepackage{xcolor}

\begin{document}

\title{Exact results of the limited penetrable horizontal visibility
  graph associated to random time series and its application}

\author{Minggang Wang}\email{magic821204@sina.com;magic82@bu.edu}
\affiliation{School of Mathematical Science, Nanjing Normal University, Nanjing 210042, Jiangsu, China}
\affiliation{Department of Mathematics, Nanjing Normal University Taizhou College, Taizhou 225300, Jiangsu, China}
\affiliation{Center for Polymer Studies and Department of Physics, Boston University, Boston,MA 02215, USA}

\author{Andr\'{e} L.M.Vilela}
\affiliation{Center for Polymer Studies and Department of Physics, Boston University, Boston,MA 02215, USA}
\affiliation{Universidade de Pernambuco, 50720-001, Recife-PE, Brazil}

\author{Ruijin Du}
\affiliation{Center for Polymer Studies and Department of Physics, Boston University, Boston,MA 02215, USA}
\affiliation{Energy Development and Environmental Protection Strategy Research Center, Jiangsu University, Zhenjiang, 212013 Jiangsu, China}

\author{Longfeng Zhao}
\affiliation{Center for Polymer Studies and Department of Physics, Boston University, Boston,MA 02215, USA}

\author{Gaogao Dong}
\affiliation{Center for Polymer Studies and Department of Physics, Boston University, Boston,MA 02215, USA}
\affiliation{Energy Development and Environmental Protection Strategy Research Center, Jiangsu University, Zhenjiang, 212013 Jiangsu, China}

\author{Lixin Tian}\email{tianlx@ujs.edu.cn}
\affiliation{School of Mathematical Science, Nanjing Normal University, Nanjing 210042, Jiangsu, China}
\affiliation{Energy Development and Environmental Protection Strategy Research Center, Jiangsu University, Zhenjiang, 212013 Jiangsu, China}

\author{H. Eugene Stanley}
\affiliation{Center for Polymer Studies and Department of Physics, Boston University, Boston,MA 02215, USA}

\begin{abstract}

\noindent
The limited penetrable horizontal visibility algorithm is a new time
analysis tool and is a further development of the horizontal visibility
algorithm. We present some exact results on the topological properties
of the limited penetrable horizontal visibility graph associated with
random series. We show that the random series maps on a limited
penetrable horizontal visibility graph with exponential degree
distribution $P(k)\sim exp[-\lambda (k-2\rho-2)], \lambda =
ln[(2\rho+3)/(2\rho+2)],\rho=0,1,2,...,k=2\rho+2,2\rho+3,...$, independent
of the probability distribution from which the series was generated. We
deduce the exact expressions of the mean degree and the clustering
coefficient and demonstrate the long distance visibility
property. Numerical simulations confirm the accuracy of our theoretical
results. We then examine several deterministic chaotic series (a
logistic map, the H$\acute{e}$non map, the Lorentz system, and an energy
price chaotic system) and a real crude oil price series to test our
results. The empirical results show that the limited penetrable
horizontal visibility algorithm is direct, has a low computational cost
when discriminating chaos from uncorrelated randomness, and is able to
measure the global evolution characteristics of the real time series.

\end{abstract}

\maketitle
\section{Introduction}

\noindent
Several methodologies for understanding the complicated behavior of
nonlinear time series have been recently developed, including chaos
analysis [1], fractal analysis [2], and complexity measurement [3]. With
the development of complex network theories [4--7], a new
multidisciplinary methodology for characterizing nonlinear time series
using complex network science has emerged and rapidly expanded
[8--24]. The main tool of these methods is to use an algorithm or
algorithms to transform a nonlinear time series into a corresponding
complex network and then use the topological structure of complex
networks to analyze the properties of the nonlinear time series.

Currently there are four ways of converting univariate time
series into complex networks. The first one is Pseudo-periodic time series transitions [8] that analyze
  pseudo-periodic time series. The second one  is the visibility graph (VG) method, which was first proposed
  by Lacasa et al.~[9--10]. To facilitate computation, Luque et
  al.~[11--12] proposed a simplified horizontal visibility algorithm
  (HVG) based on the visibility algorithm. Bezsudnov et al.~[13]
  proposed a parameter visibility method. Gao et al.~[14] proposed a
  limited penetrable visibility method (LPVG) and multiscale limited
  penetrable horizontal visibility graph (MLPHVG).  The third one  is the phase space reconstruction method
[15--16]. It begins with a phase space reconstruction of time series
analysis, maps fixed-length time series segments into nodes of a
network, and then uses the correlation coefficients (or distances)
between these nodes to determine whether they are connected or
not.  And the last one  is the coarse graining method [17--20] by which fluctuations
  of time series are transformed into signal sequences. A fixed-length
  signal sequence is treated as a network node that connects nodes of
  time series in chronological order, and a weighted complex network
  with direction is then constructed. In recent years, researchers have
  used complex network theories to study multivariate time series
  [21-24]. These four methods all effectively maintain most of the properties of
different types of time series, and they have been successfully used in
many different fields [25-30].

Although there have been abundant empirical results obtained using time
series complex network algorithms [8-24], rigorous theoretical results
are still lacking.  Only a small amount of literature [9--12] has
presented exact results on the properties of the horizontal visibility
graphs (HVG) associated with random series. Thus far no rigorous theory
other than the above algorithms has been developed. Thus our goal here
is to focus on one type of general horizontal visibility algorithm, the
limited penetrable horizontal visibility graph (LPHVG), and derive exact
results on the properties of the limited penetrable horizontal
visibility graphs associated with random series. We prove that an
independent and identically distributed random series can be mapped on a
limited penetrable horizontal visibility graph with exponential degree
distribution $P(k)\sim exp[-\lambda (k-2\rho-2)], \lambda = ln[(2\rho+3)/(2\rho+2)],
\rho=0,1,2,...,k=2\rho+2,2\rho+3,...$, which is an extension of the
result presented in Ref.~[11]. We deduce the exact mean degree and the
clustering coefficient, and we prove that the limited penetrable
horizontal visibility graph associated with any independent and
identically distributed random series has a small world
characteristic. To verify our theoretical solution, we acquire
simulation results by using several deterministic chaotic series (a
logistic map, an H$\acute{e}$non map, the Lorentz system, and the energy
price chaos system) and a real-world crude oil price series that
confirms the accuracy and usability of our exact results.

\section{Results}

\noindent
We here supply several exact results of LPHVG associated with random
time series and apply them to several deterministic chaotic series (a
logistic map, an H$\acute{e}$non map, the Lorentz system, and the energy
price chaos system) and a real-world crude oil price series.

\textbf{Degree distribution.} Let $X(t)$ be a real valued bi-infinite
time series of independent and identically distributed ($i.i.d.$) random variables
with a probability density $f(x)$ in which $x\in [a,b]$, and
consider its associated LPHVG with the limited penetrable distance $\rho
= 1$. Then
\begin{equation}\label{eq1}
P(k)\sim exp[-(k-4)ln(5/4)],k=4,5,...,\forall f(x).
\end{equation}

To prove this conclusion we first calculate the probability that an
arbitrary datum with value $x_{0}$ has a limited penetrability at
most a one-time visibility of $k$ other data. We thus list all sets of
possible configurations for data $x_{0}$ with $k=4$ (see \emph{Fig.~S1
  in Appendix}), $k=5$ (see \emph{Fig.~S2 in Appendix}), and $k=6$ (see
\emph{Fig.~S3 in Appendix}), and we calculate the probability $P(k=4)$
(see \emph{Eq.~(S4)}), $P(k=5)$ (see \emph{Eq.~(S9)}), and $P(k=6)$ (see
\emph{Eq.~(S10)}). We then deduce the rules of when a given
configuration contributes to $P(k)$ (see rules i--iv) and obtain a
general expression for $P(k)$ (see Eq.~(S12)). The detailed proof of
this result is shown in \emph{Appendix Theorem S1}. This is an exact
result for a limited penetrable horizontal visibility graph with the
limited penetrable distance $\rho= 1$. We conclude that for every
probability distribution $f(x)$, the degree distribution $P(k)$ of the
associated LPHVG has the same exponential form. In addition, from this
result we can obtain the more general result (\emph{Theorem S2} in
\emph{Appendix}) in which $X(t)$ is a real bi-infinite time series of
$i.i.d.$ random variables with a
probability distribution $f(x)$ in which $x\in [a,b]$, and can examine
its associated LPHVG with the limited penetrable distance $\rho$. Then
\begin{equation}\label{eq2}
P(k)\sim
exp\{-(k-2\rho-2)ln[(2\rho+3)/(2\rho+2)]\},\rho=0,1,2,...,k=2\rho+2,2\rho+3,...,\forall
f(x).
\end{equation}

Note that when $\rho = 0$, then $P(k)\sim exp[-(k-2)ln(3/2)]$, the result in
Ref.~[11]. In fact, when $\rho = 0$ the LPHVG becomes the HVG (see
\emph{Methods Section}). When $\rho = 1$, the result is Eq.~(1). Therefore Eq. (2) is an
extension of the previous result [11] that indicates that the degree
distribution $P(k)$ of LPHVG associated with $i.i.d.$ random time series has a
unified exponential form.

To further check the accuracy of our analytical results, we perform
several numerical simulations. We generate a random series of 3000 data
points from uniform, gaussian, and power law distributions and their
associated limited penetrable horizontal visibility graphs. Figs~1(a)
and 1(b) show plots of the degree distributions of the resulting graphs
with a penetrable distance $\rho = 1$ and $\rho=2$. Here circles indicate a series
extracted from a uniform distribution, and squares and diamonds indicate
series extracted from gaussian and power law distributions,
respectively. The solid line indicates the theoretical results of
Eq.~(2). We find that the theoretical results agree with the
numerics.  Note that a prerequisite for our theoretical results is that
the length of the time series must be infinitely long, i.e. the series size $N\rightarrow \infty$, so we can assert that the tail degree distribution of LPHVG associated to  $i.i.d.$ random series deviated from the theoretical result is only due to the effect of the finite size. In order to check the effect of the finite size, we define the relative error ($E(k)$) and the mean relative error ($ME$) to measure accurate between the numerical result under the finite size and the theoretical result, and use a cutoff value $k_{0}$ to denote the onset of finite size effects.
\begin{equation}\label{eq2}
E(k)=\frac{|P_{num}(k)-P_{the}(k)|}{P_{the}(k)},ME=\sum_{k}E(k)
\end{equation}
where, $P_{num}(k)$ and $P_{the}(k)$ represent the degree distribution of the numerical result and theoretical result respectively. We generate the random series from uniform distribution with different the series size. We have generated 10 realizations of each series size $N$. Fig. 1 (c) shows the test results of the resulting graphs with penetrable distance $\rho=1$ and Fig. 1 (d) shows the test results of the resulting graphs with penetrable distance $\rho=2$. The subplots in  Figs. 1 (c) and (d) show the relations between the mean relative error ($ME$) and the series size $N$, and the relations between the cutoff value $k_{0}$ and the series size $N$. We find that the the mean relative error ($ME$) decreases with the finite size $N$ increases, and the cutoff value $k_{0}$ increases with the finite size $N$ increases, which agreement with our previous assertion.
\begin{figure}[H]
\centering \scalebox{0.4}[0.4]{\includegraphics{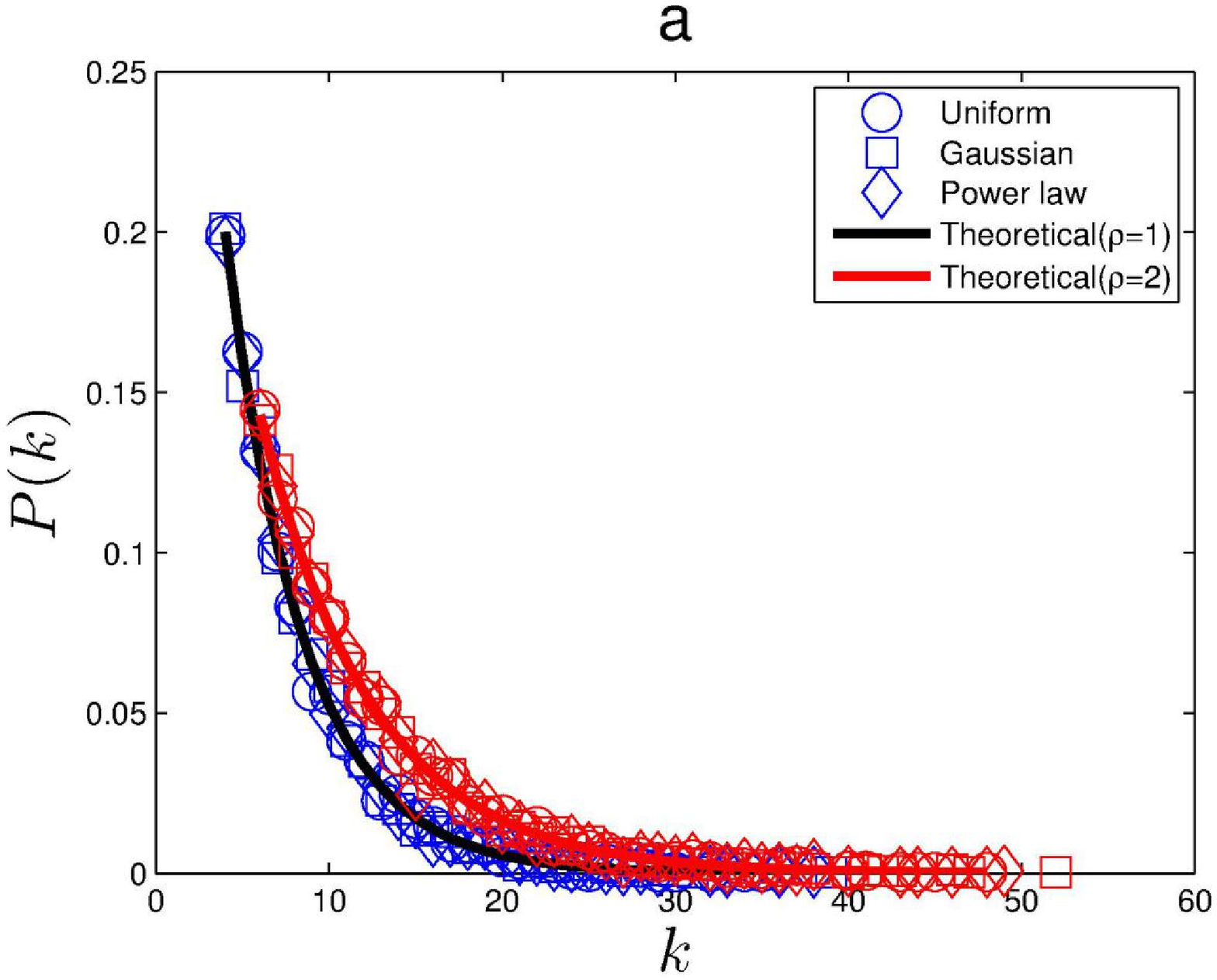}}
\scalebox{0.4}[0.4]{\includegraphics{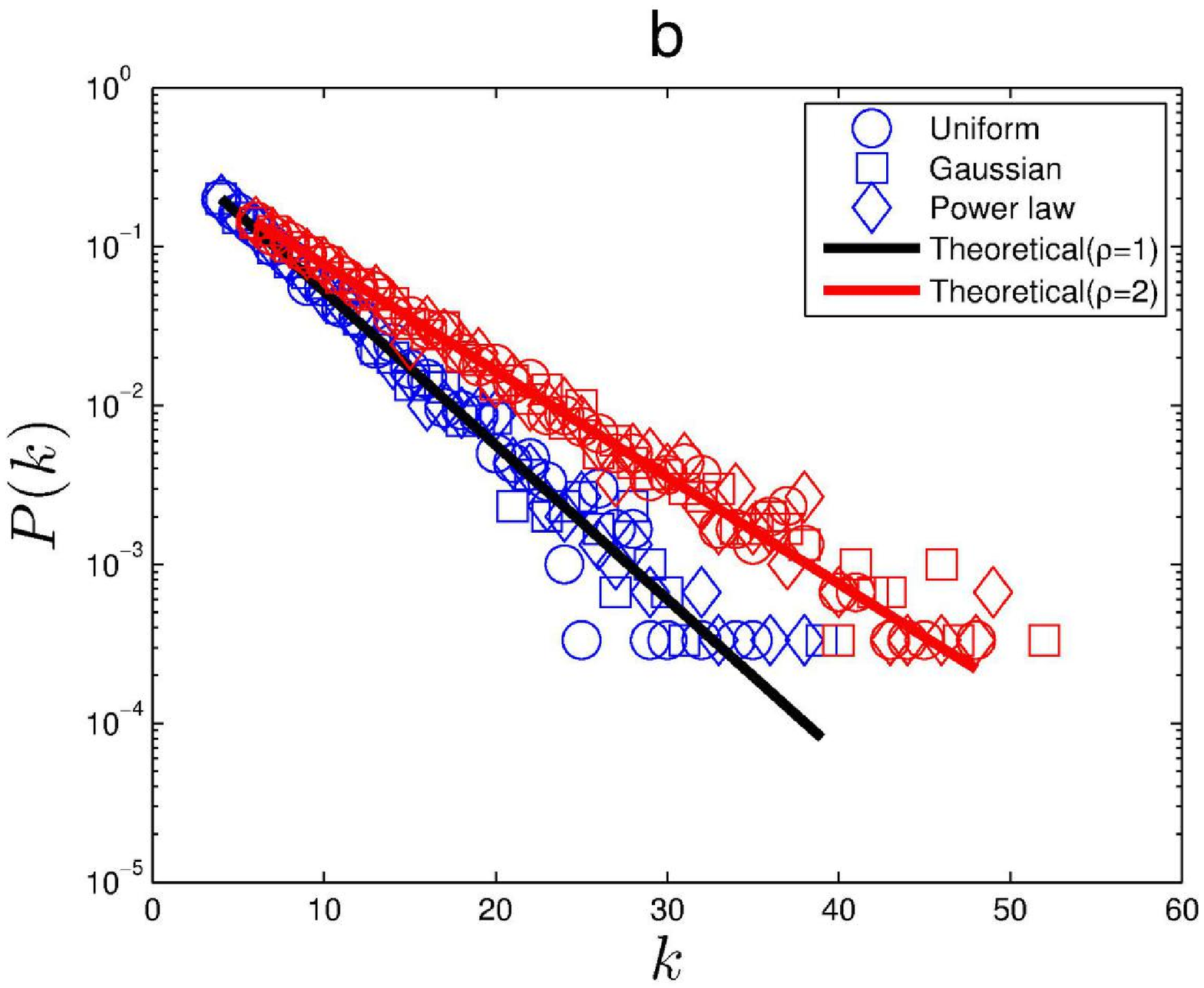}}
\scalebox{0.4}[0.4]{\includegraphics{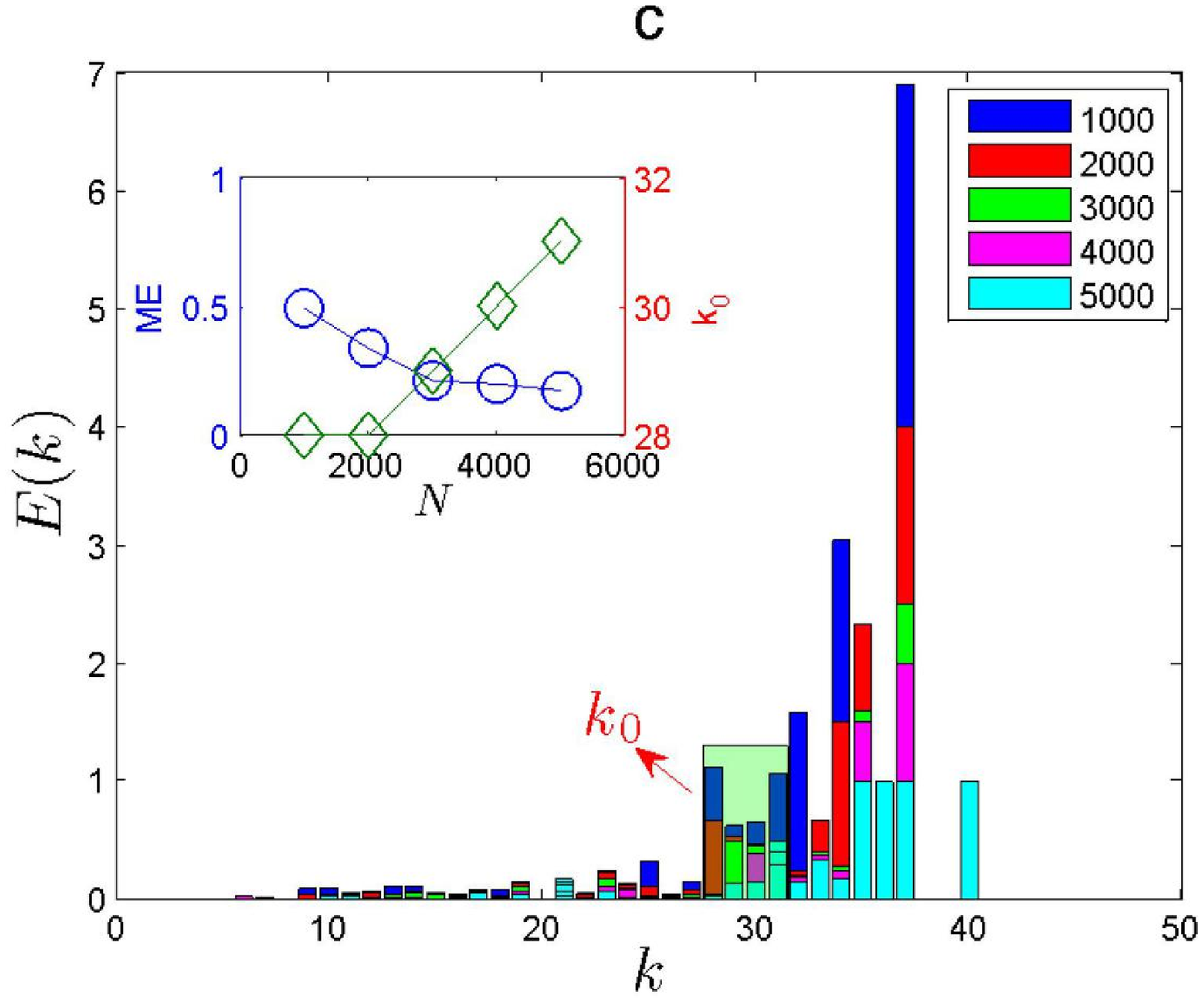}}
\scalebox{0.4}[0.4]{\includegraphics{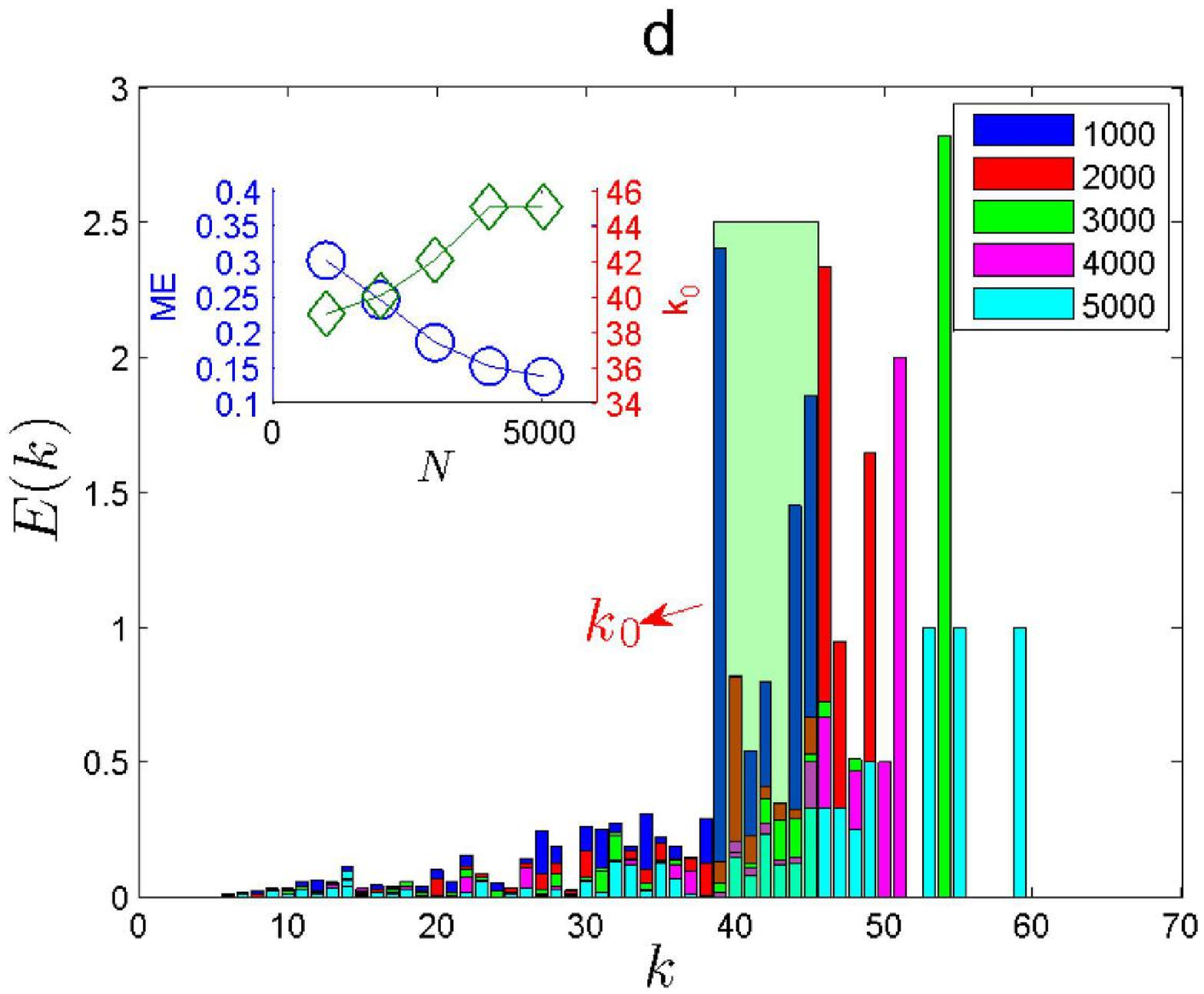}}
\end{figure}

\begin{figure}[H]
\caption{\emph{(a) Plot of the degree distribution of the resulting
    graphs with penetrable distance $\rho=1$ and $\rho=2$, (b) semi-log plot of the
    degree distribution of the resulting graphs with penetrable distance
    $\rho=1$ and $\rho=2$, (c) the test results of the resulting graphs with penetrable distance $\rho=1$ (ensemble averaged over 10 realizations), (d) the test results of the resulting graphs with penetrable distance $\rho=2$ (ensemble averaged over 10 realizations)}}.
\end{figure}

\textbf{Mean degree.} Using Eq.~(2) we calculate the mean degree
$<k>$ of the LPHVG associated with an uncorrelated random series,
\begin{equation}
\begin{array}{l}
<k> = \sum\limits_k kP(k) =
\sum\limits_{k=2(\rho+1)}^{\infty}\frac{k}{2\rho+3}(\frac{2\rho+2}{2\rho+3})^{k-2(\rho+1)}=4(\rho+1).
\end{array}
\end{equation}

We next deduce the more general expression of mean degree
$<k(T)>$. We consider an infinite periodic series of period $T$
(with no repeated values in a period) denoted
$X_{t}=\{...,x_{0},x_{1},x_{2},...,x_{T},x_{1},x_{2},...\}$, where
$x_{0}=x_{T}$. Let $\rho \ll T$ for the subseries $\tilde{X}_{t} =
\{x_{0},x_{1},x_{2},...,x_{T}\}$.  Without losing generality, we assume
$x_{0} = x_{T}$ corresponds to the largest value of the subseries, and
$x_{1},...,x_{\rho},x_{T-\rho},...x_{T-1}$ corresponds to the
$(2\rho+1)$nd largest value of the subseries. We then can construct the
LPHVG associated with the subseries $\tilde{X_{t}}$. If the LPHVG has
$E$ links and $x_{i}$ is smallest datum of $\tilde{X}_{t}$, because no
data repetitions are allowed in $\tilde{X}_{t}$, the degree of $x_{i}$ is
$2(\rho+1)$ during the construction of LPHVG, when $\rho=1$, see
Fig.~S1. We delete node $x_{i}$ and its $2(\rho+1)$ links from the
LPHVG. The resulting graph has $E-2(\rho+1)$ links and $T$ nodes. We
iterate this process $T-(2\rho+1)$ times (see Fig.~2 for a graphical
illustration of this process in the case $\rho=1,T=10$), and the total
number of deleted links is now $E_{d} = 2(\rho+1)[T-(2\rho+1)]$. The
resulting graph has $2(\rho+1)$ nodes, i.e.,
$x_{0},x_{1},...,x_{\rho},x_{T-\rho},...x_{T-1},x_{T}$, see Fig.~2(h) for $\rho=1$ and $T=10$.
Because these $2(\rho+1)$ nodes are connected by $E_{r} =
\binom{2(\rho+1)}{2}$ links, the mean degree of a limited penetrable
horizontal visibility graph associated with $X_{t}$ is
\begin{equation}\label{eq1}
<k(T)> =
2\frac{E_{d}+E_{r}}{T}
=\frac{2[(2(\rho+1))(T-(2\rho+1))+(\rho+1)(2\rho+1)]}{T}
=4(\rho+1)(1-\frac{2\rho+1}{2T})),\rho\ll T.
\end{equation}

Note that Eq. (5) holds for every periodic or aperiodic series in which
$T\rightarrow \infty$, independent of the deterministic process that
generates the series. This is the case because the only constraint in
its derivation is that data within a period are not repeated. Note that
one consequence of Eq. (5) is that every time series has an associated
LPHVG with the maximum mean degree (achieved for aperiodic series)
$<k(\infty)>=4(\rho+1)$, which agrees with Eq.~(4).

\begin{figure}[H]
\centering \scalebox{0.20}[0.18]{\includegraphics{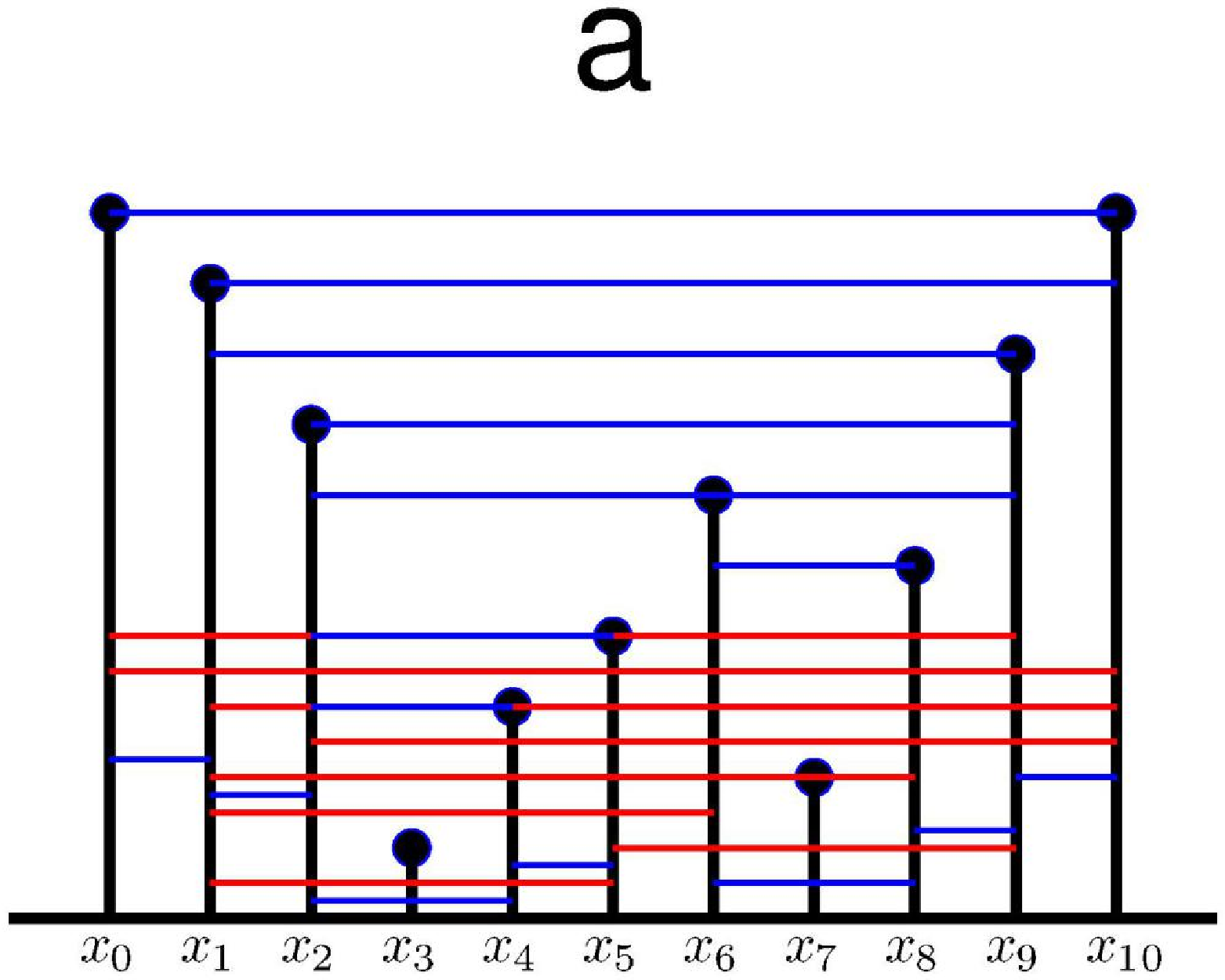}}
\scalebox{0.20}[0.18]{\includegraphics{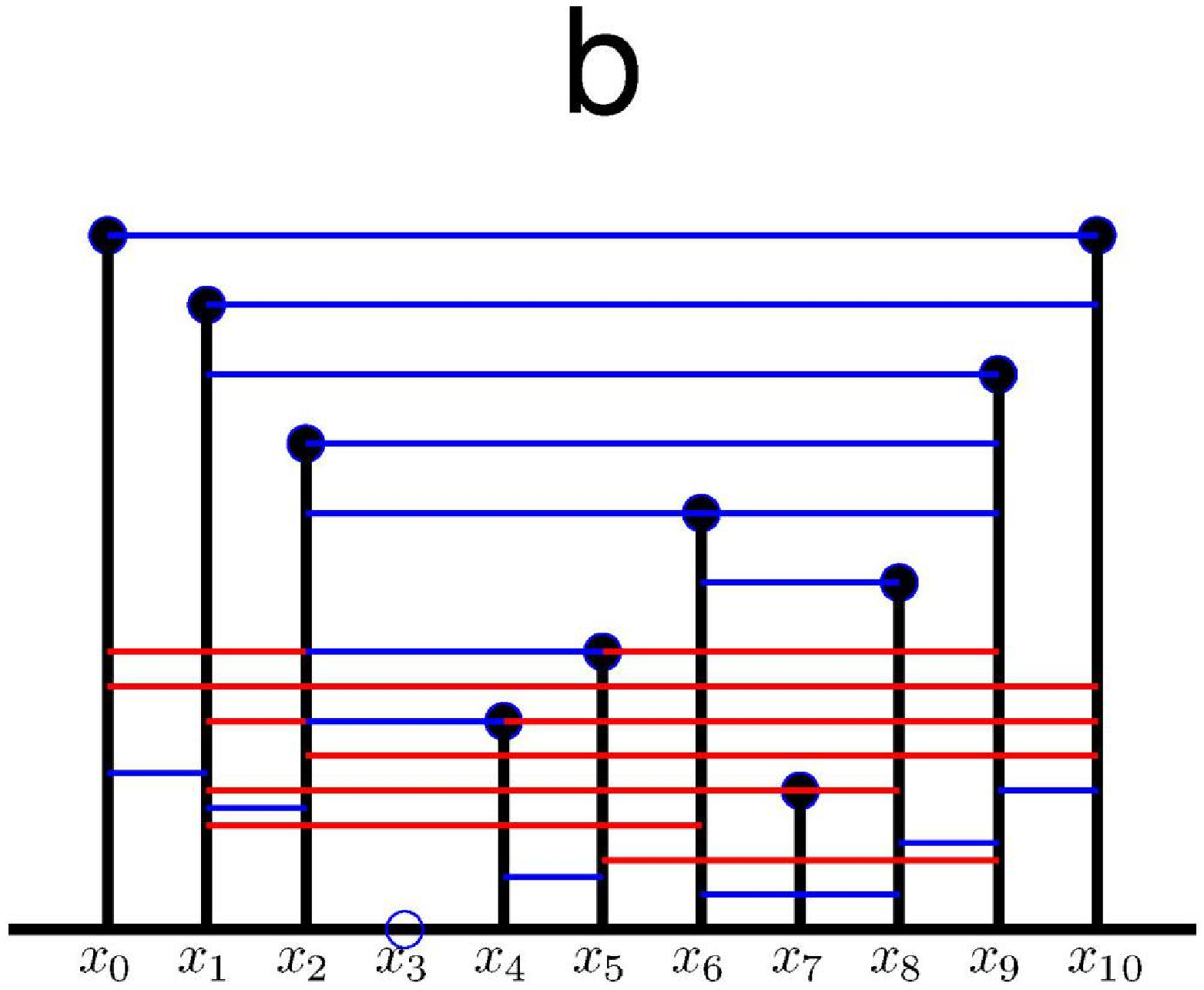}}
\scalebox{0.20}[0.18]{\includegraphics{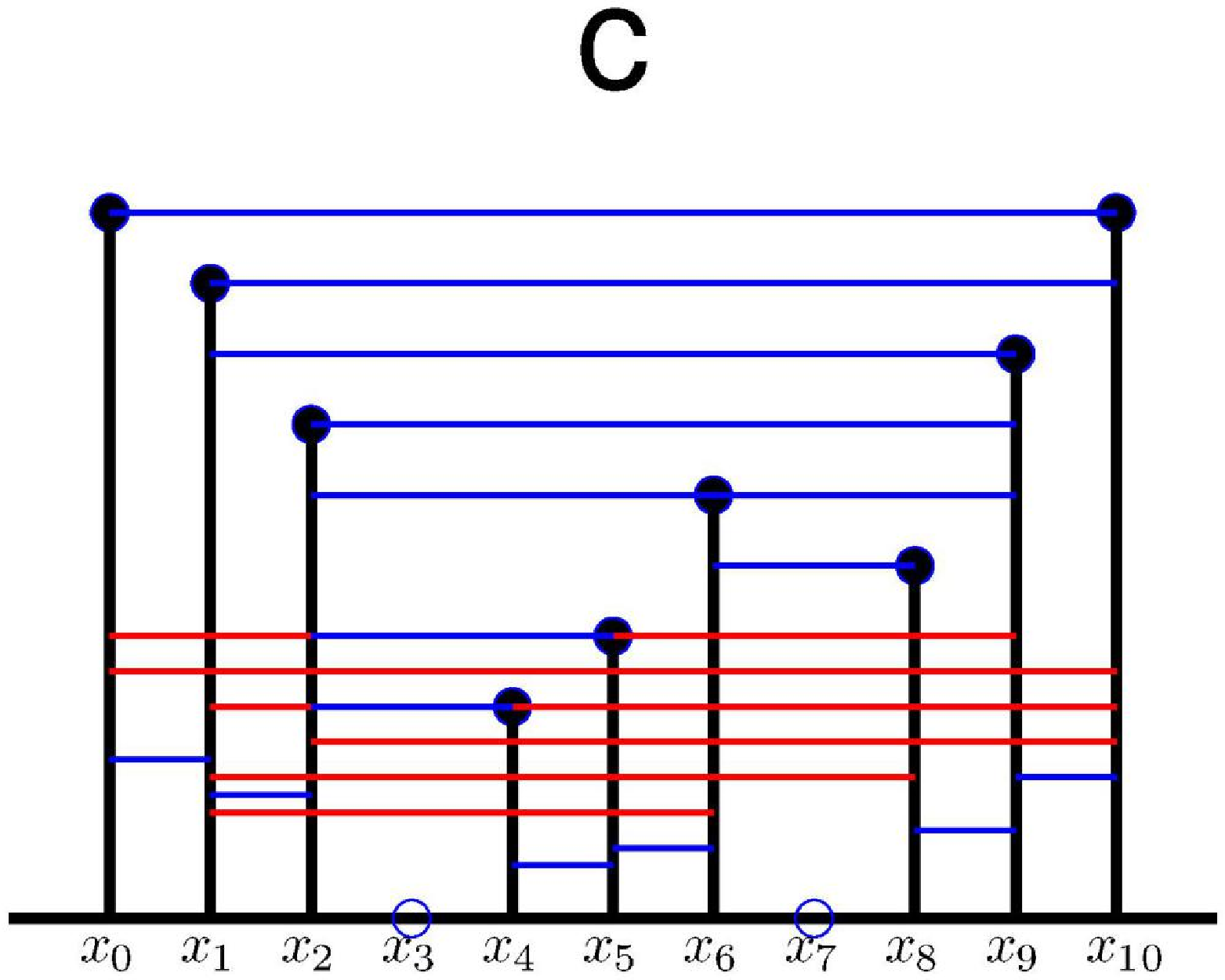}}
\scalebox{0.20}[0.18]{\includegraphics{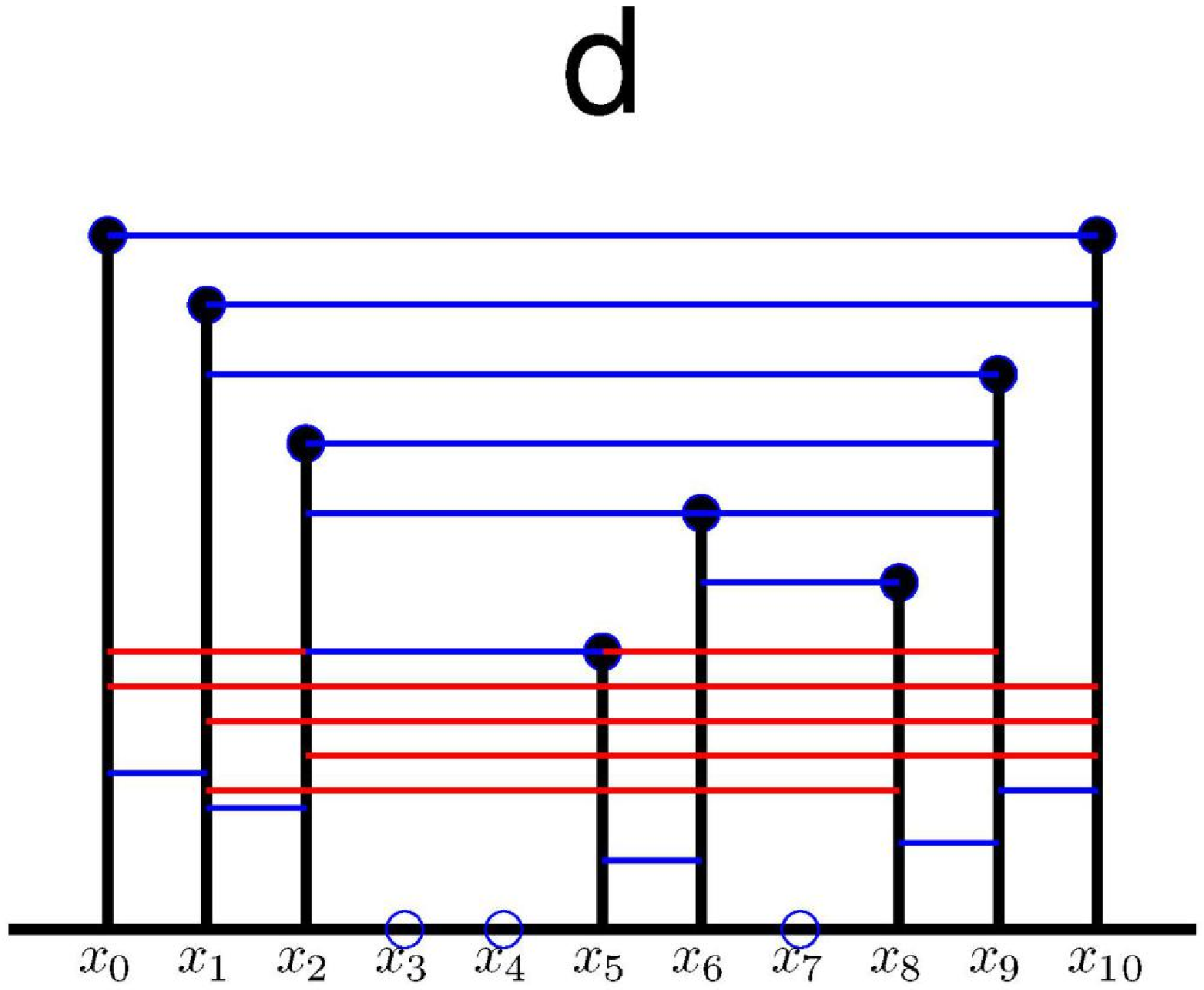}}\\
\scalebox{0.20}[0.18]{\includegraphics{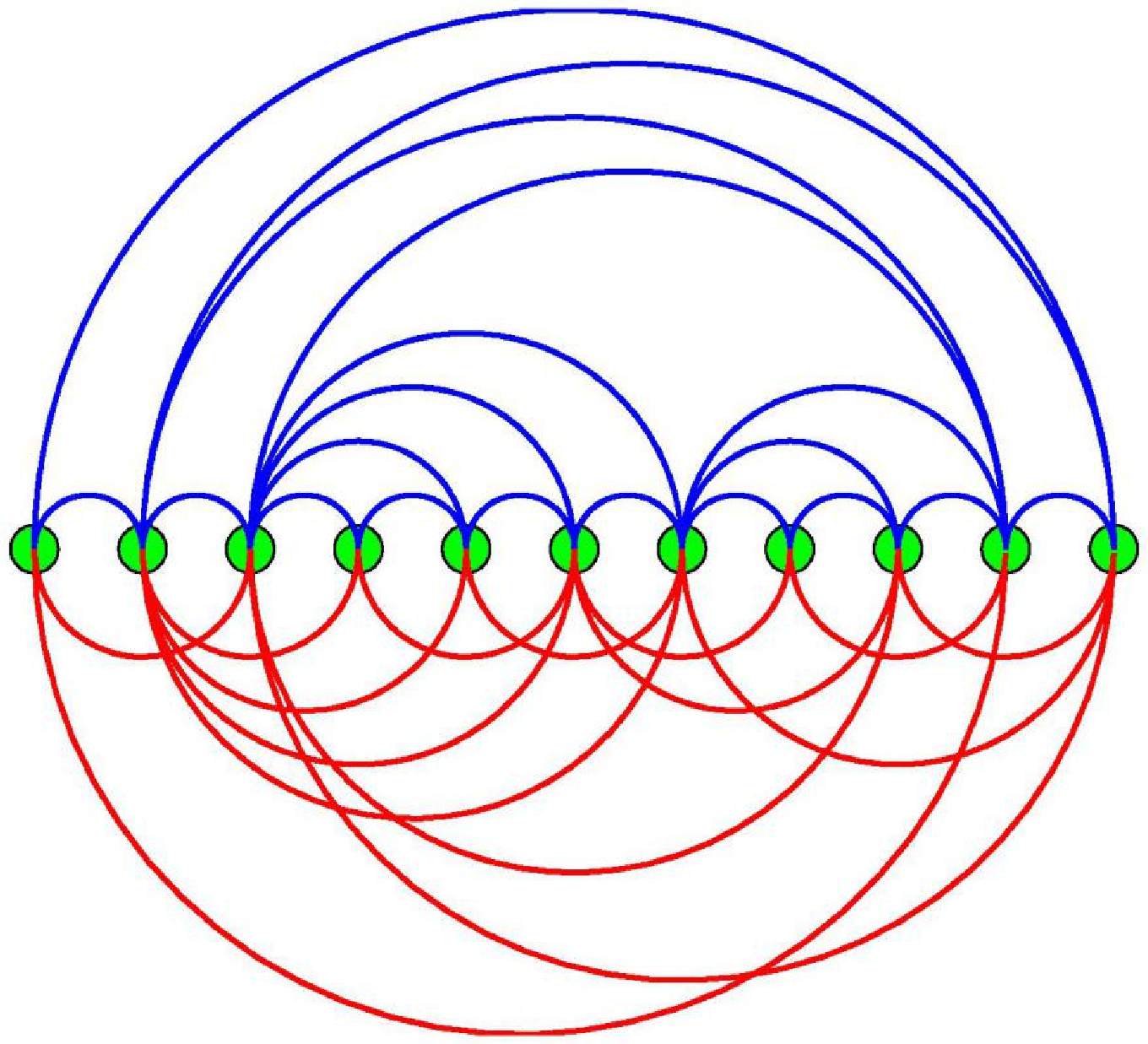}}
\scalebox{0.20}[0.18]{\includegraphics{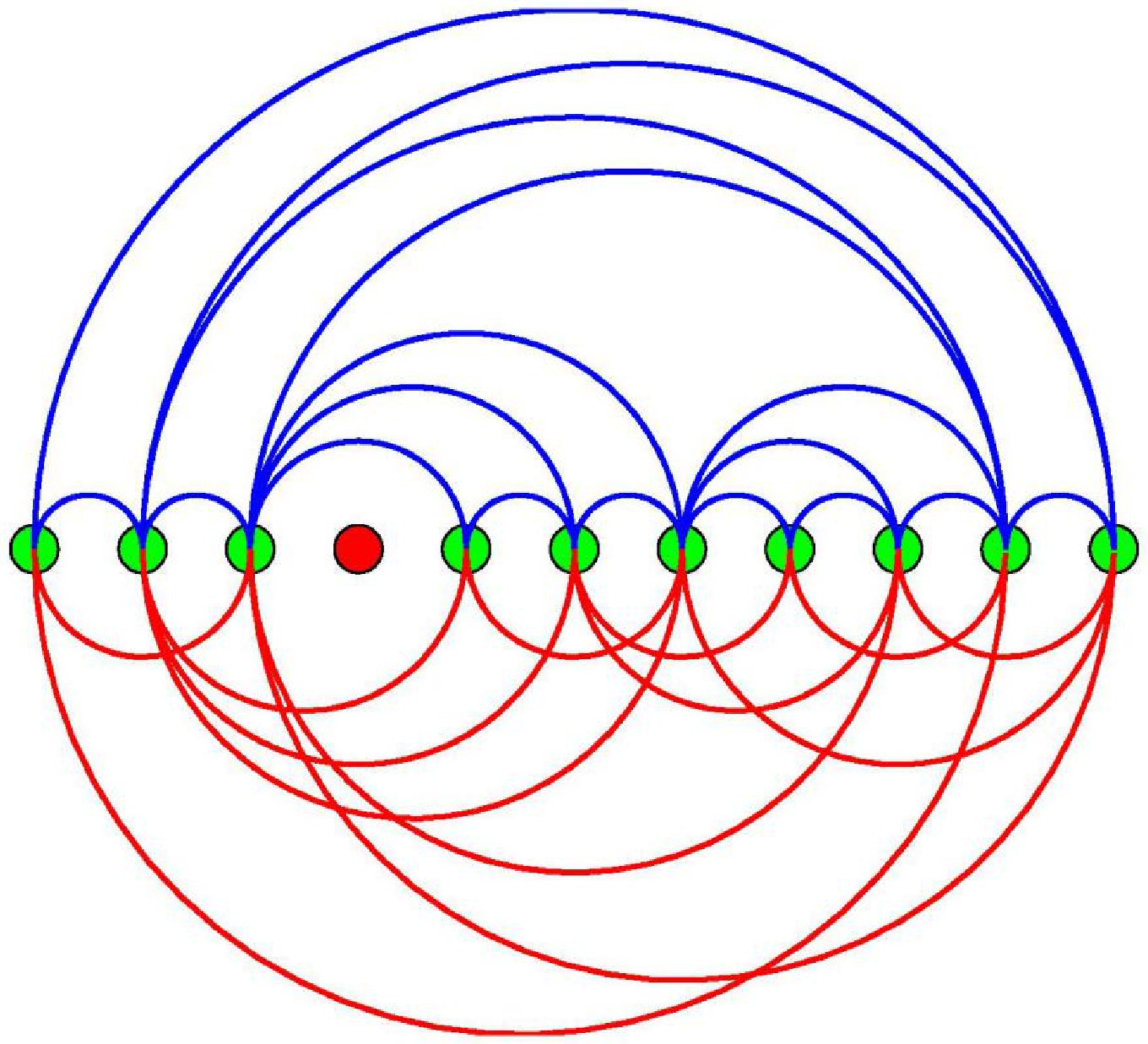}}
\scalebox{0.20}[0.18]{\includegraphics{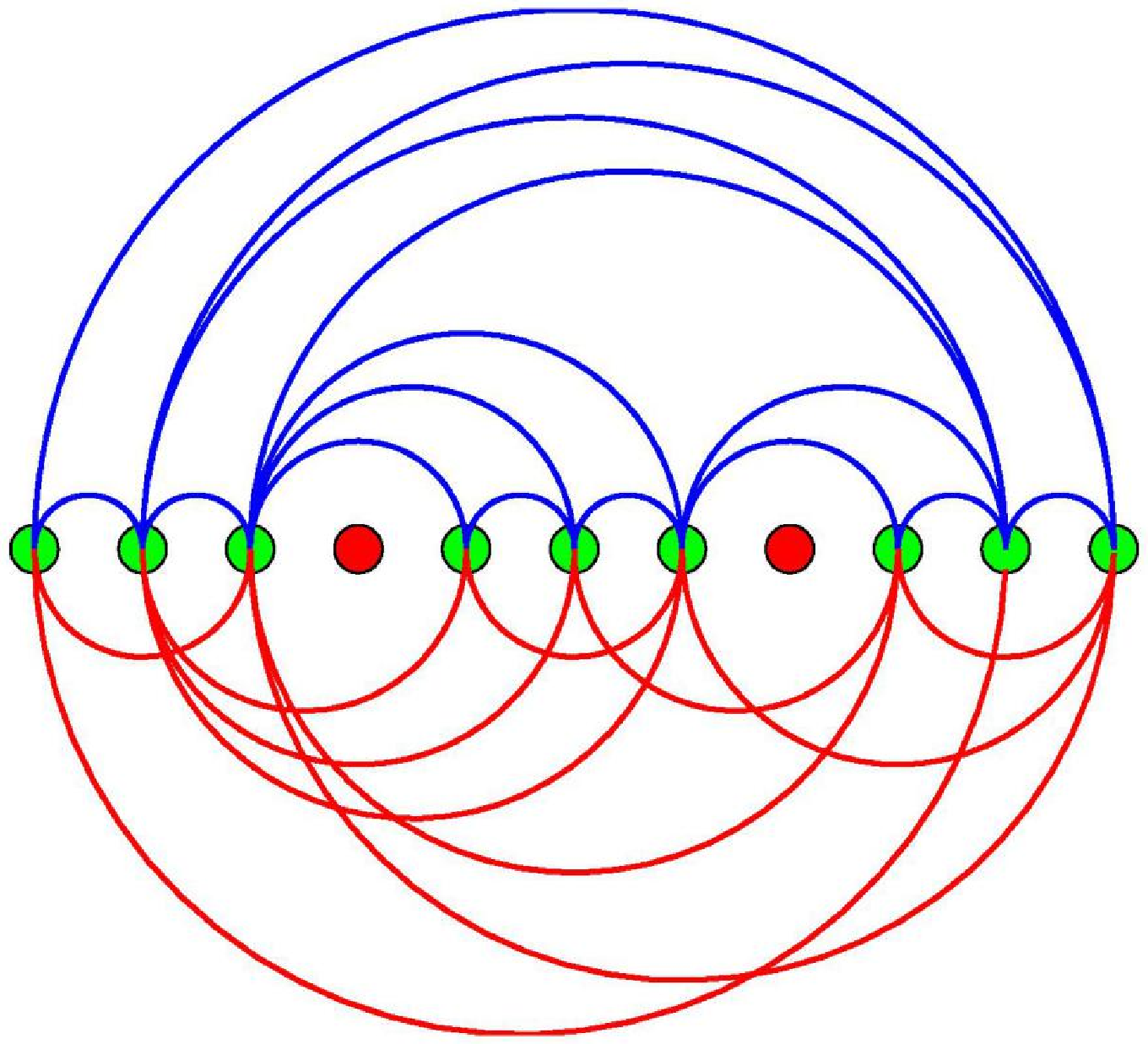}}
\scalebox{0.20}[0.18]{\includegraphics{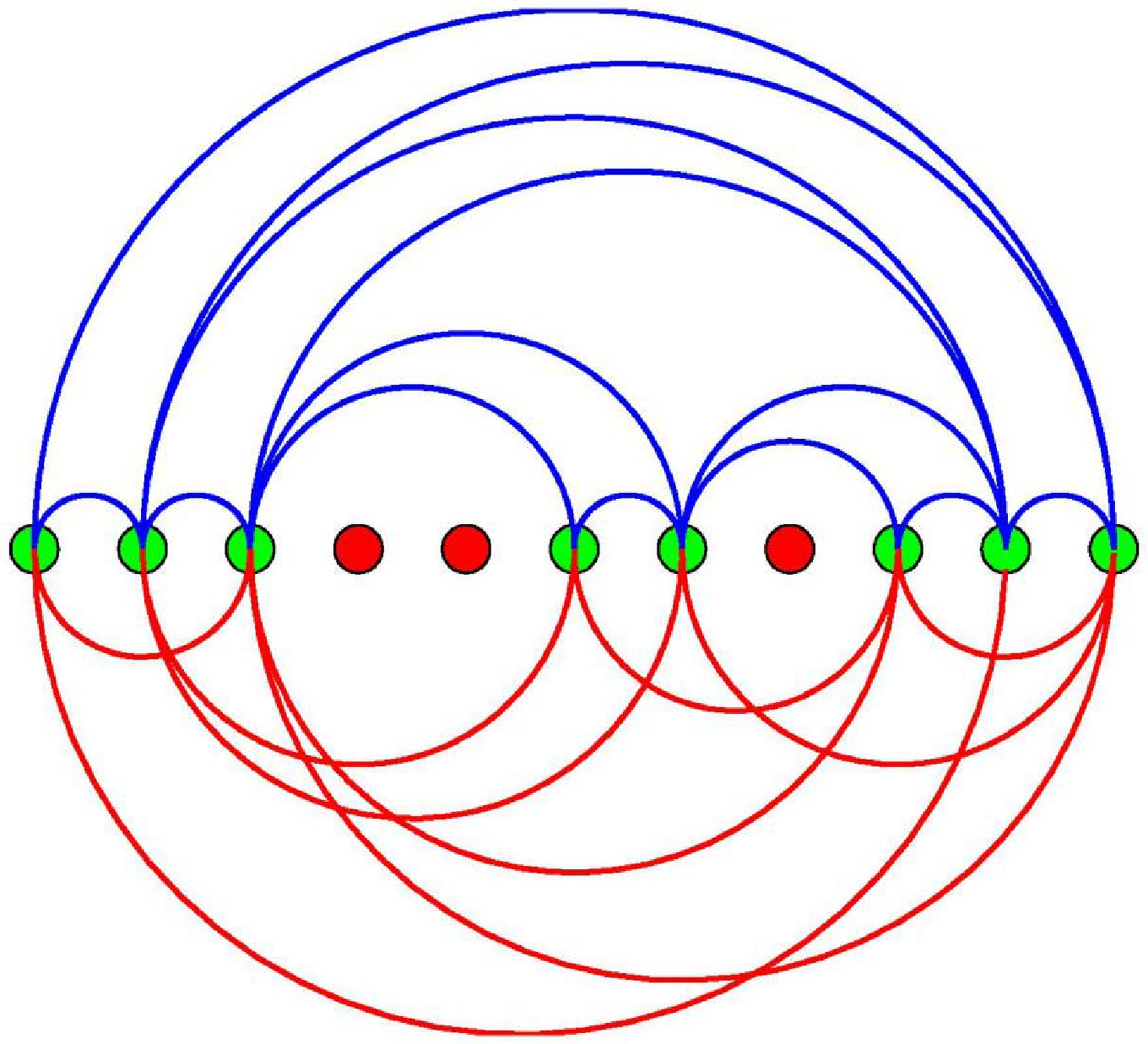}}\\
\scalebox{0.20}[0.18]{\includegraphics{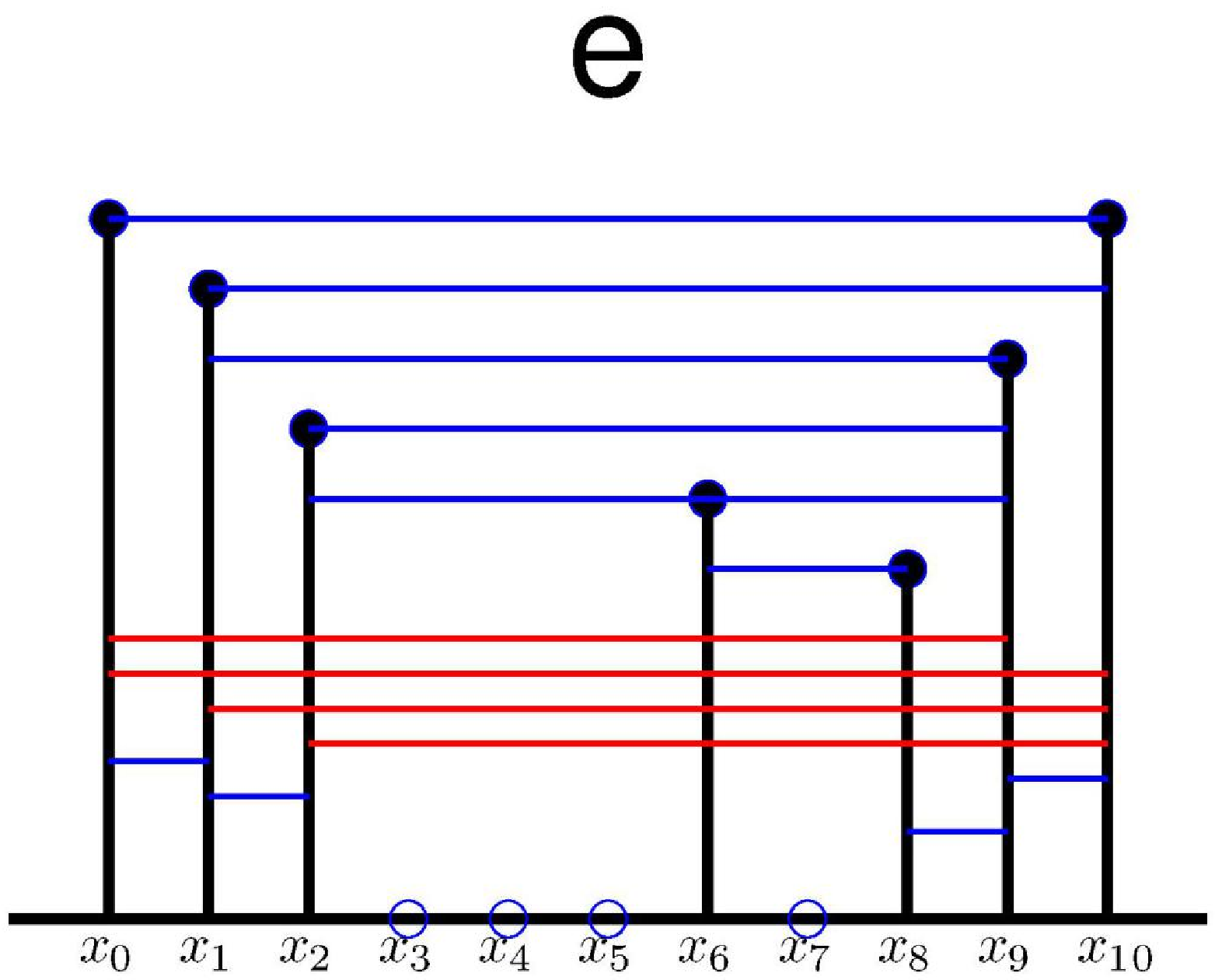}}
\scalebox{0.20}[0.18]{\includegraphics{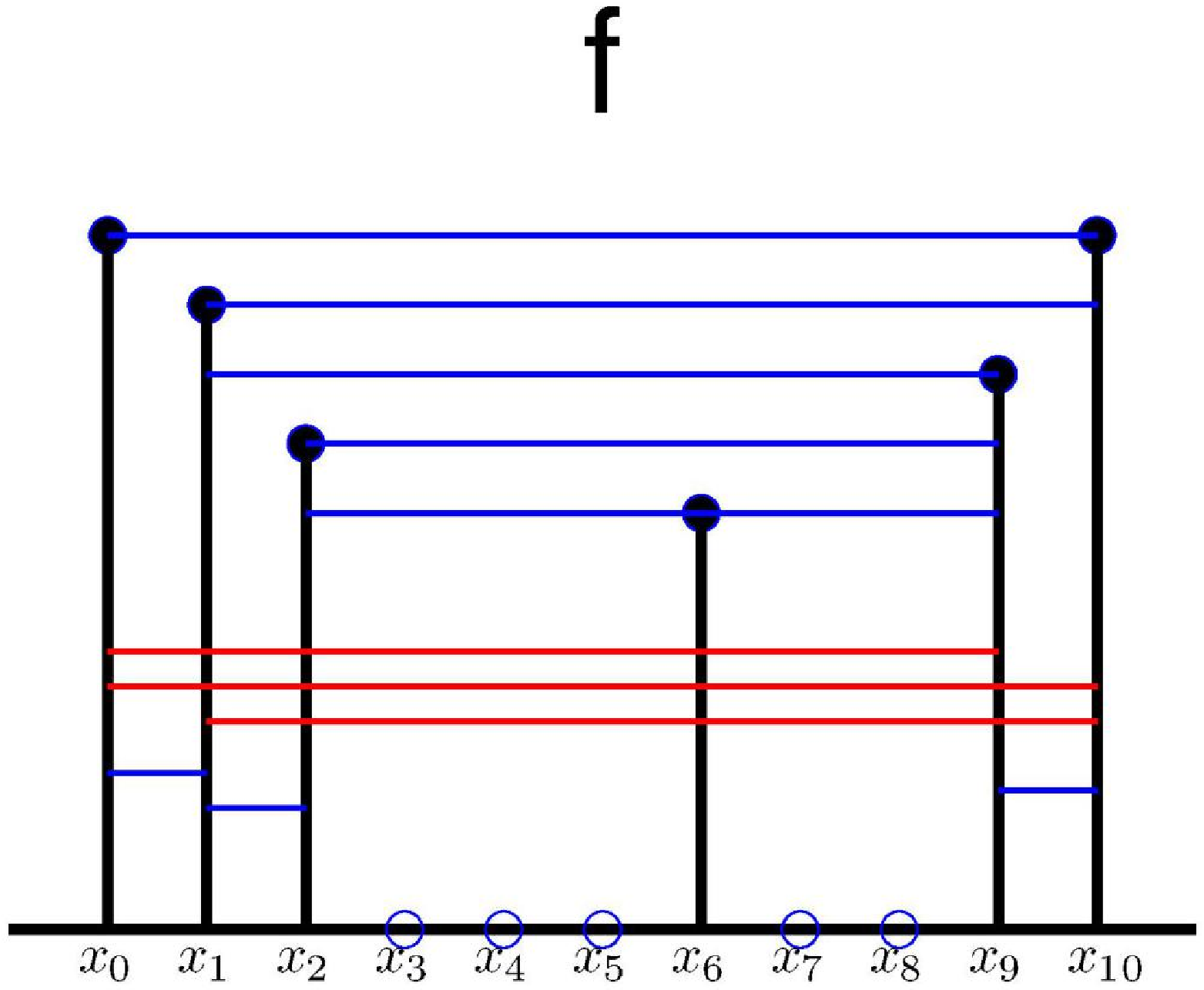}}
\scalebox{0.20}[0.18]{\includegraphics{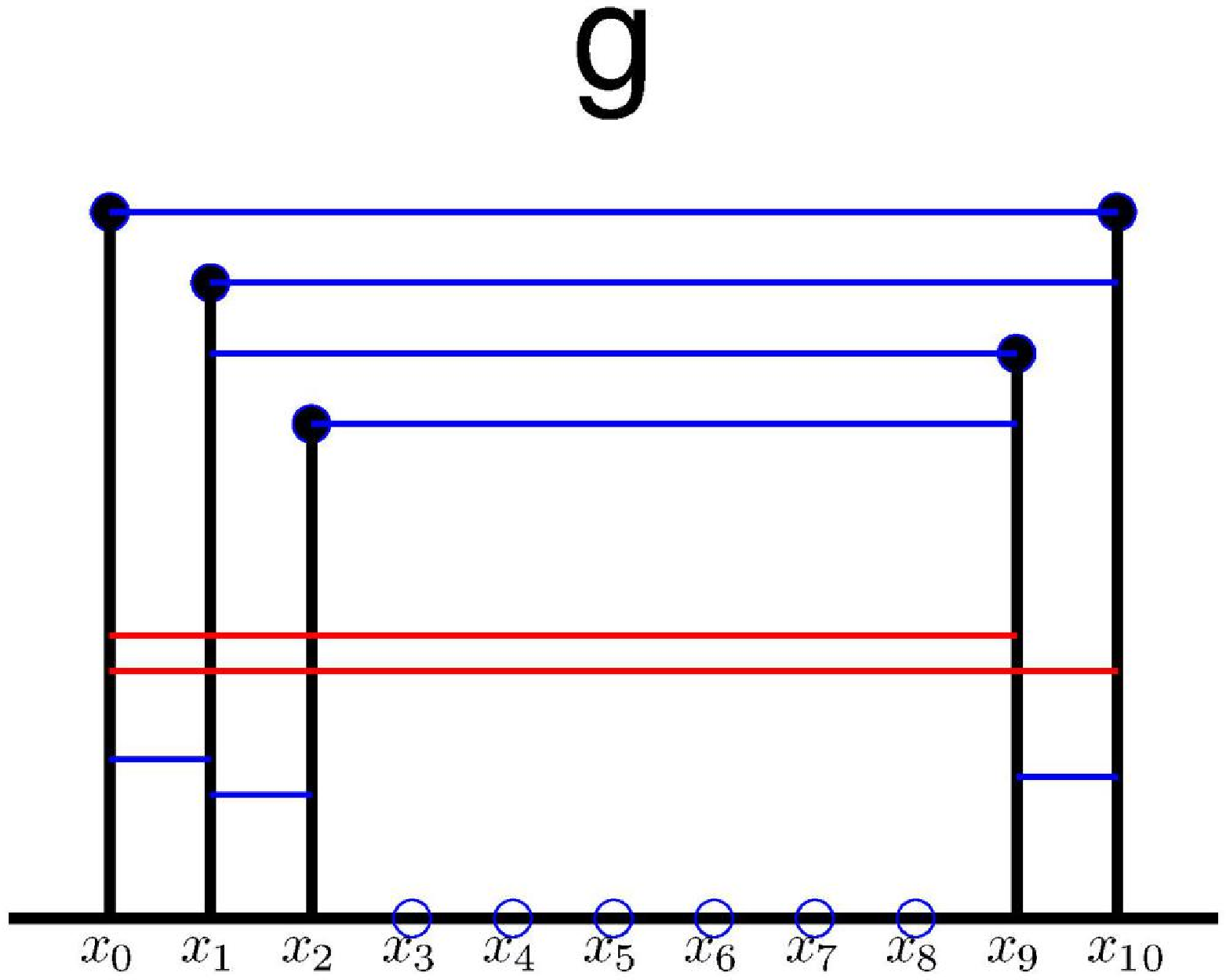}}
\scalebox{0.20}[0.18]{\includegraphics{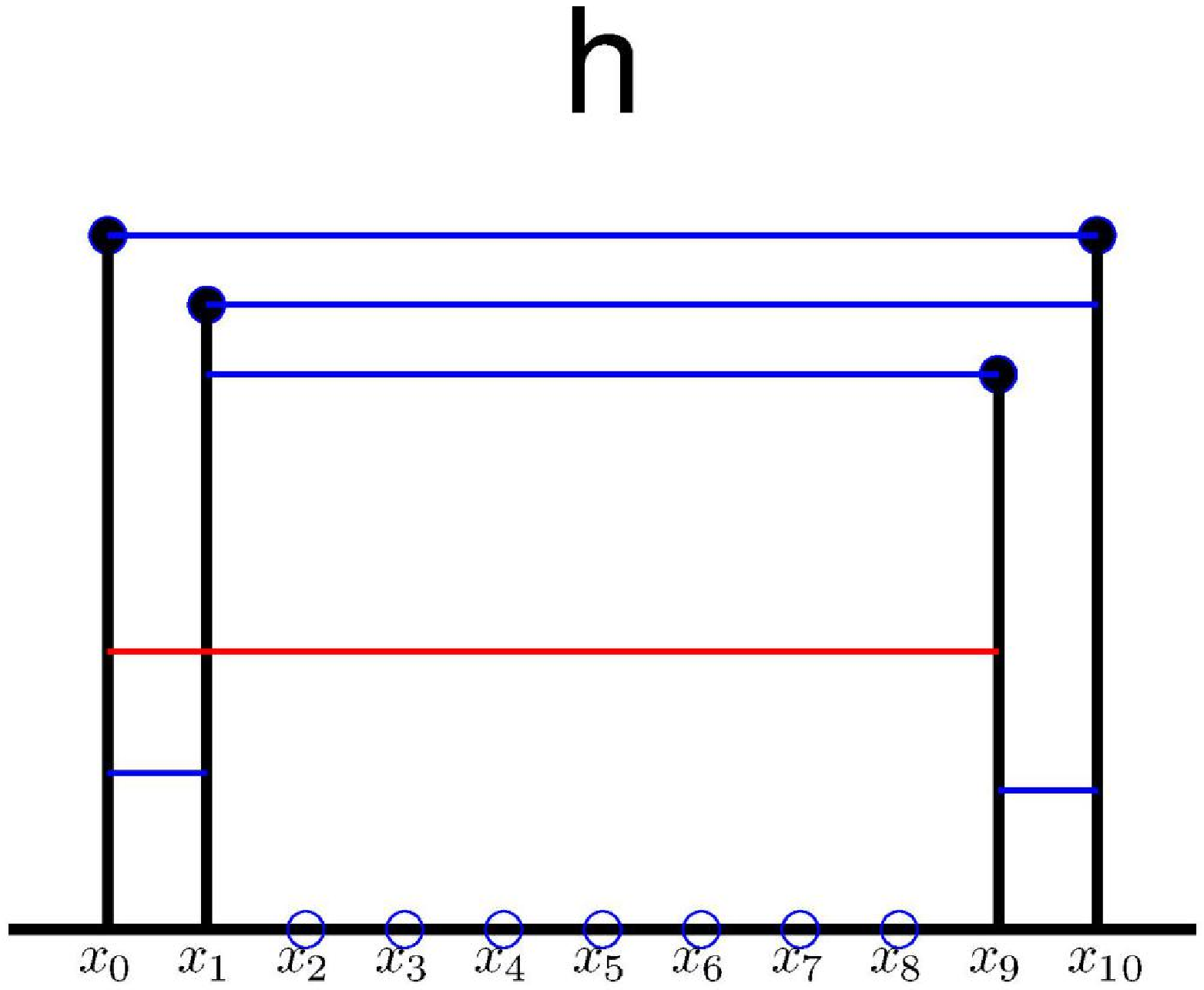}}\\
\scalebox{0.20}[0.18]{\includegraphics{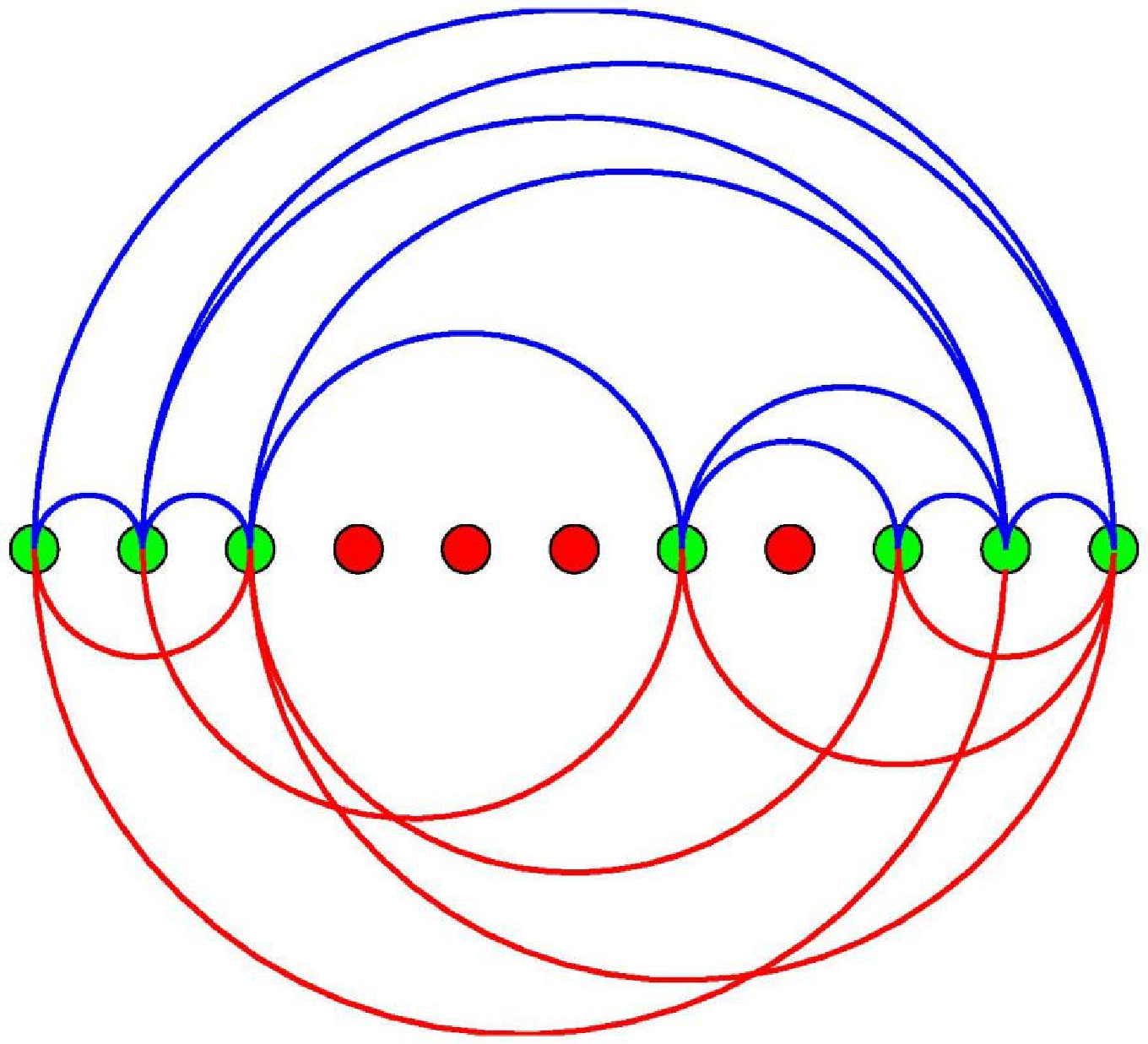}}
\scalebox{0.20}[0.18]{\includegraphics{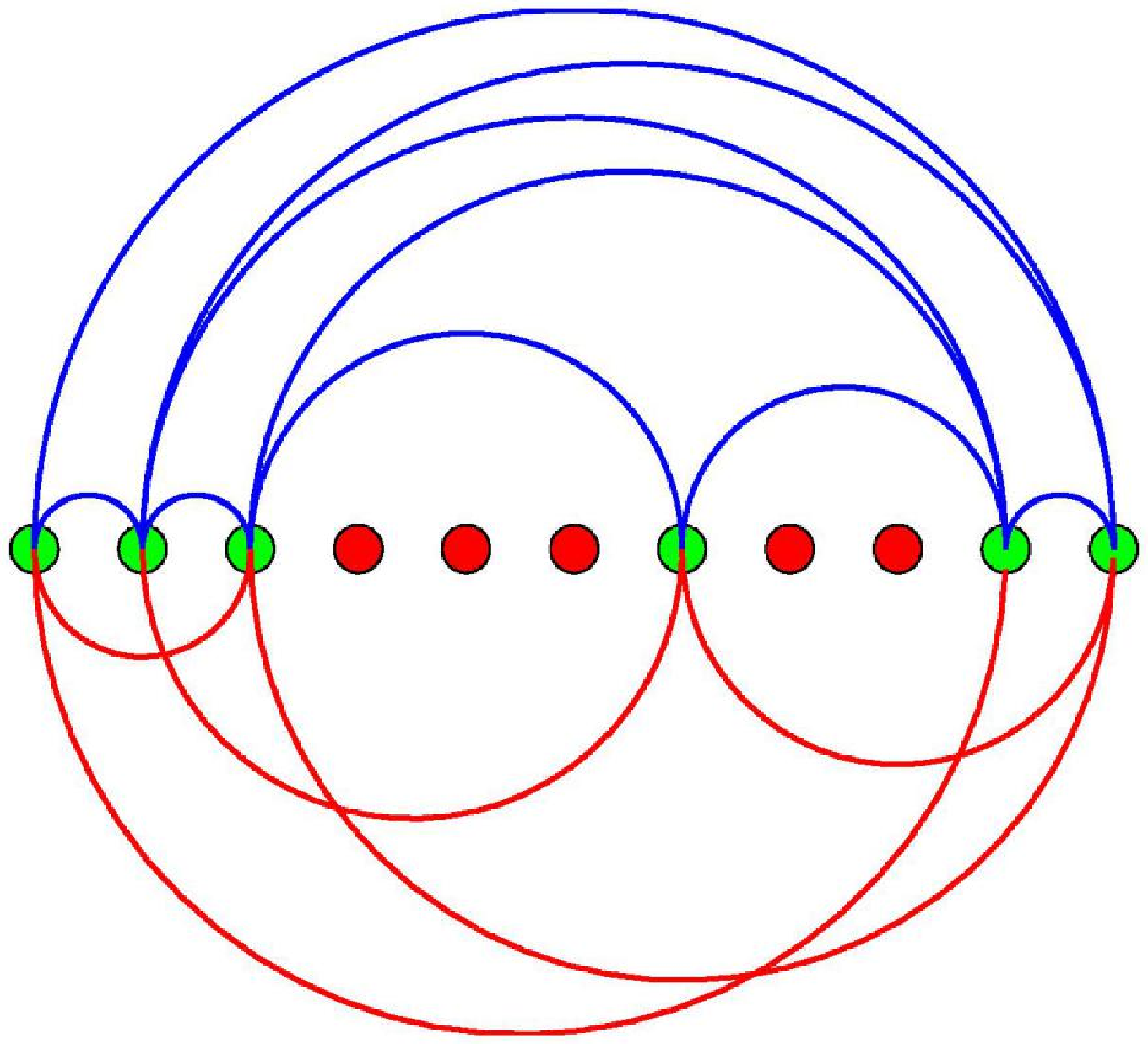}}
\scalebox{0.20}[0.18]{\includegraphics{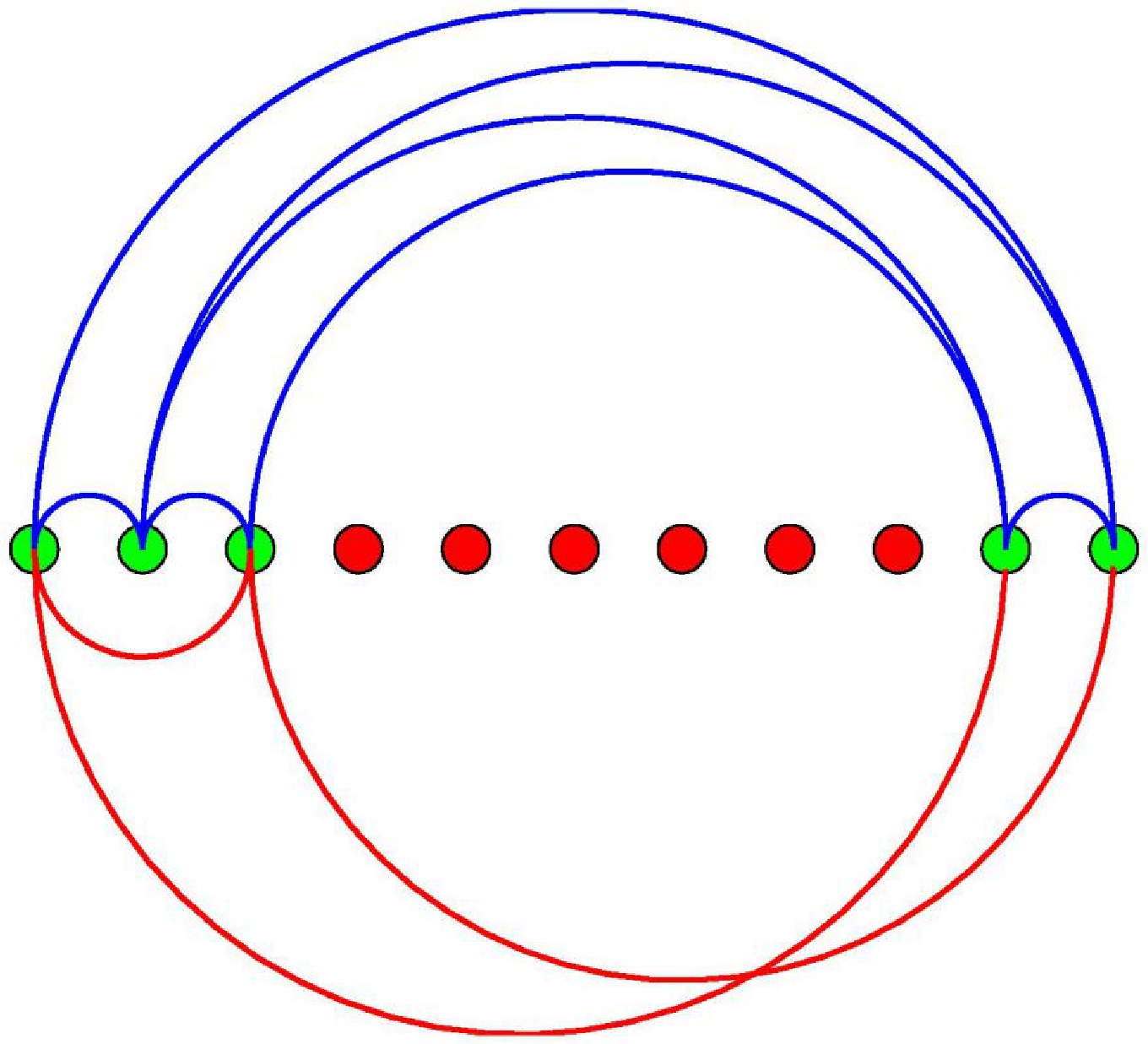}}
\scalebox{0.20}[0.18]{\includegraphics{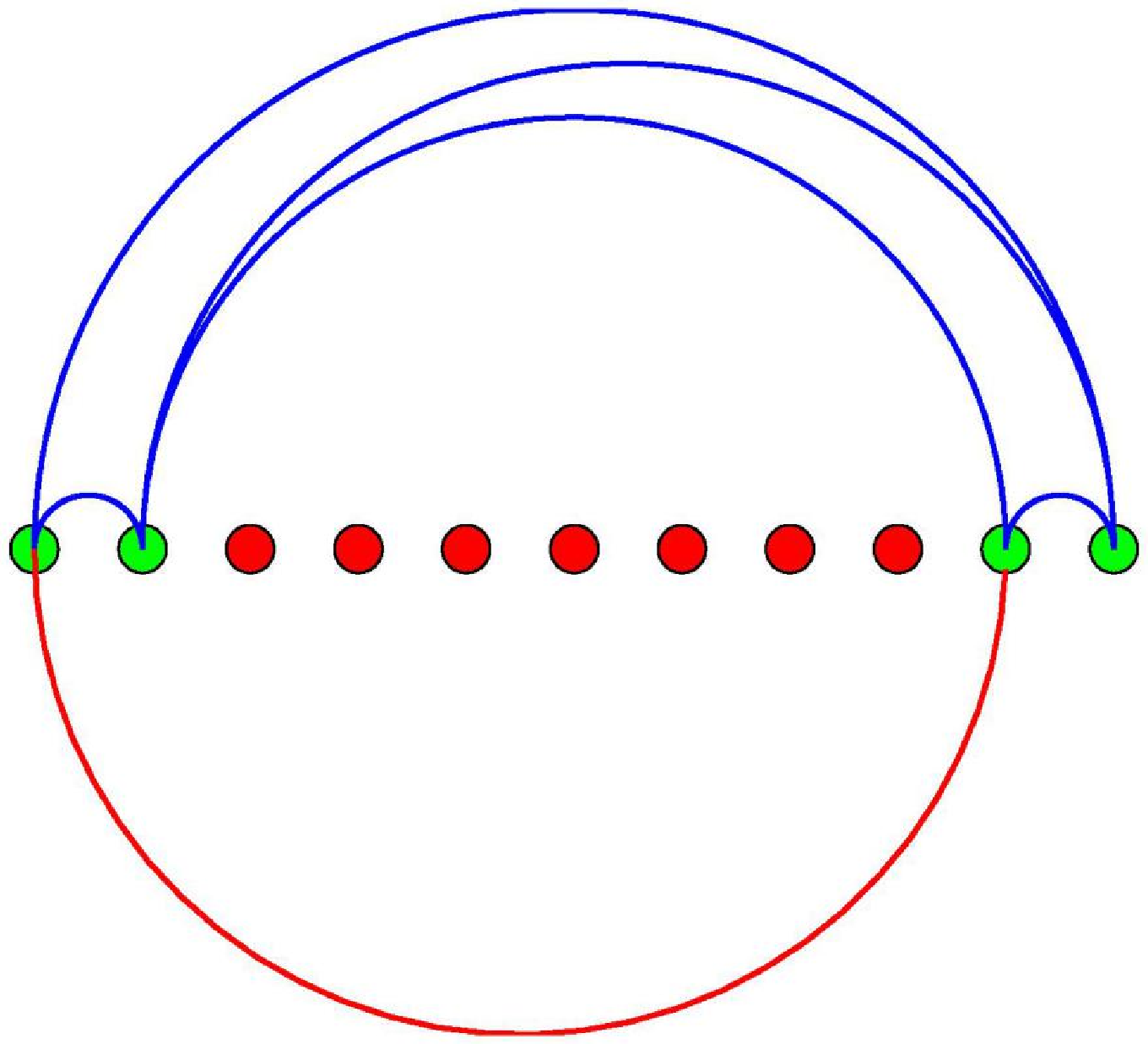}}\\

\end{figure}

\begin{figure}[H]
\caption{\emph{Graphical illustration of the constructive proof of
    $<k(T)>$, considering a LPHVG with $\rho=1$ extracted from a
    periodic series of period $T=10$}.}
\end{figure}

To check the accuracy of our analytical result, we generate simple
period-50, period-100, period-200, and period-250 time series with 1000
data points [see Fig.~3(a)]. We construct the limited penetrable
horizontal visibility graphs with the penetrable distance $\rho =
0,1,2,...,10$ associated with this periodic time series. Fig.~3(b)
shows a plot of the mean degree of the resulting LPHVGs with different
$\rho$ values, and there is an excellent agreement with the numerics
$\rho \ll T$.

\begin{figure}[H]
\centering \scalebox{0.4}[0.4]{\includegraphics{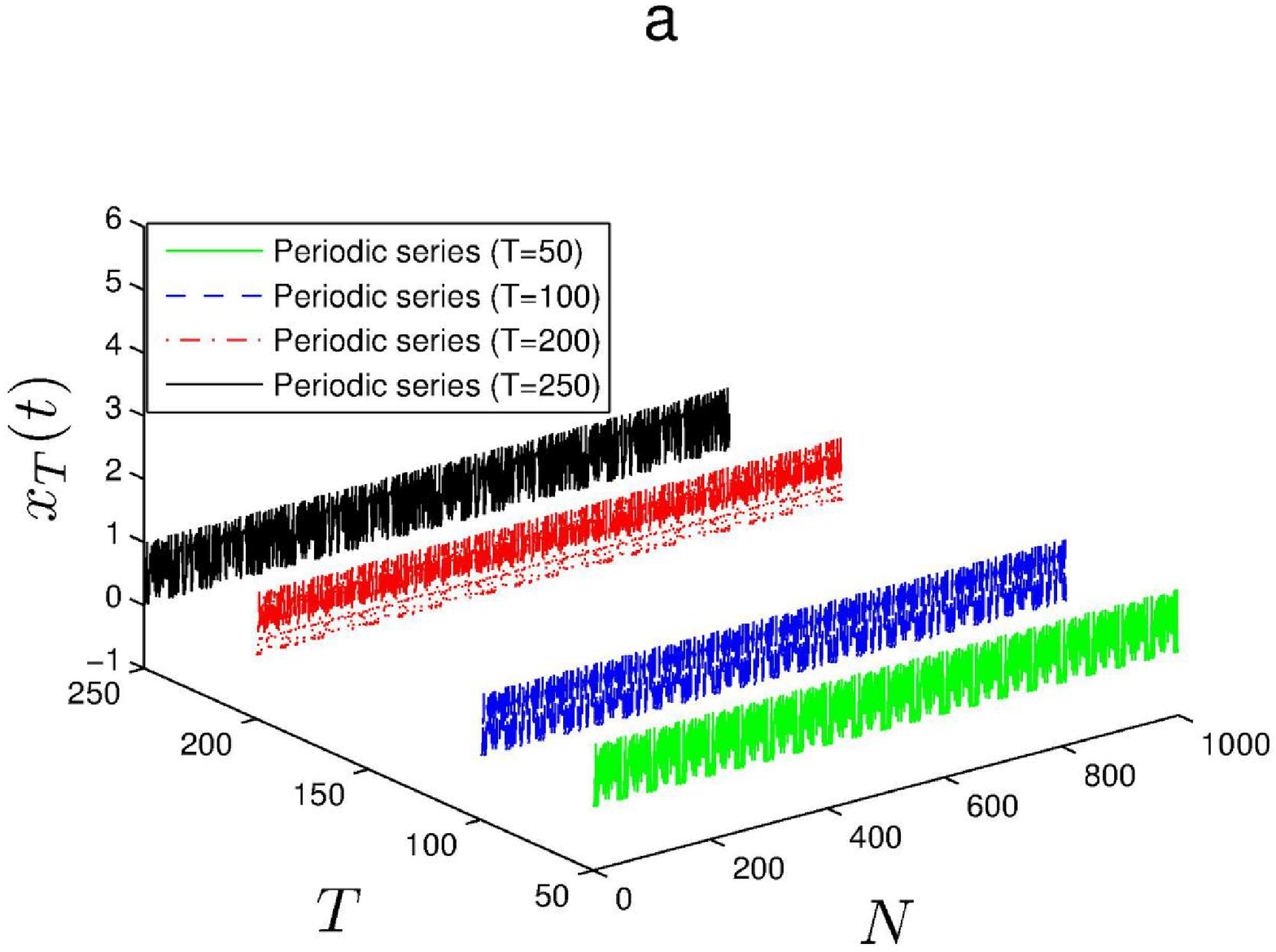}}
\scalebox{0.4}[0.4]{\includegraphics{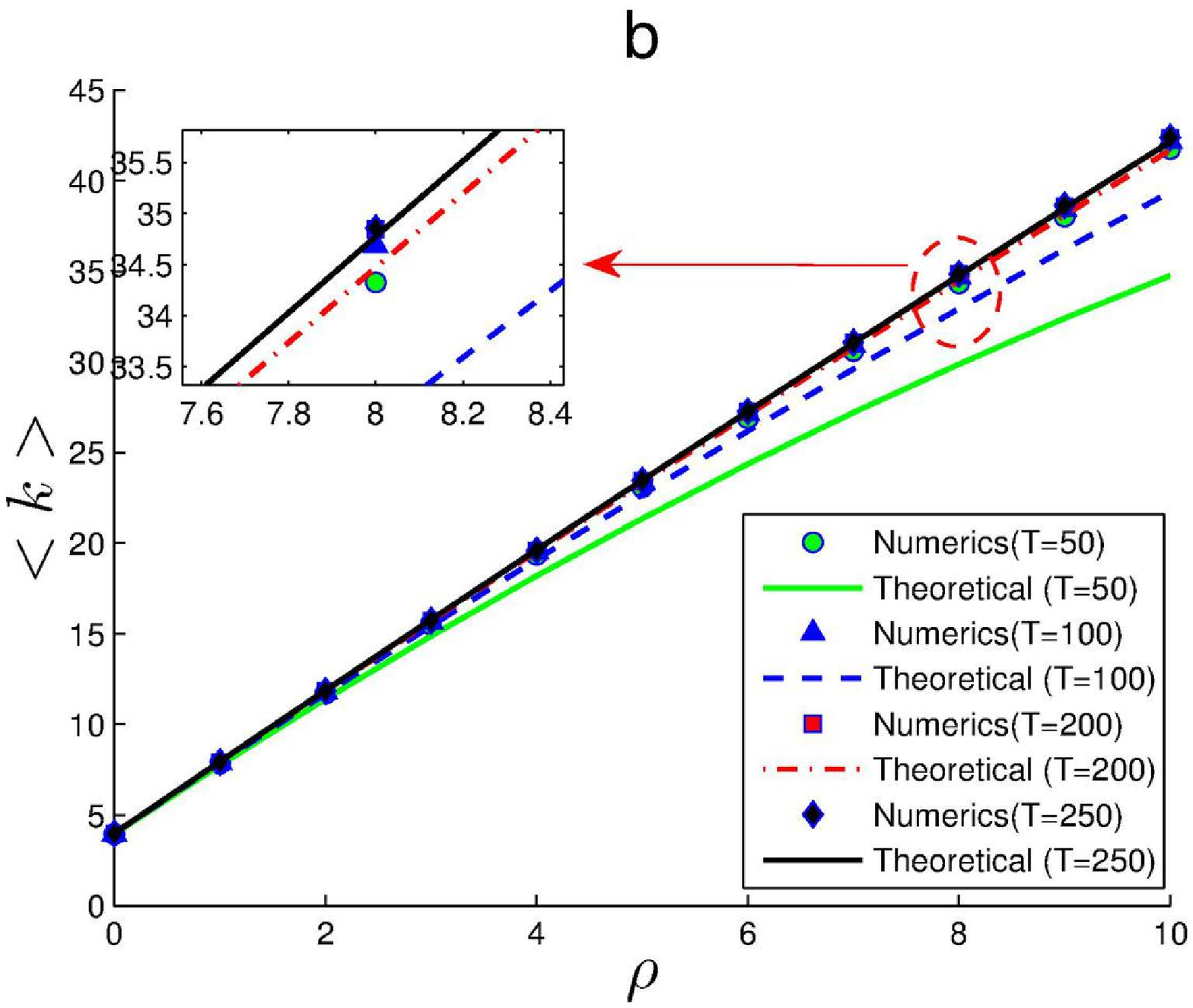}}
\end{figure}
\begin{figure}[H]
\caption{\emph{(a) A simplified period-50, period-100, period-200 and
    period-250 time series of 1000 data, (b) plotted the mean degree of
    the resulting LPHVGs with different $\rho$ (circles correspond to
    the periodic-50, triangles correspond to the periodic-100, squares
    correspond to the periodic-200 and diamonds correspond to the
    periodic-250 time series, the black, blue, red and green solid line
    correspond to the theoretical result respectively.}}
\end{figure}

\textbf{Local clustering coefficient.} In LPHVG, a given nodes with the same degree usual have the different clustering coefficients since the degree of the node contributed by the different configurations which have different structures (see the proof process of Theorem S1). By calculating the clustering coefficients of different configurations (see Theorem S3), we find that the clustering coefficients of the nodes in LPHVG are irregular, but the minimum clustering coefficient and the maximum clustering coefficient of these nodes are regular. Therefore, Based on the results of degree distribution [Eq. (2)], we can deduce the minimum local clustering coefficient $C_{min}(k)$ and the maximum clustering coefficient $C_{max}(k)$ of LPHVG associated to $i.i.d.$ random series by the following expression,

\begin{equation}\label{eq4}
C_{\rm min}(k) = \frac{2}{k}+\frac{2\rho(k-2)}{k(k-1)},\rho =
0,1,2,k\geq 2(\rho+1),
\end{equation}
\begin{equation}\label{eq5}
C_{\rm max}(k) = \frac{2}{k}+\frac{4\rho(k-3)}{k(k-1)},\rho =
0,1,2,k\geq 2(2\rho+1).
\end{equation}

Using Eqs.~(2), (6), and (7) we also obtain the local clustering
coefficient distribution $P(C_{\rm min})$ and $P(C_{\rm max})$, i.e.,
\begin{equation}
\begin{array}{l}
P(C_{\rm min}) =
\frac{1}{2\rho+3}exp\{[\frac{\varphi+\sqrt{\varphi^2-8C_{\rm
        min}(2\rho+1)}}{2C_{\rm
      min}}-2(\rho+1)]ln(\frac{2\rho+2}{2\rho+3})\},\varphi=C_{\rm
  min}+2\rho+2,
\end{array}
\end{equation}
\begin{equation}
\begin{array}{l}
P(C_{\rm max}) = \frac{1}{2\rho+3}exp\{[\frac{\phi+\sqrt{\phi^2-8C_{\rm
        max}(6\rho+1)}}{2C_{\rm
      max}}-2(\rho+1)]ln(\frac{2\rho+2}{2\rho+3})\},\phi = C_{\rm
  max}+4\rho+2.
\end{array}
\end{equation}
For a proof of this result see Theorem S3 in the Appendix. Fig.~4
shows the clustering coefficient $C_{k}$ and the clustering coefficient
distribution $P(C)$ of limited penetrable horizontal visibility graphs
associated with different random series of 3000 data points obtained
numerically. The solid black line in Figs.~4(a) and 4(b) is the
theoretical prediction of $C_{\rm min}(k)$ [see Eq.~(6)], and the solid
red line is the theoretical prediction of $C_{\rm max}(k)$ [see
  Eq.~(7)]. The solid black line in Figs.~4(c) and 4(d) is the
theoretical prediction of $P(C_{\rm min})$ [see Eq.~(8)], and the solid
red line is the theoretical prediction of $P(C_{\rm max})$ [see
  Eq.~(9)]. Fig.~4 shows that the theoretical predictions of $C_{\rm
  min}(k)$, $C_{\rm max}(k)$, $P(C_{\rm min})$, and $P(C_{\rm max})$
agree with the numerics.

\begin{figure}[H]
\centering \scalebox{0.4}[0.4]{\includegraphics{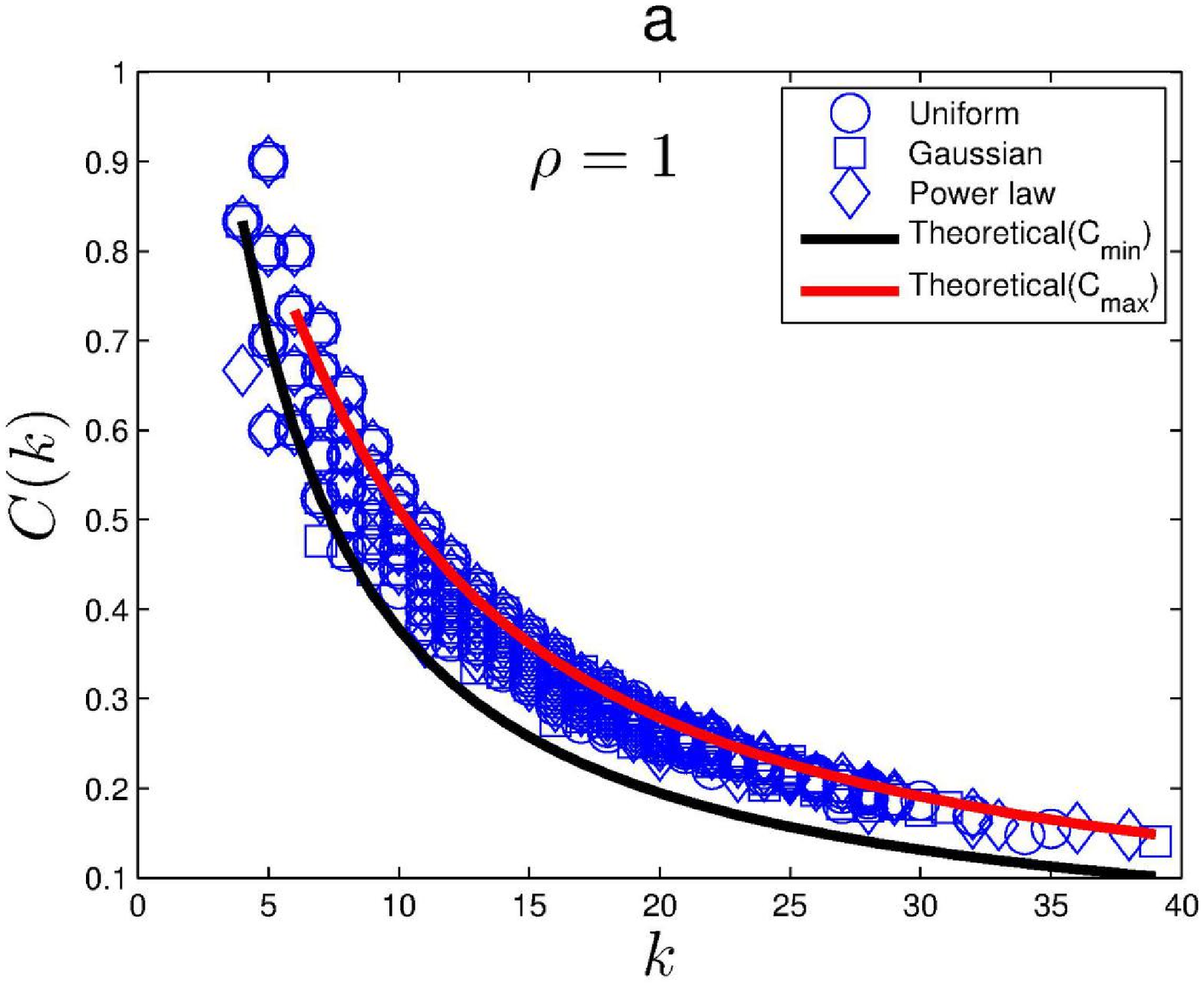}}
\scalebox{0.4}[0.4]{\includegraphics{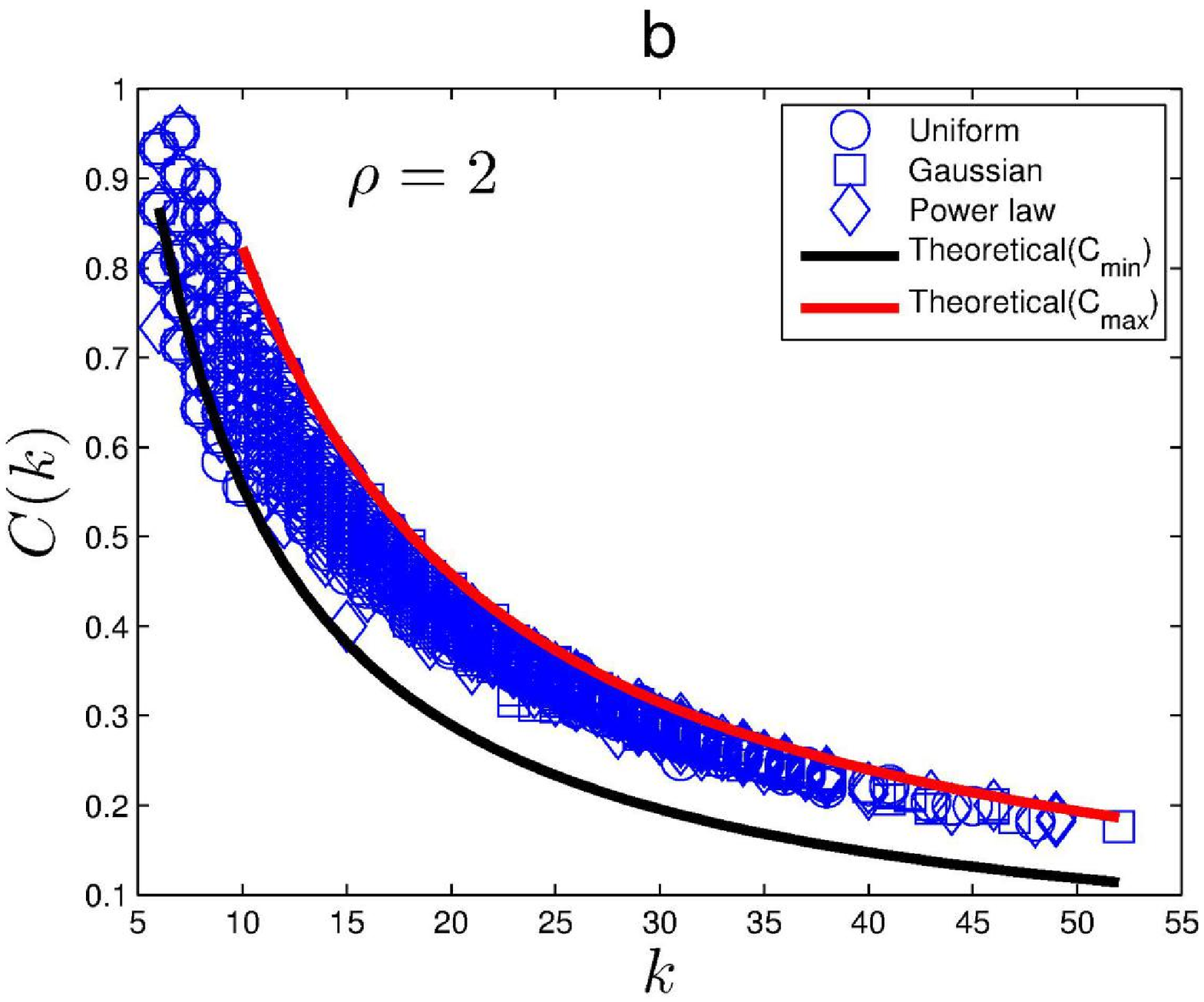}}
\scalebox{0.4}[0.4]{\includegraphics{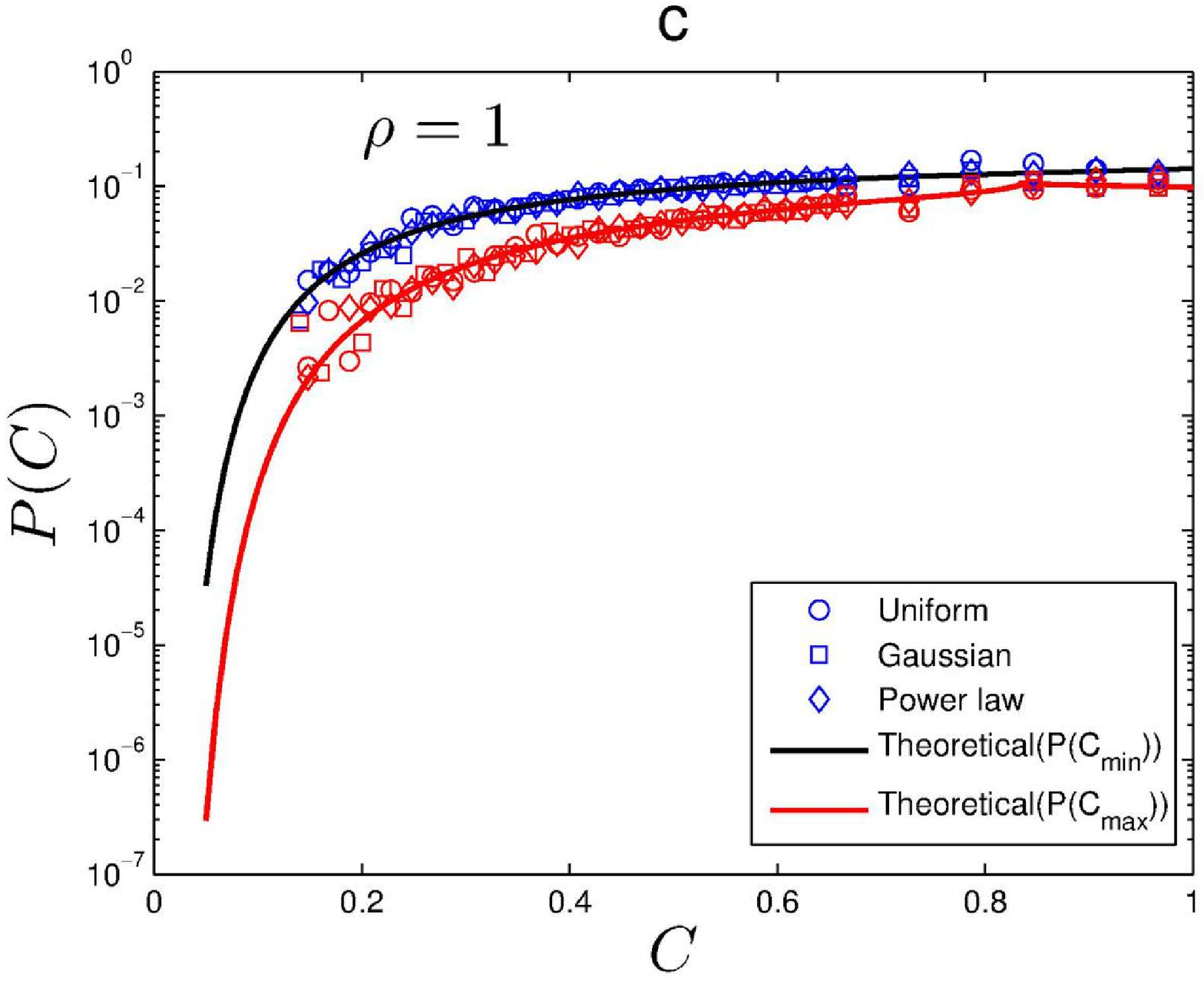}}
\scalebox{0.4}[0.4]{\includegraphics{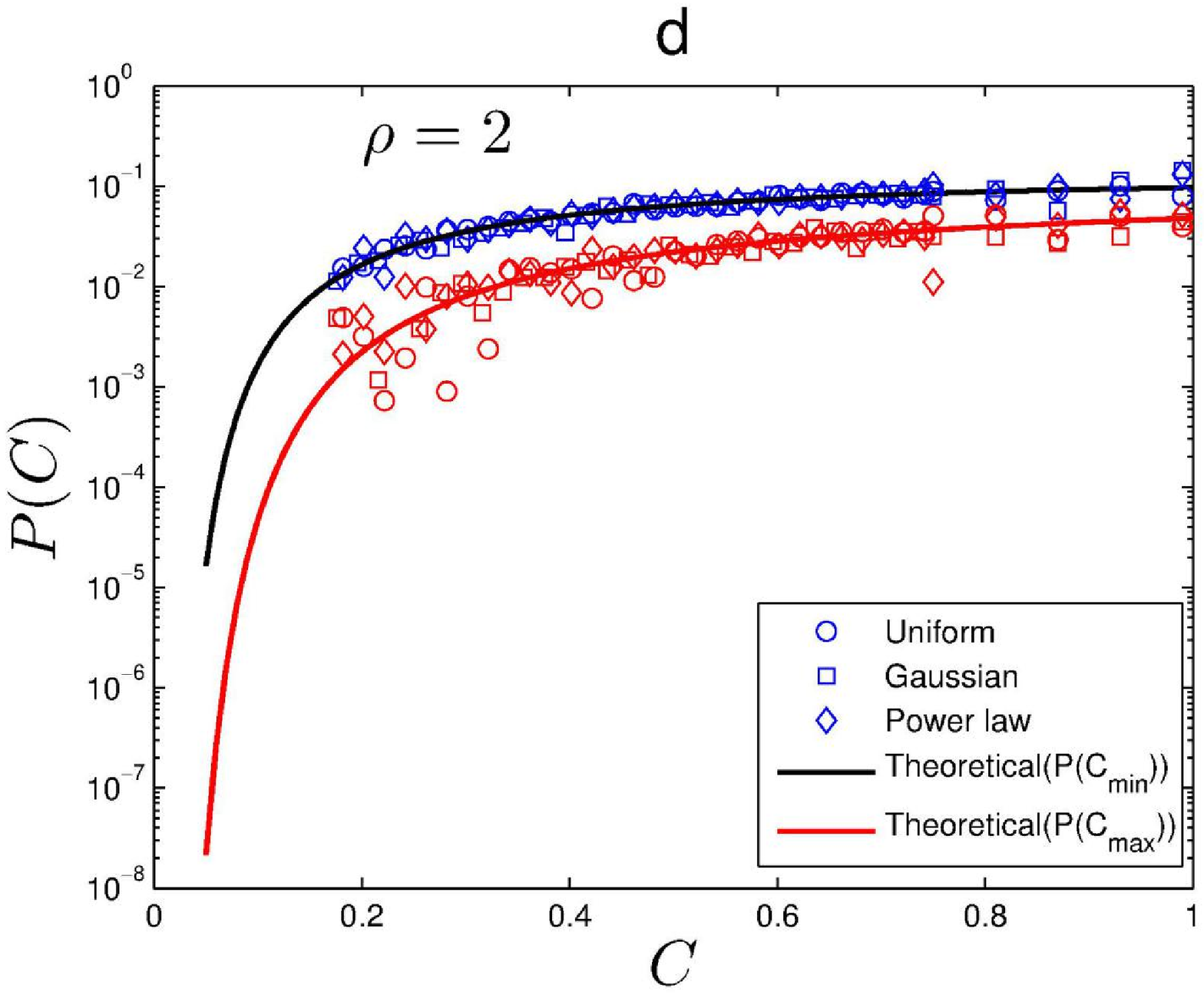}}
\end{figure}
\begin{figure}[H]
\caption{\emph{The relationship between degree and clustering
    coefficient of LPHVG (a) $\rho=1$, (b) $\rho=2$. The solid black
    line corresponds to the theoretical prediction of $C_{\rm min}(k)$
    [Eq.(6)], the solid red line corresponds to the theoretical
    prediction of $C_{\rm max}(k)$ [Eq. (7)]. The clustering coefficient
    distribution (c) $\rho=1$, (d) $\rho=2$. The solid black line
    corresponds to the theoretical prediction of $P(C_{\rm min})$
    [Eq. (8)], the solid red line corresponds to the theoretical
    prediction of $P(C_{\rm max})$ [Eq. (9)]. }}
\end{figure}

\textbf{Long distance visibility.} In a limited penetrable horizontal
visibility graph associated with a bi-finite sequence of $i.i.d.$ random variables extracted from a continuous
probability density $f(x)$, the probability $P_{\rho}(n)$ that two data
points separated by $n$ intermediate data points are connected is
\begin{equation}\label{eq6}
P_{\rho}(n) = \frac{2\rho(\rho+1)+2}{n(n+1)}.
\end{equation}

Note that $P_{\rho}(n)$ is again independent of the probability
distribution of the random variable $X$. For a detailed proof of this
result see Theorem S4 in the Appendix. Fig.~5(a) shows the adjacency
matrix $\textbf{A}$ of the limited penetrable horizontal visibility
graph associated with a random series with a different limited
penetrable distance. When $A(i,j) = 1$, we plot $\rho = 0$ (circle),
$\rho=1$ (triangle), $\rho=2$ (square), and $\rho=3$ (diamond)  at
$(i,\rho,j)$ and $(j,\rho,i)$.

\begin{figure}[H]
\centering \scalebox{0.4}[0.4]{\includegraphics{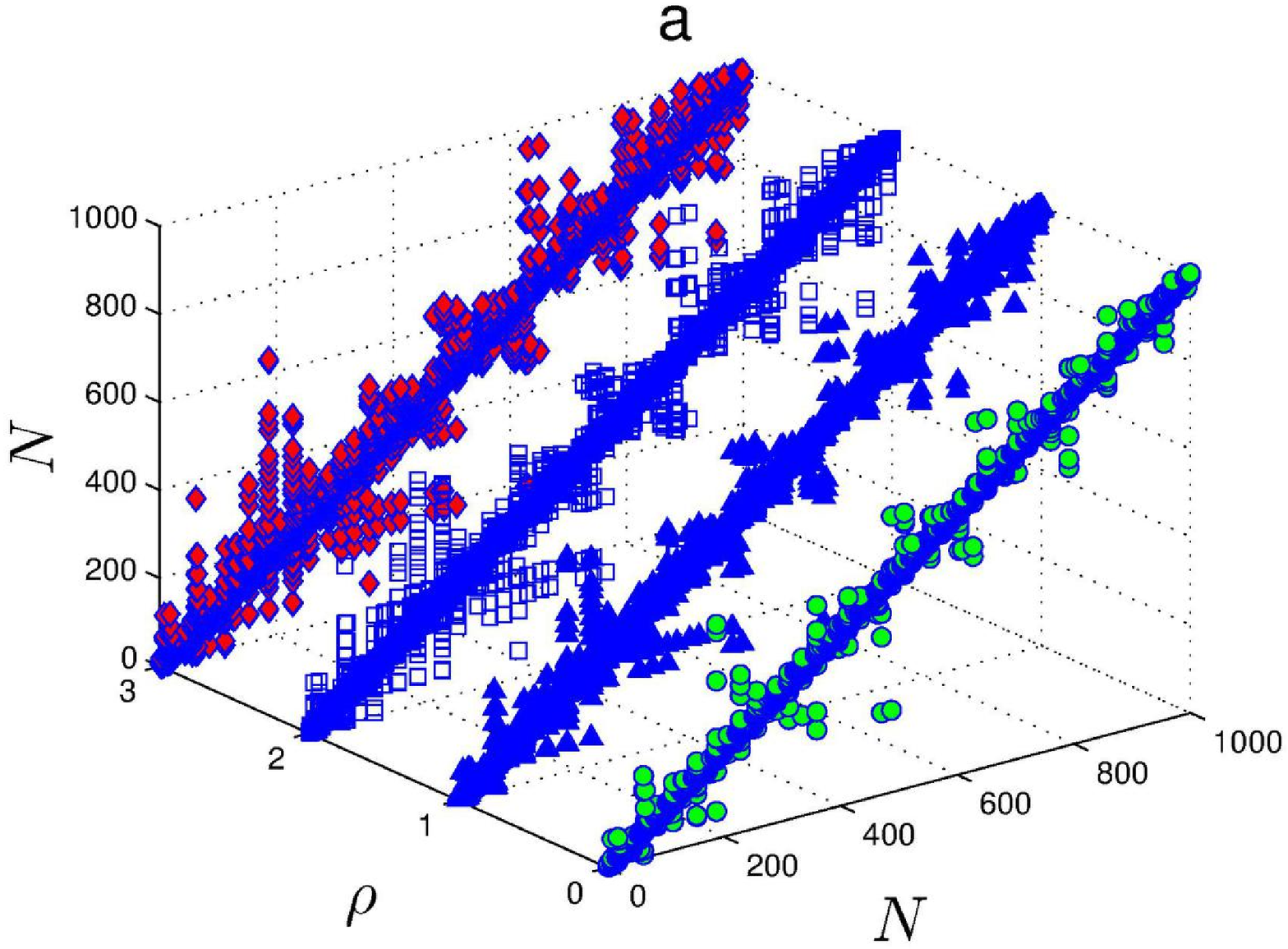}}
\scalebox{0.4}[0.4]{\includegraphics{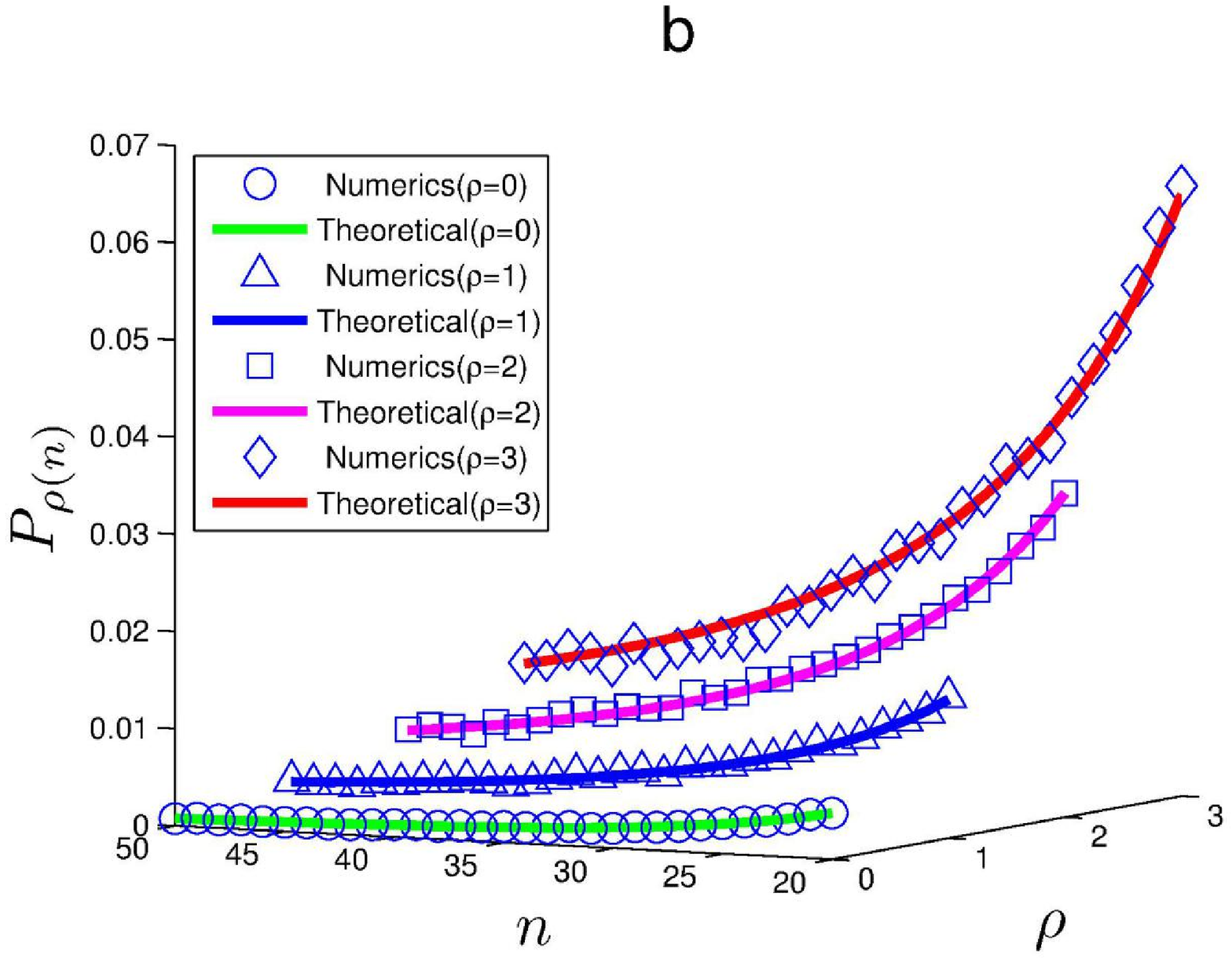}}
\end{figure}
\begin{figure}[H]
\caption{\emph{(a) Adjacency matrix of LPHVG associated to a random
    series with different $\rho$, (b) plot of the relationship of
    $\rho$,$n$ and $P_{\rho}(n)$ (the solid line correspond to the
    theoretical result [Eq. (10)], circles correspond to the numerical
    simulation result for $\rho=0$, triangles correspond to the
    numerical simulation result for $\rho=1$, squares correspond to the
    numerical simulation result for $\rho=2$, diamond correspond to the
    numerical simulation result for $\rho=3$.}}
\end{figure}

Fig.~5(a) shows a typical homogeneous structure in which the adjacency
matrix is filled around the main diagonal. In addition, the matrix
indicates a superposed sparse structure caused by the limited
penetrable visibility probability
$P_{\rho}(n)=\frac{2\rho(\rho+1)+2}{n(n+1)}$ that introduces shortcuts
into the limited penetrable horizontal visibility graph. These shortcuts
indicate that the limited penetrable horizontal visibility graph is a
small-world phenomenon. Fig.~5(b) shows that the theoretical result in
Eq.~(10) agrees with the numerics.

These results are exact with regard to the topological properties of the
limited penetrable horizontal visibility graphs associated with
$i.i.d.$ random series via the limited
penetrable horizontal visibility algorithm.

\textbf{Application to deterministic chaotic time series.} These results
can be used to discriminate between random and chaotic signals. Because
stochastic and chaotic processes share many features, discriminating
between them is difficult, and methods of identifying random processes
and discriminating between deterministic chaotic systems and stochastic
processes has received extensive study in recent decades [31--34]. Most
previous algorithms have been phenomenological and computationally
complicated. Thus new methods that can reliably distinguish stochastic
from chaotic time series are needed. Recently Lacasa et al. [11--12]
used the horizontal visibility algorithm to characterize and distinguish
between stochastic and chaotic processes, and they demonstrated that it
could easily distinguish chaotic from random series.  Here we use our
new theory to distinguish chaotic series from random series and compare with the horizontal visibility algorithm [11], and we
address four deterministic time series generated by the Logistic map
[35]
$$x_{t+1} = \mu x_{t}(1-x_{t}),\mu = 4,$$
the H$\acute{e}$non map [36],
$$x_{t+1} = 1+y_{t}-ax_{t}^{2},y_{t+1} = bx_{t},a = 1.4,b = 0.3,$$
the Lorenz chaos system [37],
$$\dot{x} = a(y-x),\dot{y} = cx-y-xz, \dot{z} = xy-bz, a = 10, b = 8/3,
c = 28,$$
and the energy price-supply-economic growth system [27],
$\dot{x} =
a_{1}x+a_{2}(C-y)+a_{3}(z-K_{1}),\dot{y}=-b_{1}y+b_{2}x-b_{3}z(1-z/K_{2}),\dot{\
z}=c_{1}z(1-z/L)+C_{2}yz
a_{1}=0.3,C=27,a_{2}=0.5563,a_{3}=0.15,b_{1}=0.4,b_{2}=0.6073,b_{\
3}=0.3,K_{1}=15,K_{2}=15,c_{1}=0.3,c_{2}=0.006,L=19.$

Fig. 6 shows the limited penetrable horizontal visibility graphs of
3000 data points extracted from two different chaotic maps and two
different chaotic system with $\rho=0$ to the left, $\rho=1$ in the
middle, and $\rho=2$ to the right. We calculate their degree
distribution numerically (top panel) and the relationship between degree
and clustering coefficient (bottom panel). In every case $P(k)$ deviates
from Eq.~(2) and $C(k)$ deviates from Eqs.~(6) and (7). We also find
that the degree distributions of the LPHVGs associated with these
chaotic maps and chaotic systems can be approximated using the
exponential function $P(k)\sim exp(-\hat{\lambda})k$, but
$\hat{\lambda}\neq \lambda = ln[(2\rho+3)/(2\rho+2)]$ in each case, and we
conjecture that there is a functional relationship between the random
and chaos dimensions [11]. Thus the parameter $\lambda =
ln[(2\rho+3)/(2\rho+2)]$ is the frontier between random series and chaotic
series and serves to distinguish randomness from chaos. Fig.~6 shows
(bottom panel) that the limited penetrable horizontal visibility graphs
of $\rho=1$ in the middle and $\rho=2$ on the right serve as better
discriminators from the perspective of $C(k)$ than the horizontal
visibility graph of $\rho=0$ on the left.

\begin{figure}[H]
\centering \scalebox{0.28}[0.28]{\includegraphics{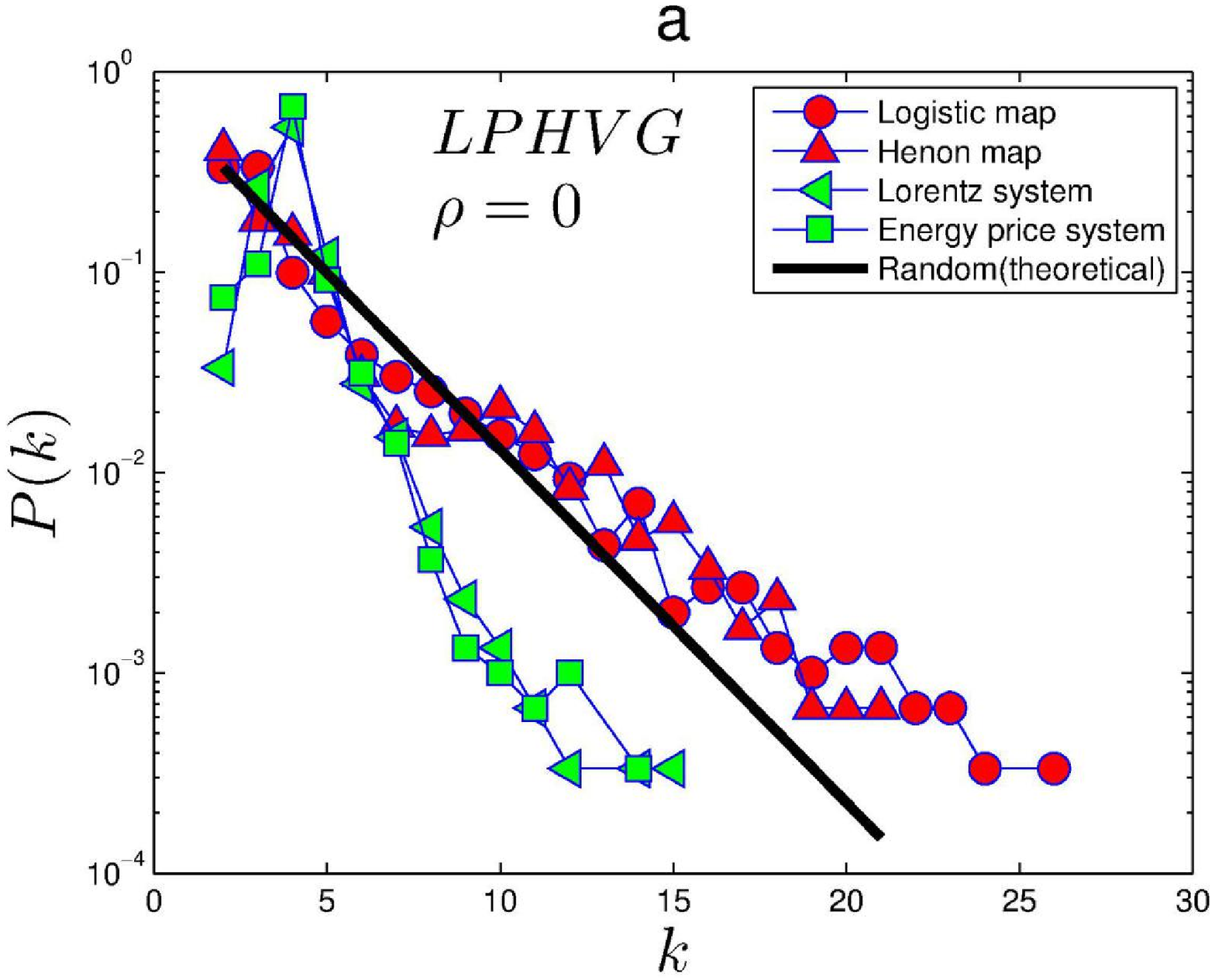}}
\scalebox{0.28}[0.28]{\includegraphics{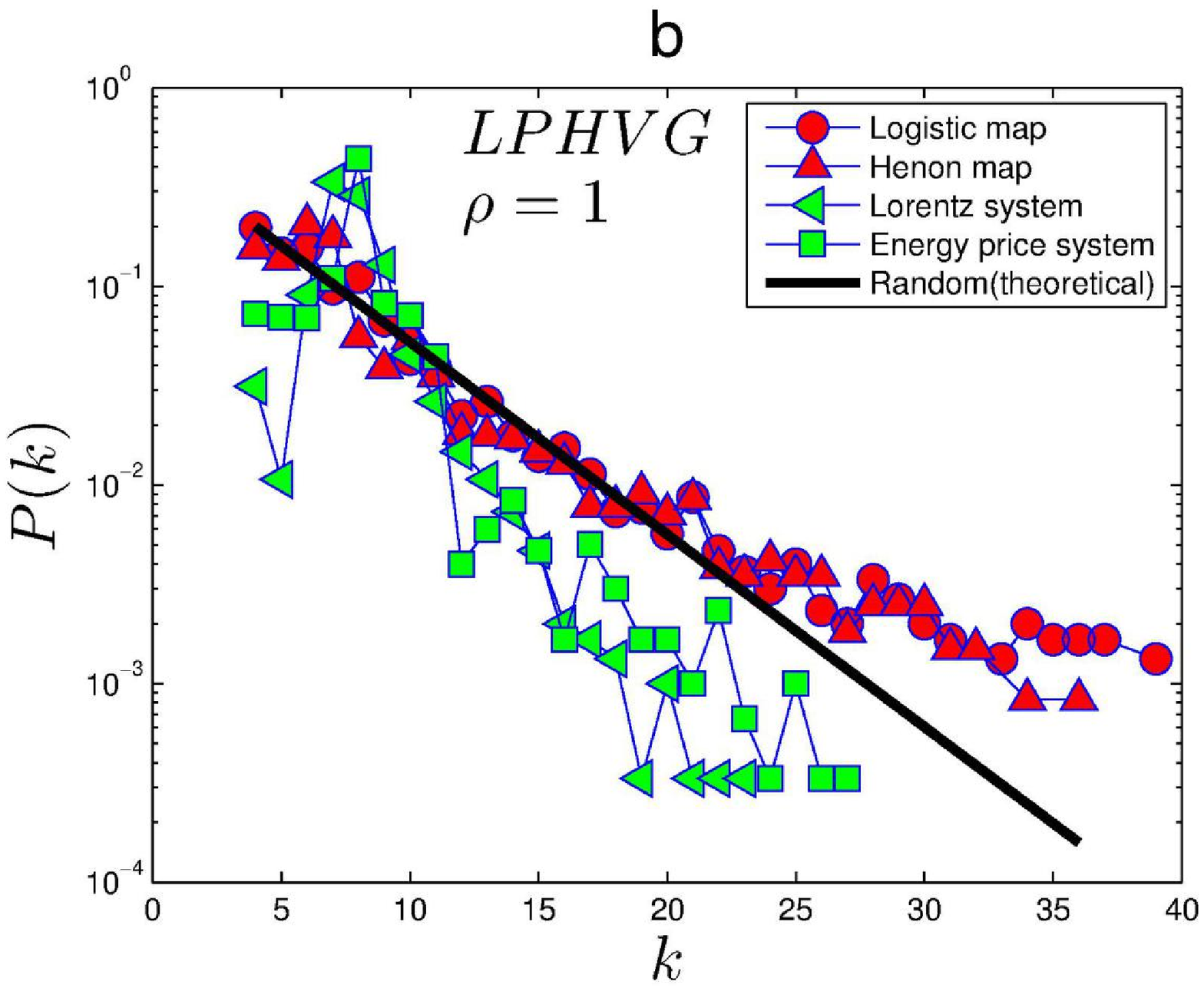}}
\scalebox{0.28}[0.28]{\includegraphics{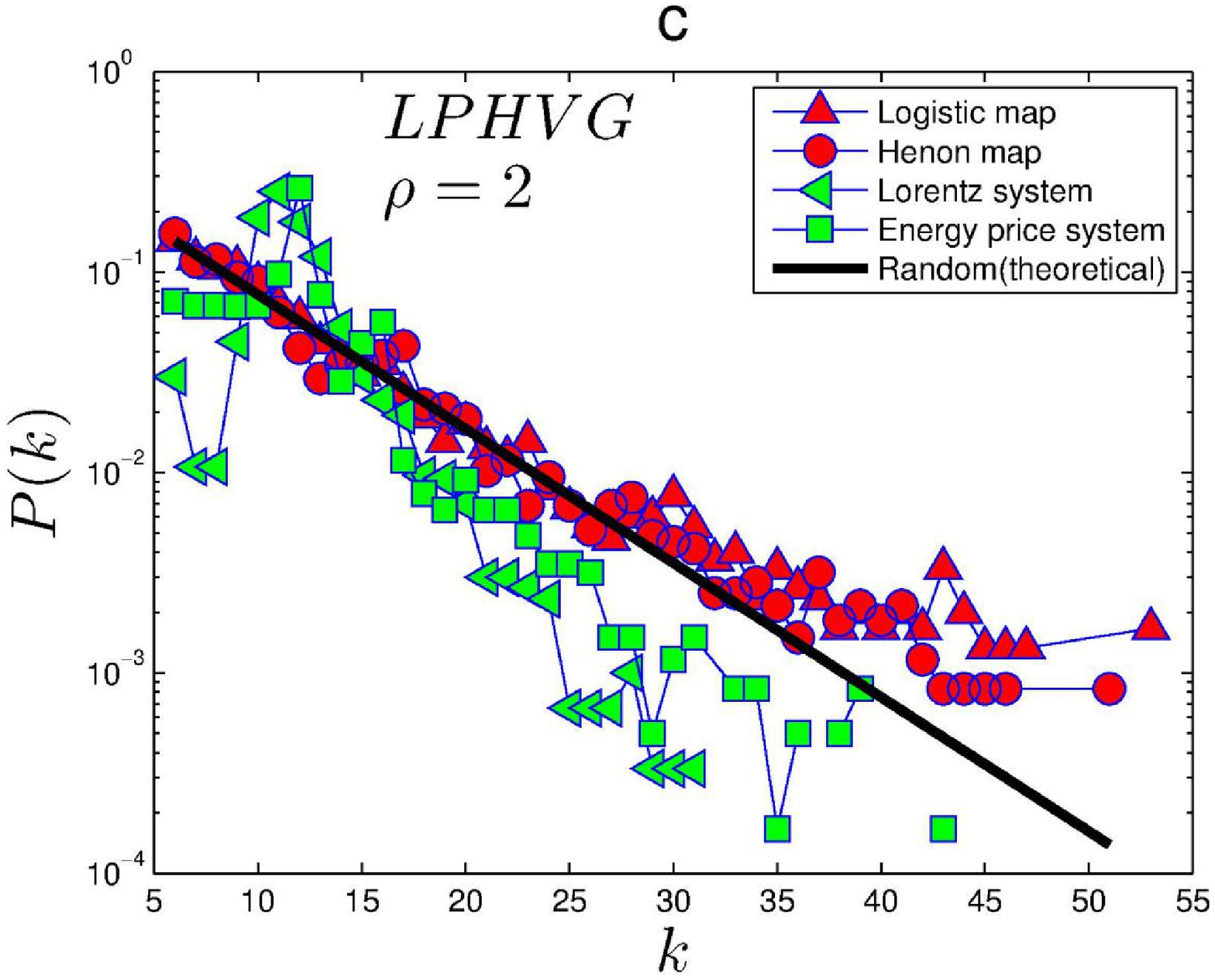}}\\
\scalebox{0.28}[0.28]{\includegraphics{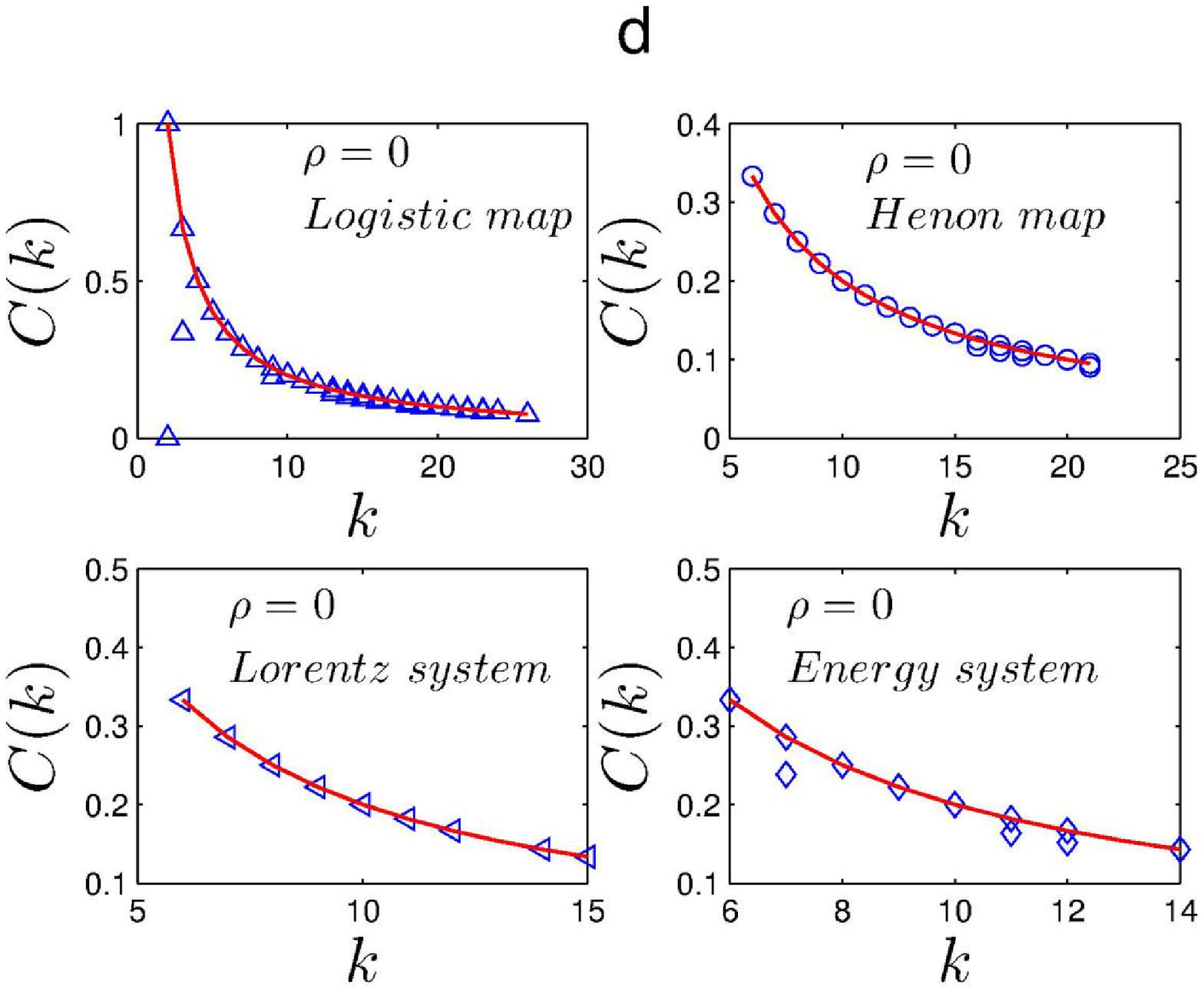}}
\scalebox{0.28}[0.28]{\includegraphics{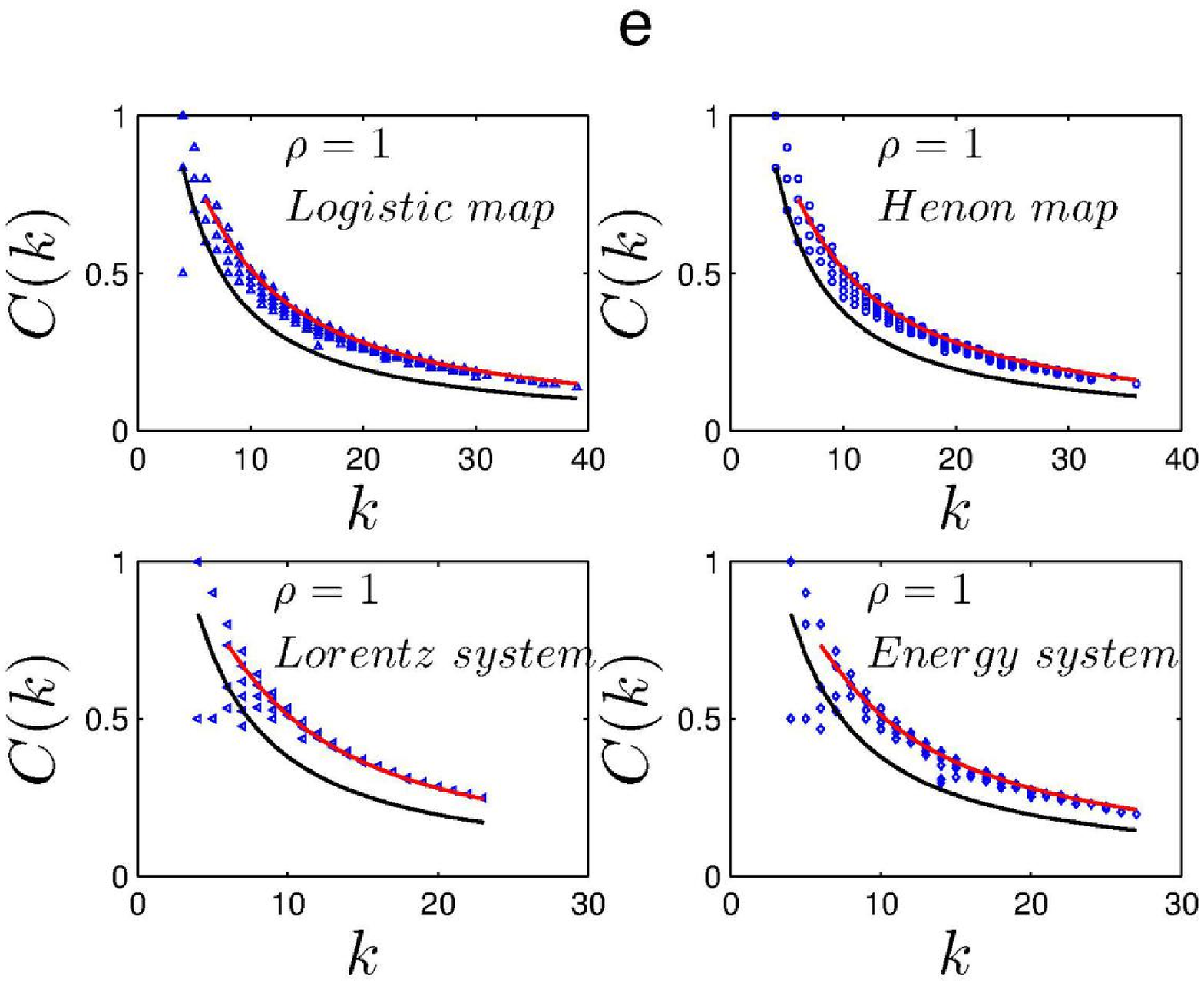}}
\scalebox{0.28}[0.28]{\includegraphics{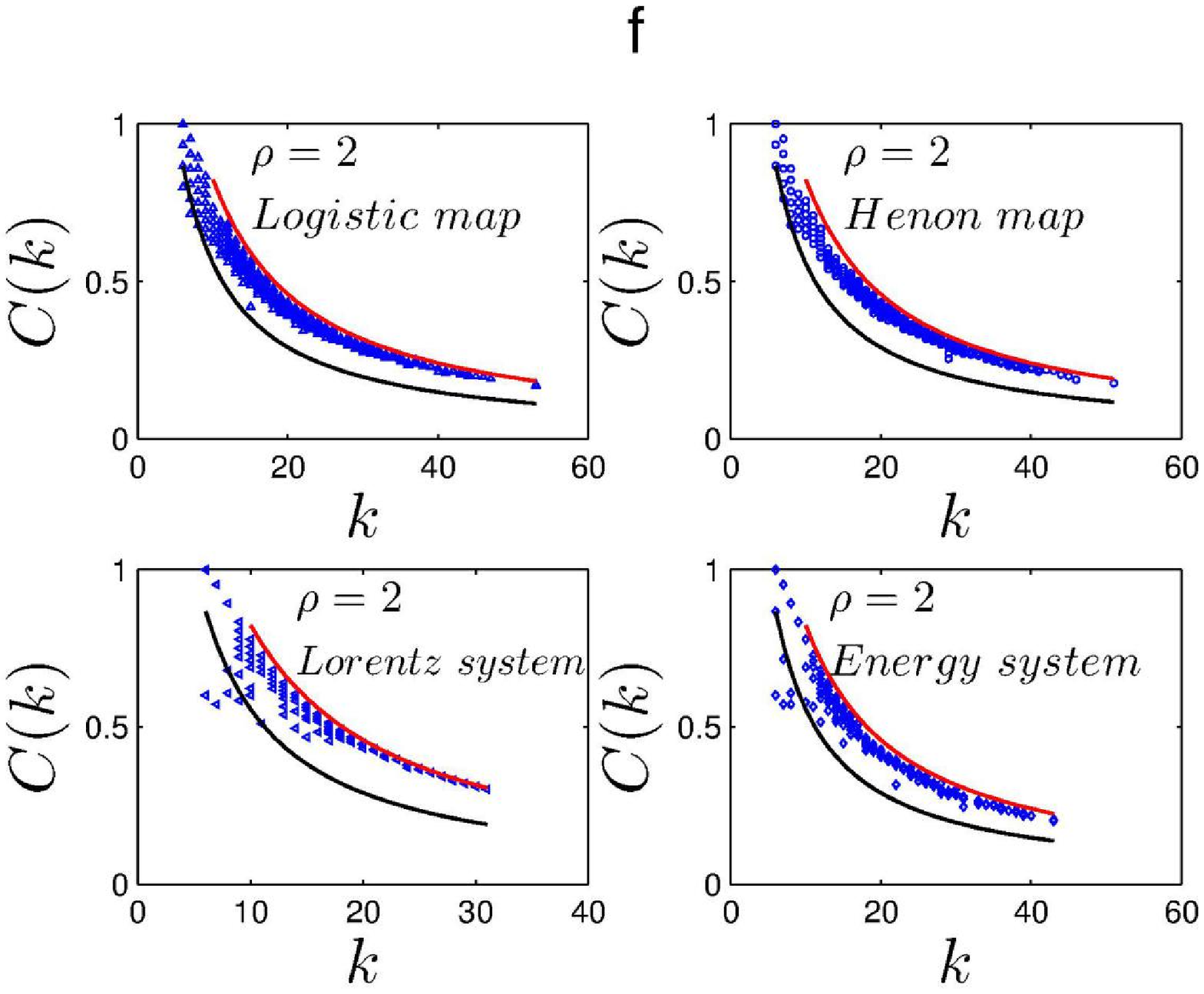}}
\end{figure}

\begin{figure}[H]
\caption{\emph{The upper part: Semilog plot of the degree distributions
    of Limited penetrable horizontal visibility graphs associated to
    series generated through Logistic map, Henon map, Lorenz chaotic
    system and Energy price- supply-economic growth system. The bottom
    part: The relationship between degree and clustering coefficients.}}
\end{figure}

\textbf{Application to real crude oil future price series.} As a further
example, we use data from the U.S. Energy Information Administration on
the crude oil future contract 1 (Dollars per Barrel) from 4 April 1983
to 28 March 1985, and find that they exhibit chaotic and long-range
correlations [38--39]. We select 500 sample data points and demonstrate
that we can use our method to distinguish chaotic series from random
series when the data sample is small (although for theoretical results
we need infinite data). Fig.~7 shows the results of the horizontal
visibility graph [see Fig. 7(a)] and the limited penetrable
horizontal visibility graph [see Fig. 7(b)] of 500 data points extracted from crude oil
futures. We find that the degree distributions both deviate from
Eq.~(2), which means the crude oil future price sequence is not random
but chaotic. Comparing the results of HVG ($\rho=0$ in LPHVG) and LPHVG,
we find that here LPHVG works better than HVG because the selected crude
oil price series is too short and there are fewer links in HVG. An
advantage in this case is that we can choose a suitable parameter $\rho$
when constructing LPHVG.

\begin{figure}[H]
\centering \scalebox{0.4}[0.4]{\includegraphics{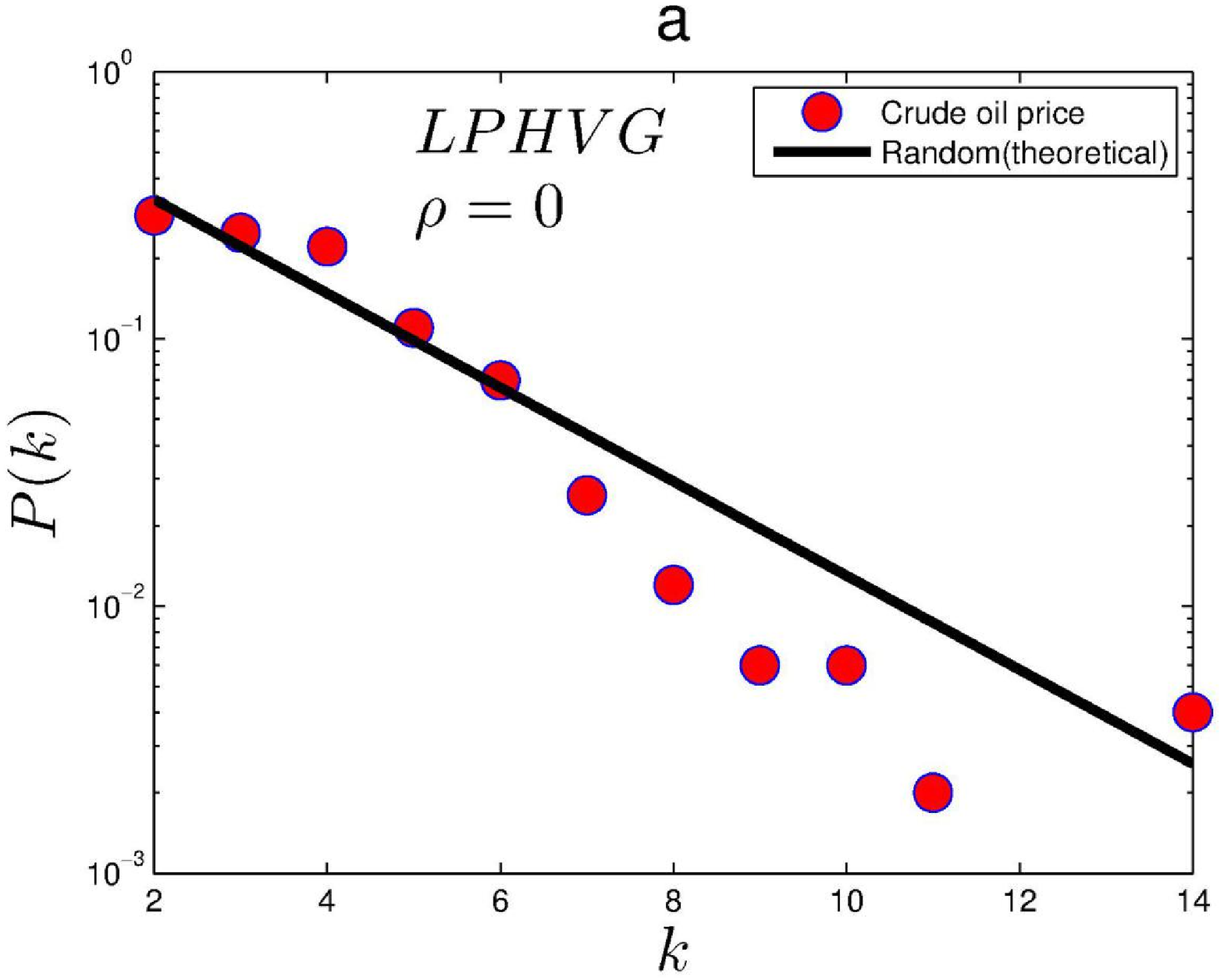}}
\scalebox{0.4}[0.4]{\includegraphics{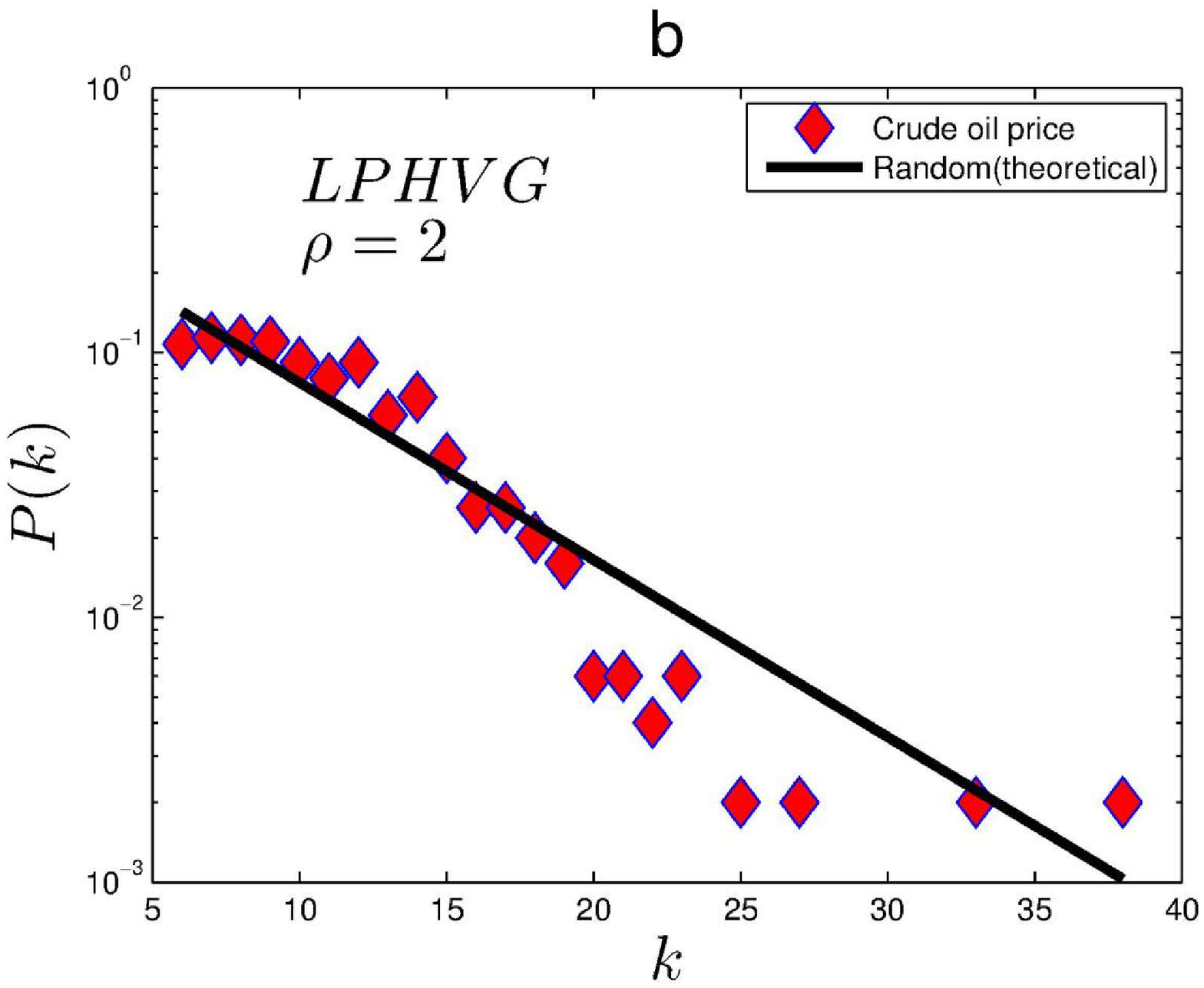}}
\end{figure}
\begin{figure}[H]
\caption{\emph{Semilog plot of the degree distributions of LPHVGs
    associated to crude oil price series.}}
\end{figure}

To further illustrate the application, we draw on the above analysis and
use LPVHG ($\rho=2$) to describe the global evolution of crude oil
future prices (for the calculation process see Methods). Our sample data
is from the crude oil future contract 1 (in dollars per barrel) from 4
April 1983 to 15 August 2017 [see Fig.~8(a)]. Because fluctuations in
crude oil future prices differ in different time periods, we separate
our data into two periods, a more stable period from 4 April 1983 to 10
February 2004 and a period of sharp fluctuations from 11 February 2004
to 15 August 2017 [19]. Using our calculation method (see the Method
section) we establish eighty-two 100-week time series windows (i.e.,
$L=500$), the first of which is from 4 April 1983 to 28 March
1985. Because each window moves 20 weeks to generate the next window
(i.e., $l=100$), two adjacent windows have overlaps of 80 weeks. This
enables information from one window to move to the next in
succession. Each window contributes 500 nodes to building the local
limited penetrable visibility graph network. Figs~8(b) and 8(c) show
the evolution of the adjacency matrix of LPHVGs associated with a random
series extracted from a uniform distribution and from crude oil price
series, respectively.  Note that adjacent matrices in the random time
series and the crude oil price time series significantly differ, but
their respective adjacent matrices in different time windows are
similar. Figures~8(d)--8(f) show the evolution of the mean degree, mean
clustering coefficient and mean path length, respectively. We find that
the mean degree, mean clustering coefficient and mean path length of the
LPHVG associated with the random series agree with the theoretical
values, but these three quantities of LPHVG associated with the crude
oil price series do not. The levels of mean degree of the LPHVG
associated with the crude oil price series are smaller than the
theoretical values, but the mean clustering coefficient and mean path
length are larger. They also show different trends in different
fluctuation periods. Values in the sharp fluctuation period are larger
than values in the stable fluctuation period. Figs~8(g)--(i) show the
global evolution of random series and crude oil price series (see Eqs (12---(16) in Methods). Note that the random time
series has neither short-range nor long-range correlations, but the
crude oil price time series has both. Thus using LPHVG enables us to
describe the global time series evolution.

\begin{figure}[H]
\centering \scalebox{0.28}[0.28]{\includegraphics{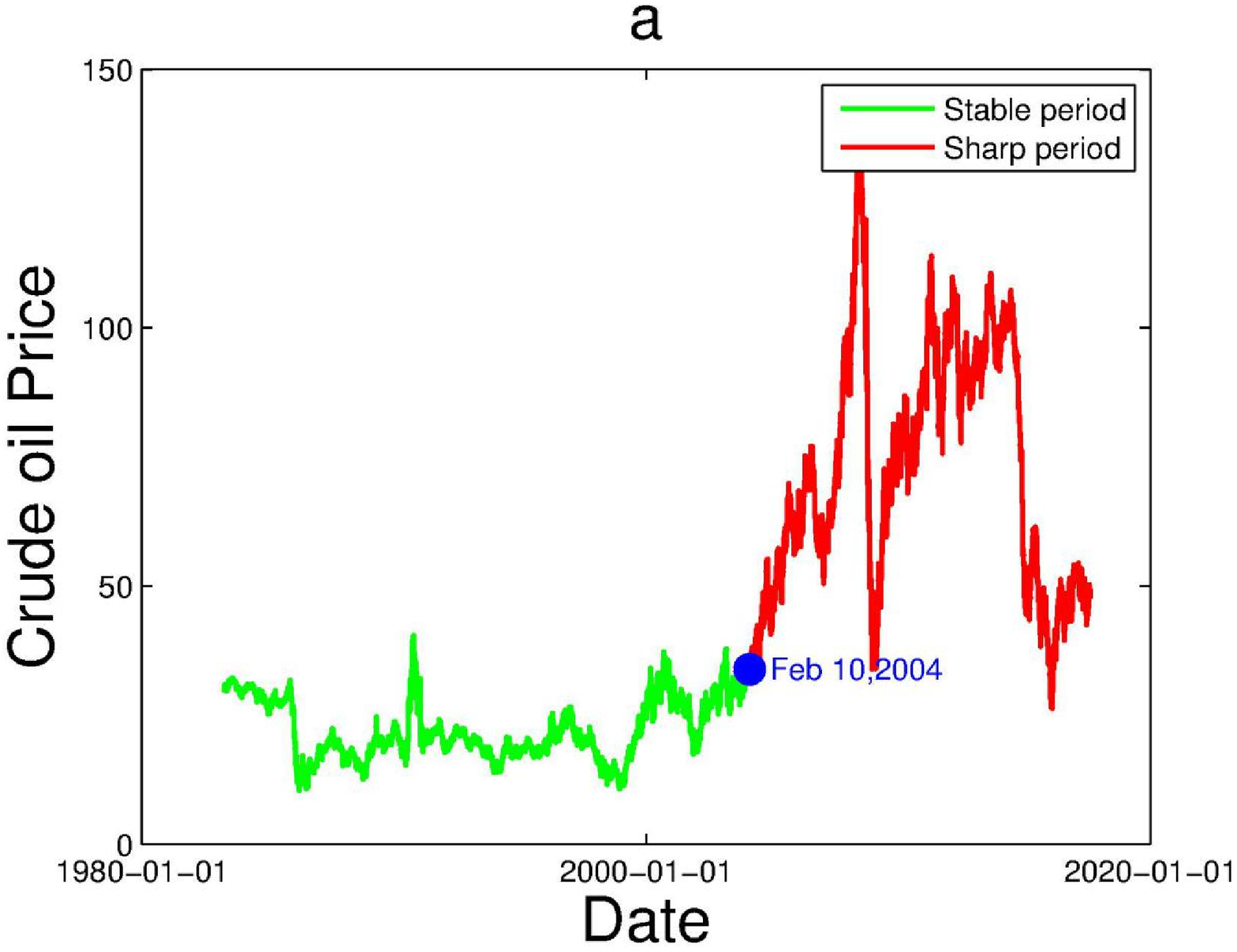}}
\scalebox{0.28}[0.28]{\includegraphics{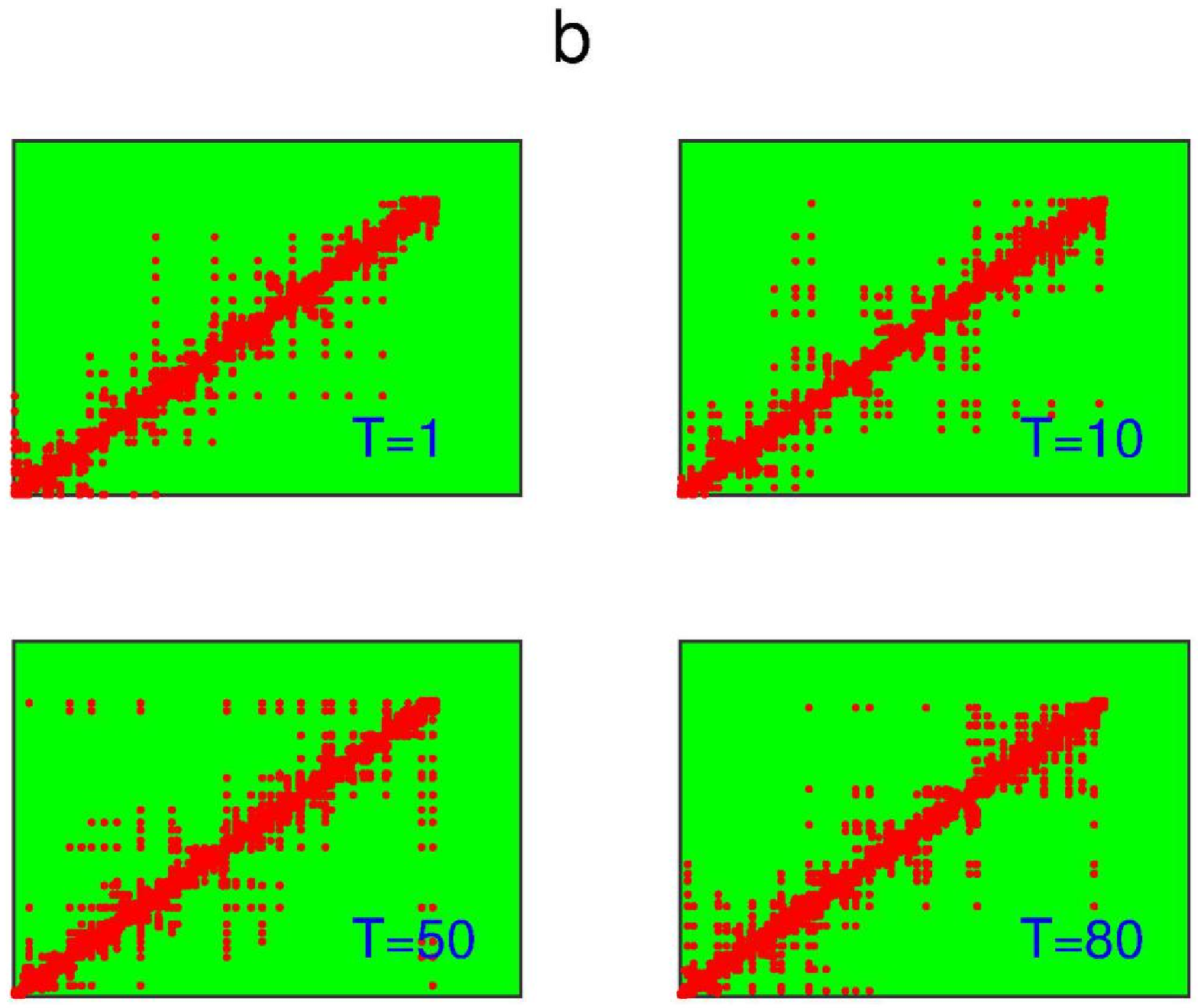}}
\scalebox{0.28}[0.28]{\includegraphics{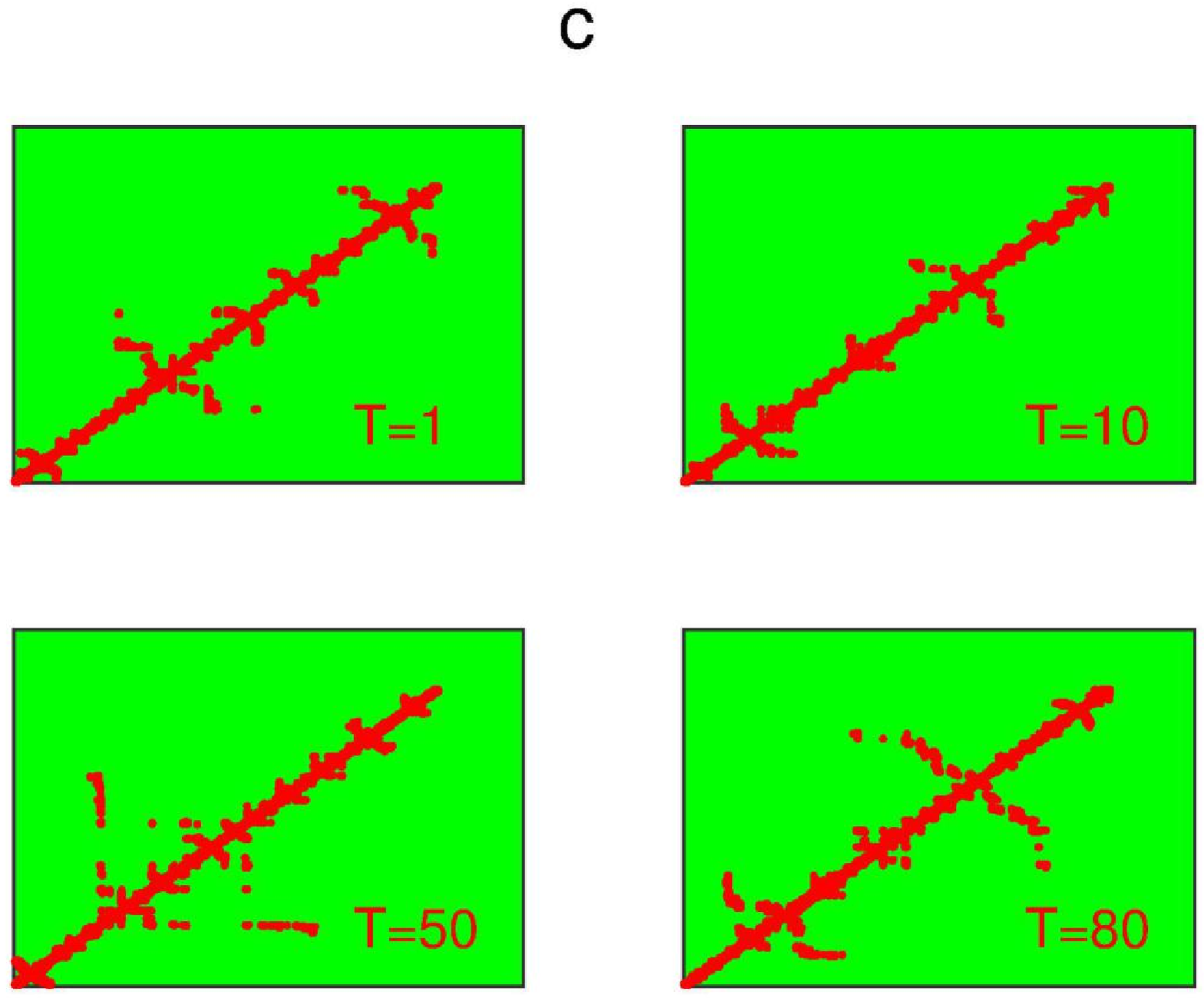}}\\
\scalebox{0.28}[0.28]{\includegraphics{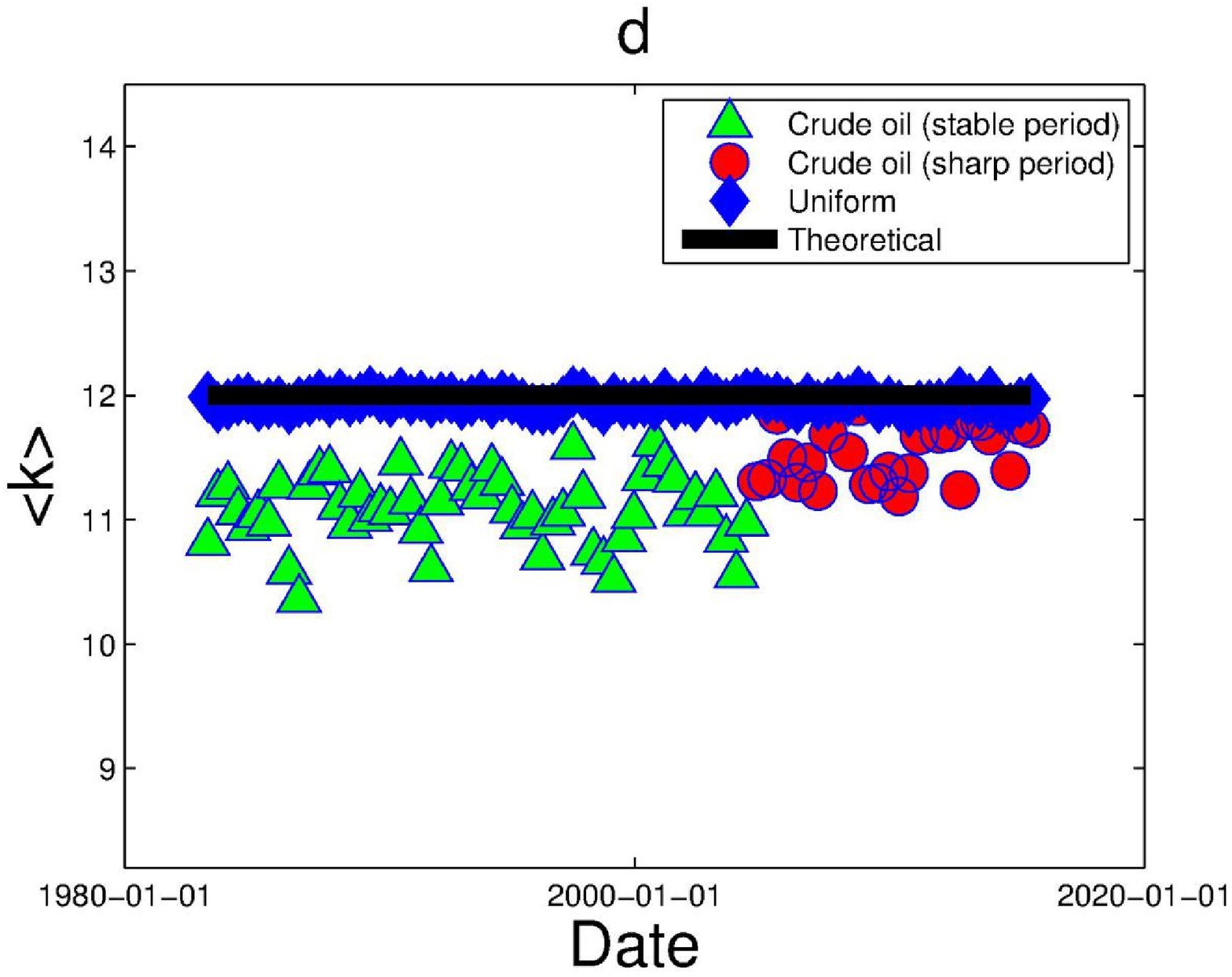}}
\scalebox{0.28}[0.28]{\includegraphics{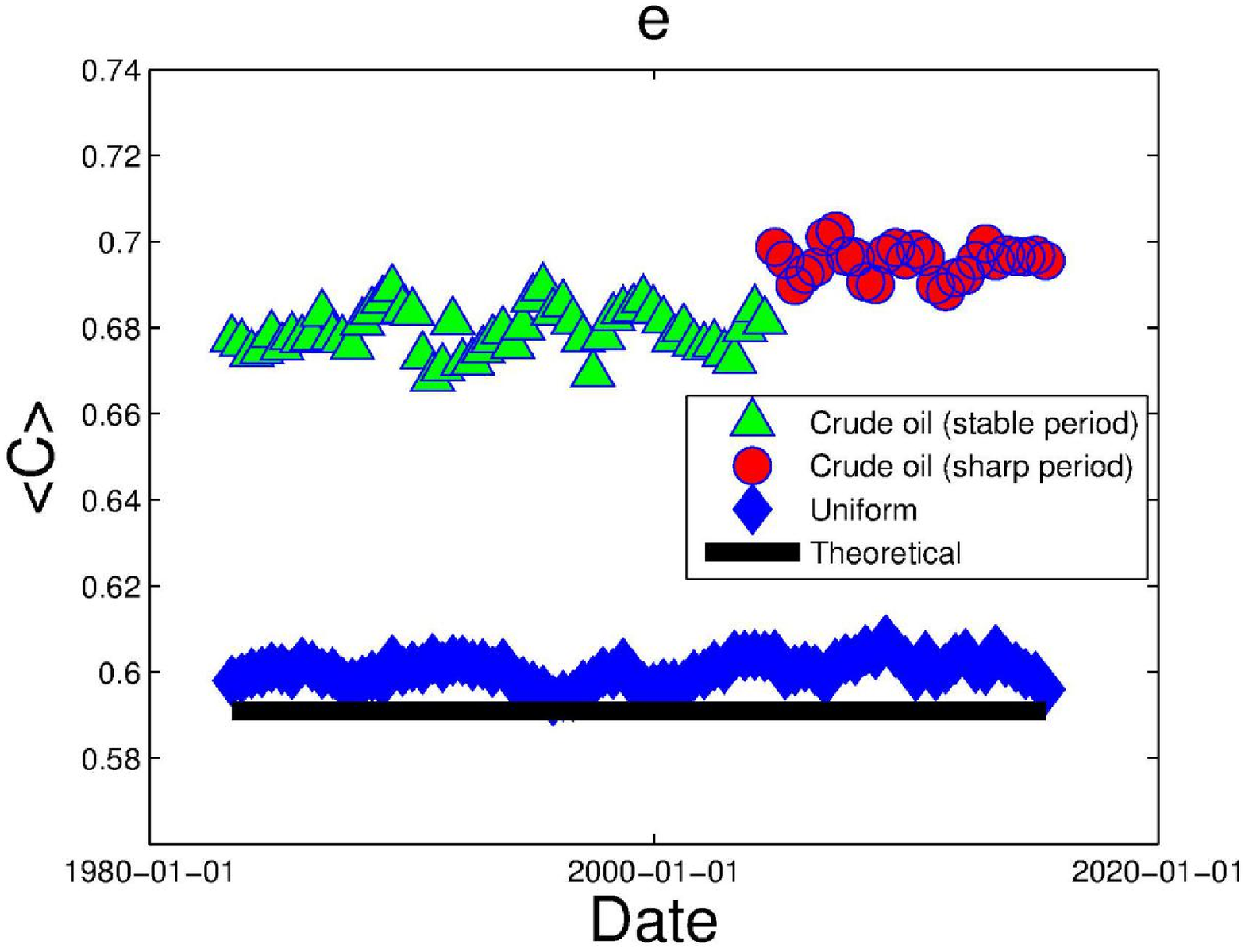}}
\scalebox{0.28}[0.28]{\includegraphics{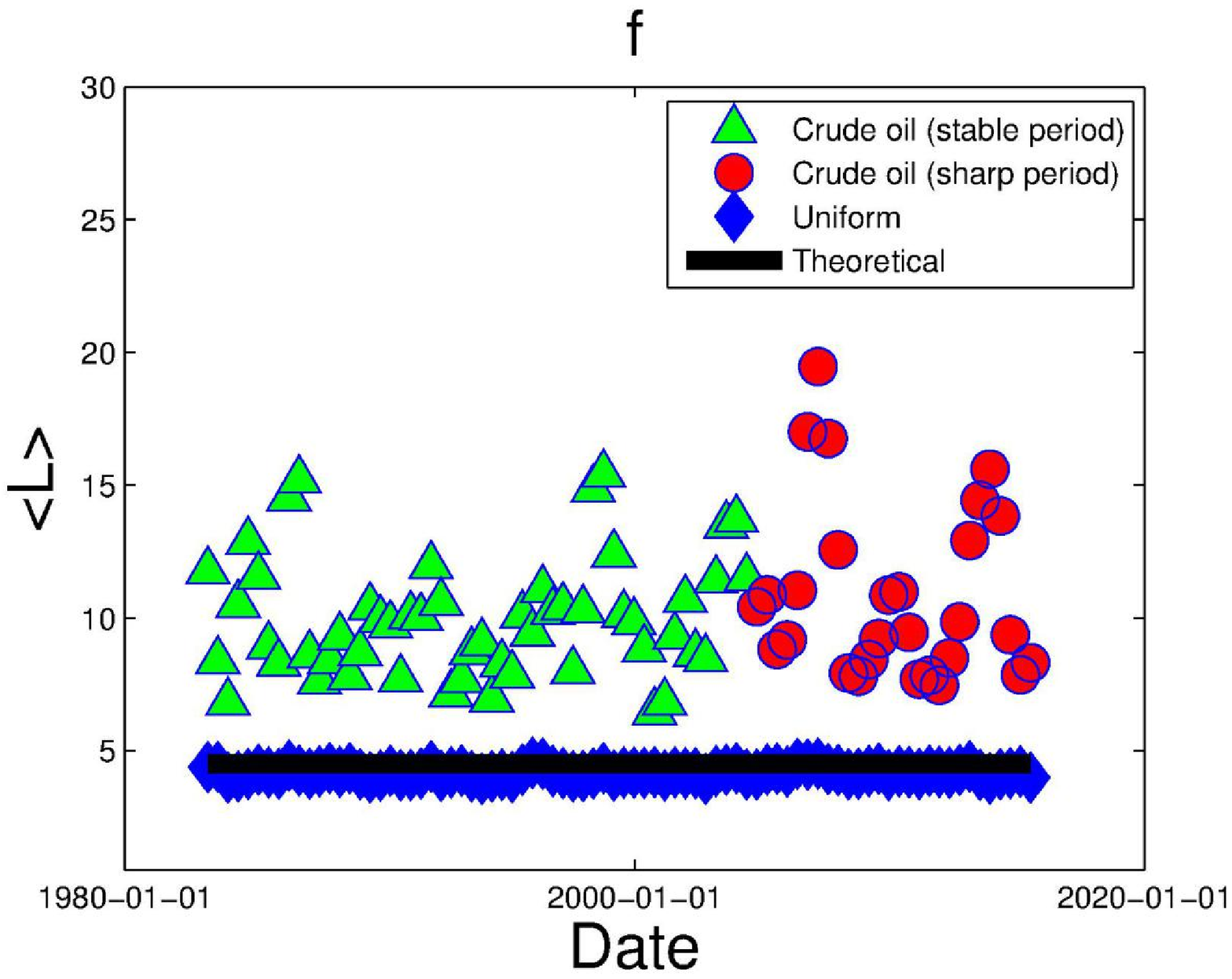}}\\
\scalebox{0.28}[0.28]{\includegraphics{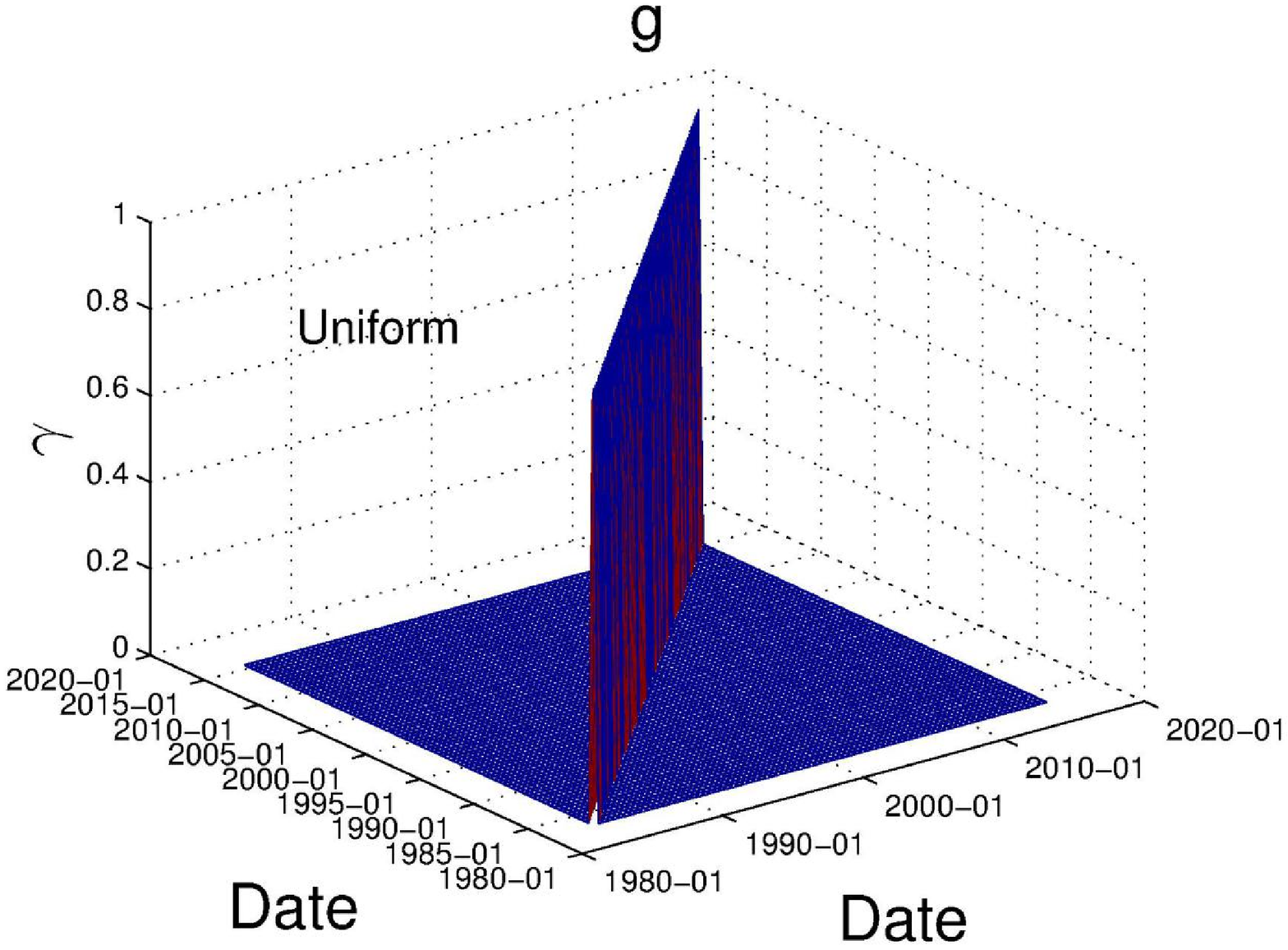}}
\scalebox{0.28}[0.28]{\includegraphics{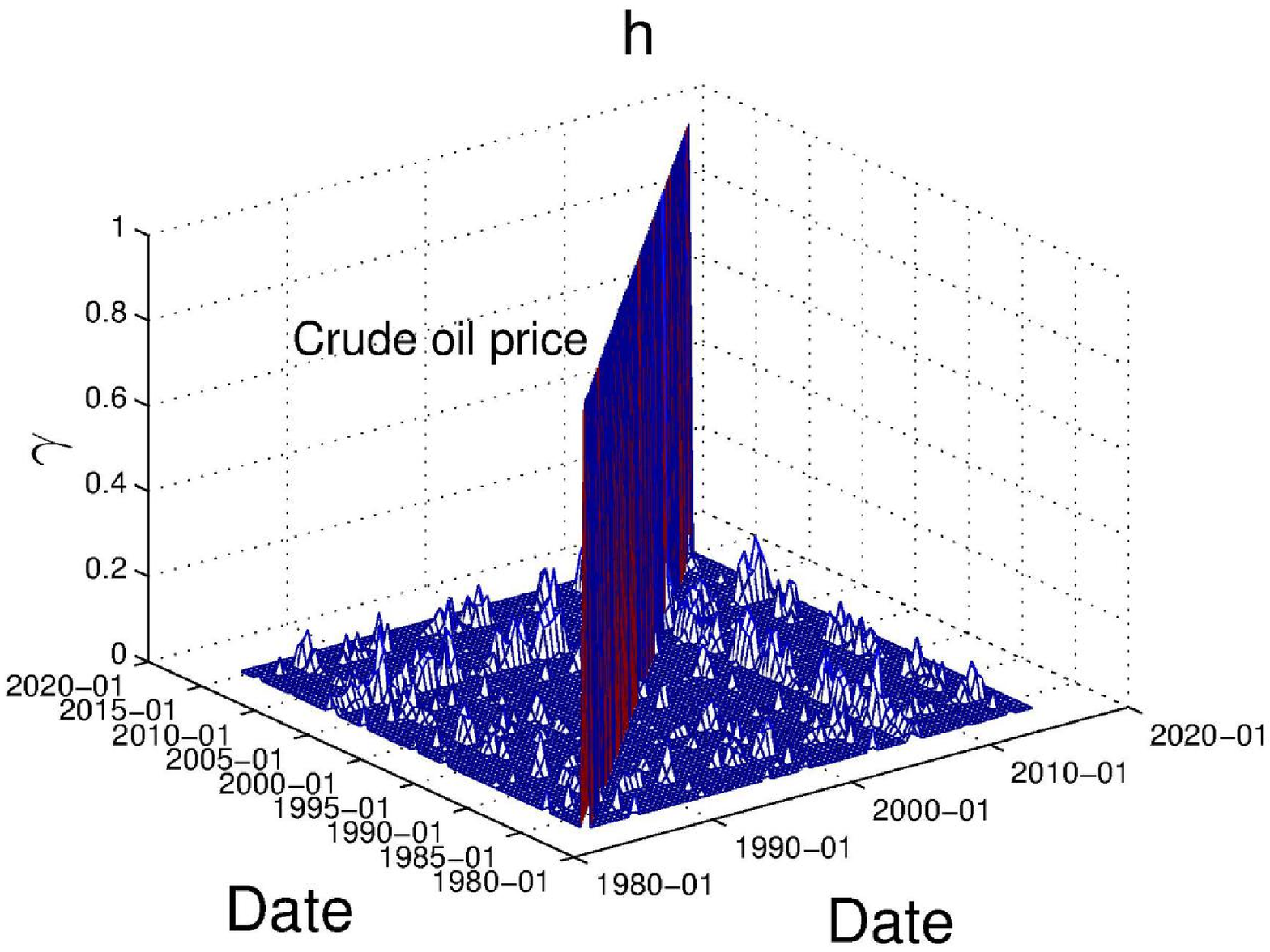}}
\scalebox{0.28}[0.28]{\includegraphics{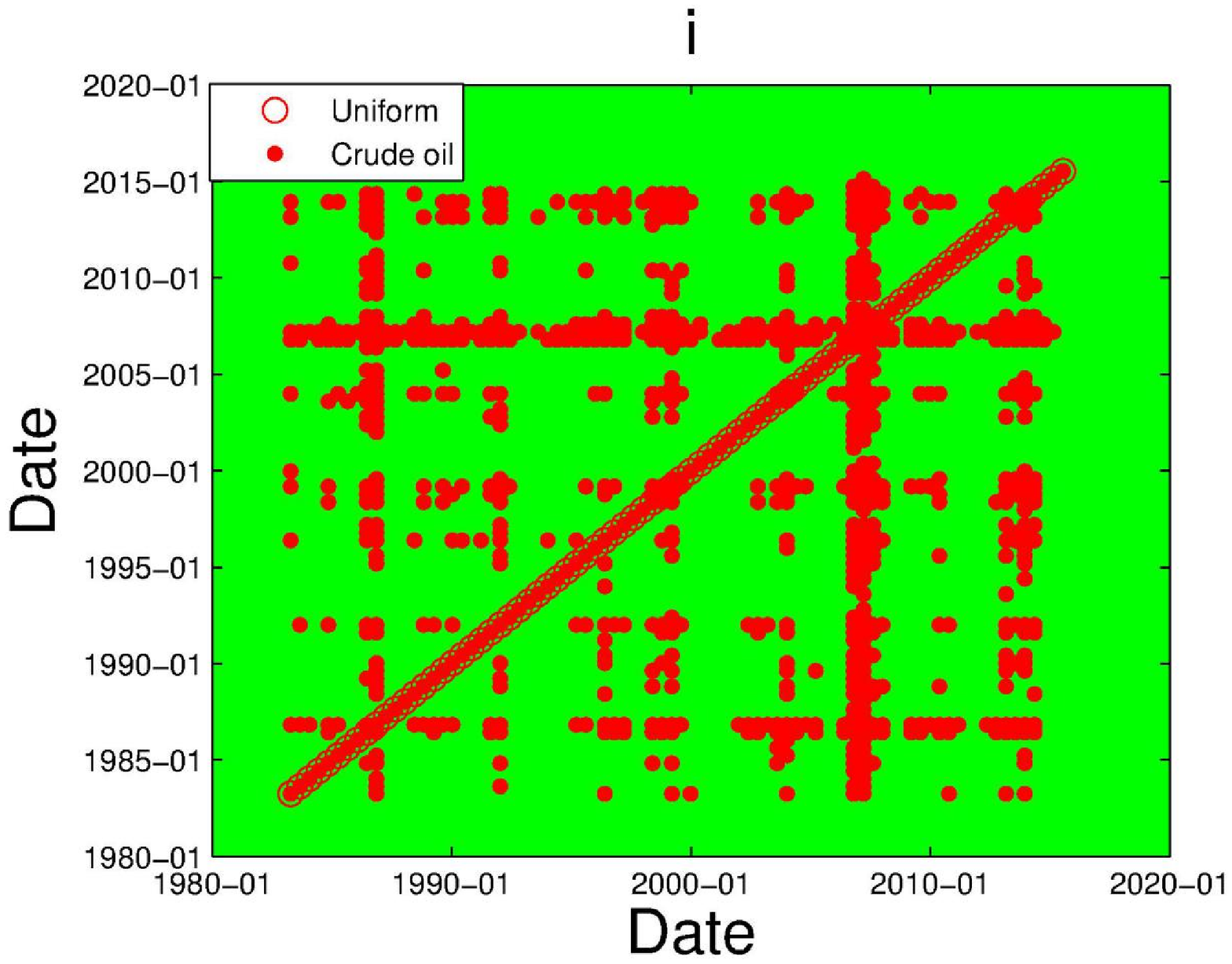}}
\end{figure}
\begin{figure}[H]
\caption{\emph{(a) Crude oil price series, the green line represents the
    stable period and the red line represents the sharp period. (b)
    Evolution of the adjacency matrix of LPHVGs associated to the random
    series extracted from a uniform distribution. (c) Evolution of the
    adjacency matrix of LPHVGs associated to crude oil price series. (d)
    Evolution of the mean degree of the LPHVGs associated to random
    series, crude oil price series and the theoretical value. (e)
    Evolution of the mean clustering coefficient of the LPHVGs
    associated to random series, crude oil price series and the
    theoretical value. (f) Evolution of the mean path length of the
    LPHVGs associated to random series, crude oil price series and the
    theoretical value. (g) The correlation index distribution of random
    series. (h) The correlation index distribution of crude oil price
    series. (i) The recursive graph of correlation index associated to
    random series and crude oil price series.}}
\end{figure}

\section{Discussion}

We have introduced a limited penetrable horizontal visibility algorithm,
a more generalized case of the horizontal visibility algorithm [11--12]
in which the limited penetrable distance is $\rho=0$. We obtain exact
results on several properties of the limited penetrable horizontal
visibility graph associated with a general uncorrelated random series,
the reliability of which has been confirmed by numerical simulations. In
particular, the degree distribution of the graph has the exponential
form $P(k)\sim exp[-\lambda (k-2\rho-2)],\lambda =
ln[(2\rho+3)/(2\rho+2)],\rho=0,1,2,...,k=2\rho+2,2\rho+3,...$. The
calculated expression of mean degree $<k>
=4(\rho+1)(1-\frac{2\rho+1}{2T}))$ holds for every periodic or aperiodic
series $T\rightarrow \infty$, independent of the deterministic process
that generates them. The clustering coefficient $C$ has a relationship
with degree $k$, i.e., $C_{\rm min}(k) =
\frac{2}{k}+\frac{2\rho(k-2)}{k(k-1)},\rho = 0,1,2,k\geq
2(\rho+1)$, $C_{\rm max}(k) = \frac{2}{k}+\frac{4\rho(k-3)}{k(k-1)},\rho
= 0,1,2,k\geq 2(2\rho+1)$.  The probability $P_{\rho}(n) =
\frac{2\rho(\rho+1)+2}{n(n+1)}$ introduces shortcuts to the limited
penetrable horizontal visibility graph that exhibit a small-world
phenomenon. Because these results are independent of the distribution
from which the series was generated, we conclude that all uncorrelated
random series have the same limited penetrable horizontal visibility
graph and, in particular, the same degree distribution, mean degree,
clustering coefficient distribution, and small world
characteristics. This algorithm can thus be used as a simple test for
discriminating uncorrelated randomness from chaos. We show that the
method can distinguish between random series that follow the theoretical
predictions and chaotic series that deviate from them. In addition, we
employ the method to measure the global evolution characteristics of
time series by using LPHVG, and the empirical results confirm its
validity.

Our exact results presented here are extension of previous work [11]. We
adjust the limited penetrable parameter $\rho$ to the actual situation
in order to distinguish chaos from uncorrelated randomness. The method
can serve as a preliminary test for locating deterministic fingerprints
in time series. If we determine that $P(k)$ has an exponential tail that
deviates from Eq.~(2), or that $C(k)$ deviates from Eqs.~(6) and (7), we
apply embedding methods to the series.  Topics of further research could
include whether this algorithm is also able to quantify chaos, the
relationship between such standard chaos indicators as Lyapunov
exponents and the correlation dimension, how to tune the limited
penetrable parameter $\rho$, how to use the limited penetrable horizontal visibility graph to handle two-dimensional manifolds, the topological properties of the
visibility graphs (VG) and limited penetrable visibility graphs (LPVG),
and expanded applications of LPHVG.

\section{Methods}

\textbf{Limited Penetrable Horizontal Visibility Graph (LPHVG).} The
limited penetrable visibility graph (LPVG) [30] and the multiscale
limited penetrable horizontal visibility graph (MLPHVG) [14] are a
recent extension of the VG [9] and HVG [11--12] used to analyze
nonlinear time series. The limited penetrable horizontal visibility
graph (LPHVG) is a geometrically more simple and analytically solvable
version of LPVG [30] and MLPHVG [14]. To define it we let
$\{x_{i}\}_{i=1,2,...,N}$ be a time series of $N$ real numbers. If we
set the limited penetrable distance to $\rho$, LPHVG maps the time
series into a graph with $N$ nodes and an adjacency matrix
$\textbf{A}$. Nodes $i$ and $j$ are connected through an undirected edge
($A_{ij} = A_{ji} = 1$) when $x_{i}$ and $x_{j}$ have limited penetrable
horizontal visibility (see Fig.~9), i.e., if at most $\rho$ intermediate
data $x_{q}$ is
\begin{equation}\label{eq7}
x_{q}\geq inf\{x_{i},x_{j}\},\forall q\in (i,j).
\end{equation}
This mapping is a limited penetrable horizontal visibility graph
(LPHVG). When we set the limited penetrable distance $\rho=0$, LPHVG
degenerates into HVG [11]. When $\rho\neq 0$, there are more connections
between any two nodes in LPHVG than in HVG. Fig.~9(b) shows the new
established connections (red lines) when we infer the LPHVG on the basis
of HVG with a limited penetrable distance $\rho=1$. Note that the
limited penetrable horizontal visibility graph of a given time series
has all the properties of its horizontal visibility graph, e.g., it is
connected and invariant under all affine transformations of the series
data [9, 11].

\begin{figure}[H]
\centering \scalebox{0.6}[0.5]{\includegraphics{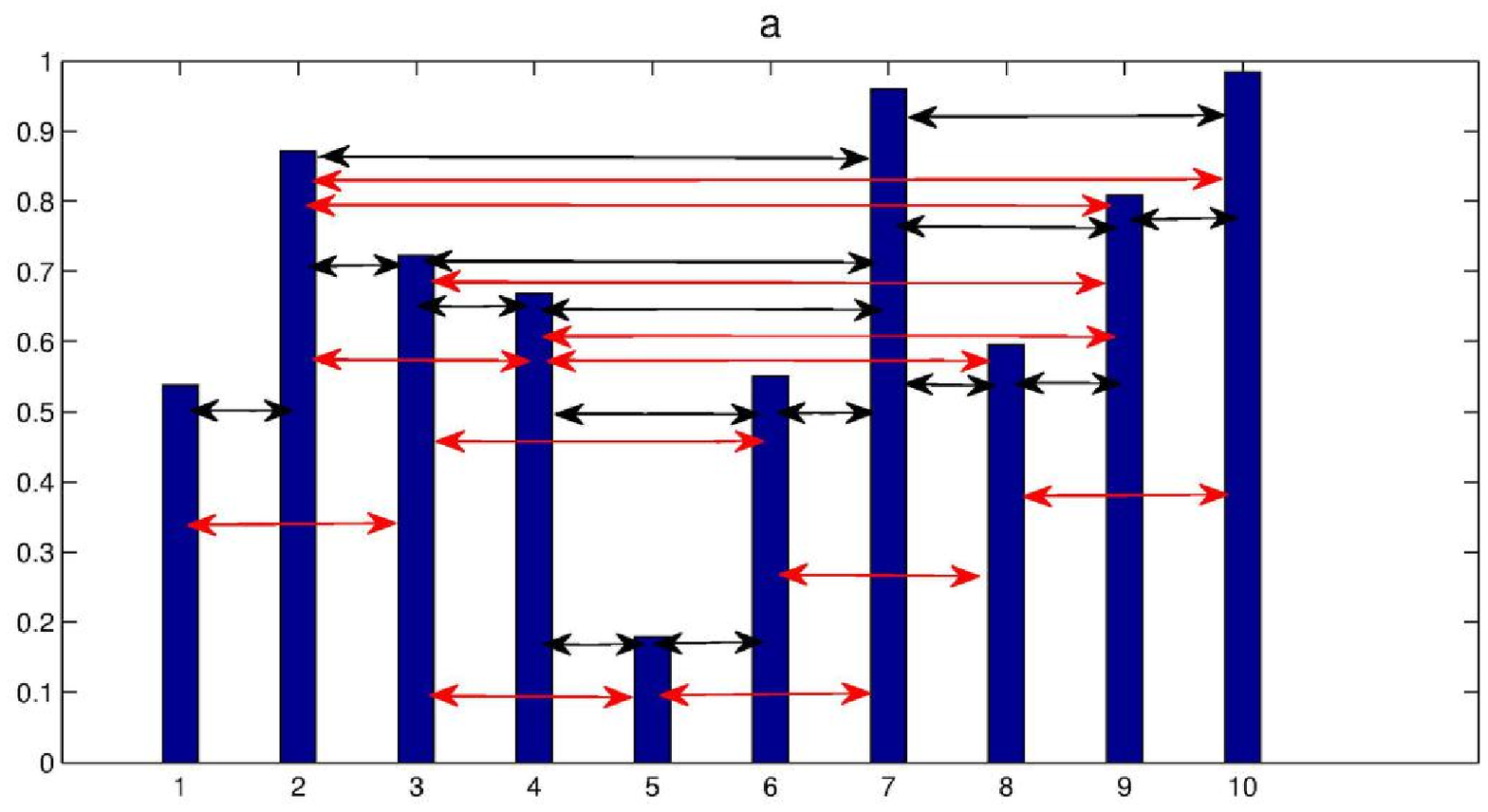}}
\scalebox{0.5}[0.4]{\includegraphics{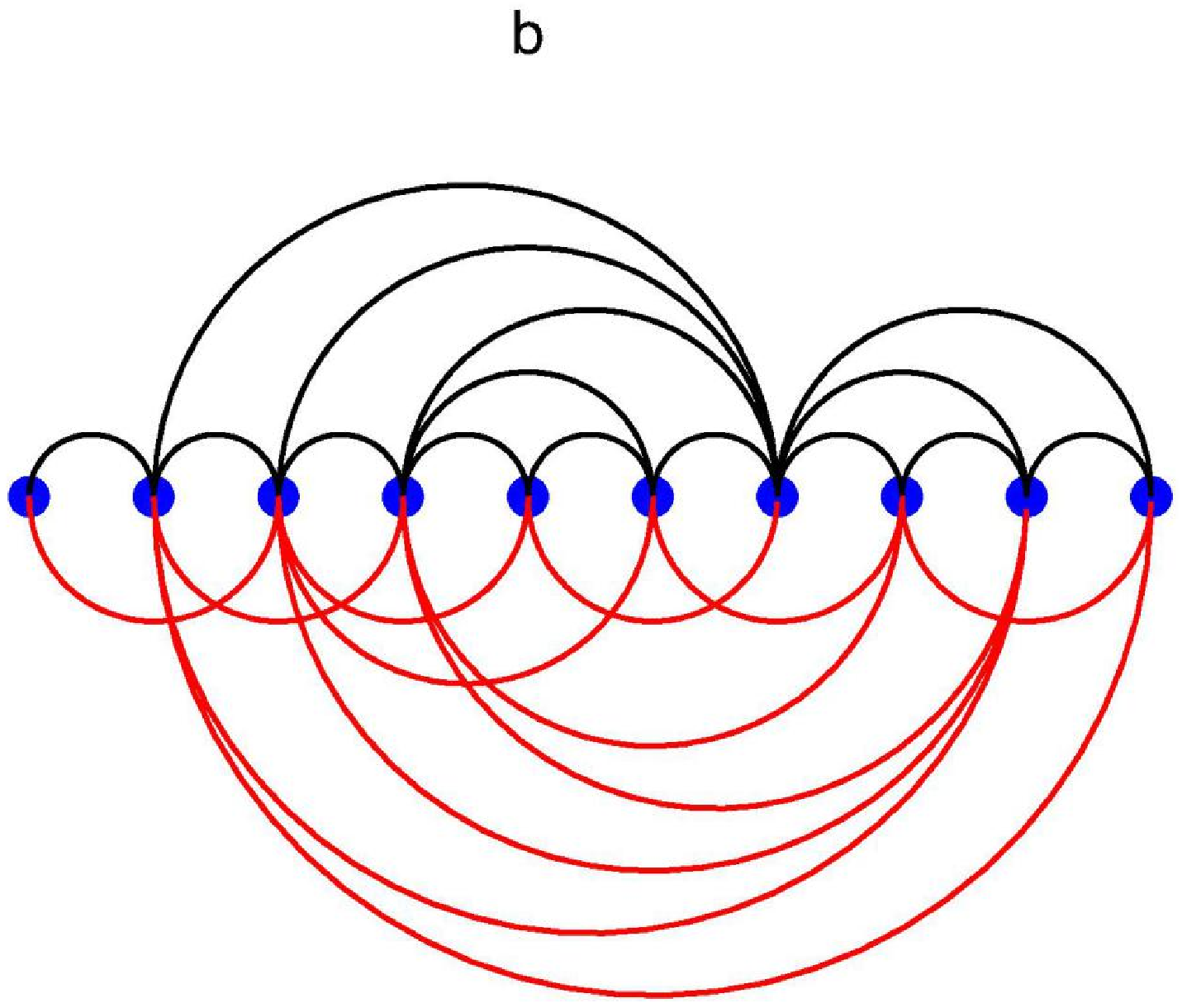}}
\end{figure}
\begin{figure}[H]
\caption{\emph{Example of (a) a time series (10 data values) and (b) its
    corresponding LPHVG with the limited penetrable distance $\rho$
    being 1, where every node corresponds to time series data in the
    same order. The limited penetrable horizontal visibility lines
    between data points define the links connecting nodes in the
    graph.}}
\end{figure}

\textbf{Measurement of the Global Evolution Characteristics of Time
  Series using LPHVG.} A time series is defined $\textbf{X} =
\{x(t)\},t=1,2,...,N$.  To characterize the evolution of the time series
using LPHVG, we divide the time series of the entire scale of the time window into equal small-scale segments and assume that the
length of the sliding window is $L$. We define $l$ the step length
between sliding time windows. To ensure that small-scale segments of the
time series are continuous, we require that $l<L$. This allows us to
obtain $T = [(N-L)/l +1]$ small-scale time windows, where $[...]$ is the
rounding function. For every small-scale time window $t$, we transform
time series into the $LPHVG(t)$ using the limited penetrable horizontal
visibility algorithm. The topological structure of LPHVG changes with
time $t$. To describe this process from the global perspective, we use
the Euclidean distance to measure the relationship between LPHVGs. We
define the Euclidean distance between $LPHVG(t_{m})$ and $LPHVG(t_{n})$
to be
\begin{equation}\label{eq8}
d(LPHVG(t_{m}),LPHVG(t_{n})) =
\sqrt{\sum_{i=1}^{L}\sum_{j=1}^{L}(a_{ij}^{(t_{m})}-a_{ij}^{(t_{n})})},a_{ij}^{(t_{m})}\in
\textbf{A}^{t_{m}},a_{ij}^{t_{n}}\in \textbf{A}^{t_{n}}.
\end{equation}
We then determine the distance matrix
\begin{equation}\label{eq9}
\textbf{D}_{T\times T} = \{d_{t_{m},t_{n}}\}_{t_{m}=1,2,...,T,t_{n} =
  1,2,...,T},
\end{equation}
and assign a threshold value to $\theta$
\begin{equation}\label{eq10}
\theta = min\{d_{t{m},t_{n}}^{\rm rand}\}_{t_{m}\neq
  t_{n}},d_{t_{m},t_{n}}^{\rm rand}\in \textbf{D}_{T\times T}^{\rm rand}.
\end{equation}
Here $\textbf{D}_{T\times T}^{\rm rand}$ is the distance matrix
associated with the independent and identically distributed random time
series. Using the threshold $\theta$, we define the correlation index
$\gamma$,
\begin{equation}\label{S11}
\gamma_{t_{m},t_{n}} =
\begin{cases}
 0,d_{t_{m},t_{n}}\geq \theta. \\
 1-d_{t_{m},t_{n}}/\theta,d_{t_{m},t_{n}}< \theta.
\end{cases}
\end{equation}
Here $\gamma_{t_{m},t_{n}}$ is the correlation degree of LPHVG at time
$t_{m}$ and time $t_{n}$, and $\gamma_{t_{m},t_{n}}$ can be visualized
using a recursive graph constructed using the formula
\begin{equation}\label{eq12}
\Re(t_{m},t_{n}) = \Theta (\theta -
d(LPHVG(t_{m},LPHVG(t_{n})))),\Theta(x) =
\begin{cases}
1,x> 0,\\
0,x\leq 0,
\end{cases}
\end{equation}
where $\Theta(x)$ is the Heaviside function. We use the formula to plot
the relationship between LPHVGs in two-dimensional coordinates in which
both the abscissa and the ordinate are at time $t$. In the recursive
graph when the Euclidean distance between $LPHVG(t_{m})$ and
$LPHVG(t_{n})$ is sufficiently close, i.e., when $\Re(t_{m},t_{n}) = 1$,
we plot the red dot at $(t_{m},t_{n})$ and $(t_{n},t_{m})$.
Note that at $(t_{m},t_{m})$ and $(t_{n},t_{n})$, i.e., at the main
diagonal, the red dots remain throughout, and we can use
it to characterize the global dynamic changes in correlation.

\section{Acknowledgments}

\noindent
The Research was supported by the following foundations: The National Natural Science Foundation of China (71503132, 71690242, 91546118, 11731014, 71403105, 61403171,),Qing Lan Project of Jiangsu Province (2017), University Natural Science Foundation of Jiangsu Province (14KJA110001), Jiangsu Center for Collaborative Innovation in Geographical Information Resource Development and Application, CNPq, CAPES, FACEPE and UPE.

\section{Appendix}

\textbf{Theorem S1.} Let $X(t)$ be a real bi-infinite time series
of $i.i.d.$ random variables with
probability density $f(x)$, with $x\in [a,b]$, and consider its
associated LPHVG with a limited penetrable distance $\rho=1$. Then
$$P(k)\sim exp[-(k-4)ln(5/4)],k=4,5,...,\forall f(x).$$

\textbf{Proof:} Using a method similar to that presented in
Refs.~[10,11], we select a generic datum $x_{0}$ to be the seed. We
calculate the probability that an arbitrary datum with value $x_{0}$ has
a limited penetrable visibility of exactly $k$ other data. From the
definition of LPHVG, when $x_{0}$ has penetrable visibility of $k$ data
there will be at least two penetrable data and two bounding data, one
penetrable and one bounding datum on the right-hand side of $x_{0}$ and
one on the left-hand side, such that the $k-4$ remaining visible and
penetrable visible data are located inside two bounding data. Note that
$k = 4$ is the minimum possible degree (see Fig.~S1).

\begin{center}
\includegraphics[width=28em, height=23em]{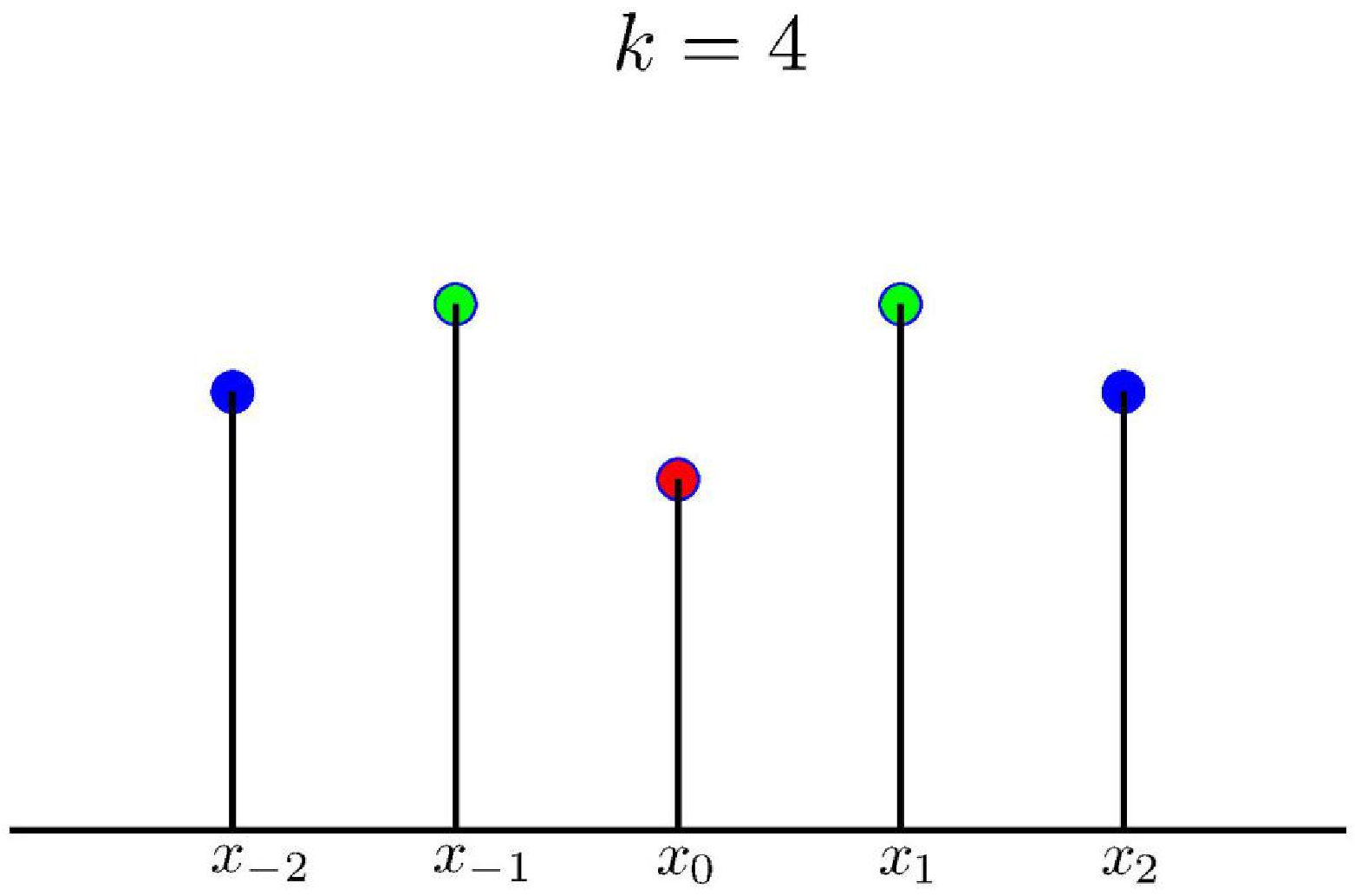}
\vspace{0em}
\end{center}
\begin{center}
\noindent {\small {\bf Fig.S1.} \emph{Set of possible configuration for
    a seed data $x_{0}$ with $k=4$. The green dots are penetrable data,
    the blue dots are bounding data.}}
\end{center}

To derive the degree distribution of the associated LPHVG, we first
compute some easy terms. Fig.~S1 shows the simplest case $P(k=4)$ in
which there are two penetrable data $(x_{-1}, x_{1})$ and two bounding
data $(x_{-2}, x_{2})$. To assure that $k=4$, we set the height of both
the penetrable and bounding data greater than $x_{0}$, i.e., $x_{-1}\geq
x_{0},x_{1} \geq x_{0}$ and $x_{-2}\geq x_{0},x_{2} \geq x_{0}$. Then
\begin{equation*}
\begin{array}{l}
P(k=4)=  Prob(x_{-2},x_{-1},x_{1},x_{2}\geq x_{0})\\
      =\int_{a}^{b}f(x_{0})dx_{0}\int_{x_{0}}^{b}f(x_{-2})dx_{-2}
\int_{x_{0}}^{b}f(x_{-1})dx_{-1}\int_{x_{0}}^{b}f(x_{1})dx_{1}
\int_{x_{0}}^{b}f(x_{2})dx_{2}.
\end{array}
\eqno(S1)
\end{equation*}

In order to simplify Eq. (S1), we define the cumulative probability distribution function $F(x)$ of
any probability density $f(x)$ to be
\begin{equation*}
\begin{array}{l}
F(x) = \int_{a}^{x}f(t)dt,
\end{array}
\eqno(S2)
\end{equation*}
where $dF(x)dx = f(x)$,$F(a) = 0$ and $F(b) = 1$. With a loss of
generality, we assume $a = 0, b = 1$ , i.e., $F(0) = 0$ and $F(1) =
1$. Here the relation between $f$ and $F$ holds, i.e.,
\begin{equation*}
\begin{array}{l}
\frac{dF^(n)(x)}{dx} = nf(x)F^{n-1}(x).
\end{array}
\eqno(S3)
\end{equation*}

Using Eqs.~(S2) and (S3), we rewrite Eq.~(S1) to be
\begin{equation*}
\begin{array}{l}
P(k=4) = \int_{0}^{1}f(x_{0})[1-F(x_{0})]^{4}dx_{0} = \frac{1}{5},\forall f(x).
\end{array}
\eqno(S4)
\end{equation*}

When  $P(k=5)$ (see Fig.~S2 ), four configurations emerge: Case 1: $C_{0}^{1}$, in which $x_{0}$ has penetrable variables
  $x_{-1}$ and $x_{1}$, bounding variables $x_{-2}$ and $x_{3}$, and a
  right-hand side inner variable $x_{2}$. Case 2: $C_{0}^{2}$, in which $x_{0}$ has penetrable variables
$x_{-1}$ and $x_{2}$, bounding variables $x_{-2}$
and $x_{3}$, and a right-hand side inner variable $x_{1}$. Case 3: $C_{1}^{1}$, in which $x_{0}$ has penetrable variables
$x_{-2}$ and $x_{1}$, bounding variables $x_{-3}$ and
$x_{2}$, and a left-hand side inner variable $x_{-1}$. Case 4: $C_{1}^{2}$, in which $x_{0}$ has penetrable variables
$x_{-1}$ and $x_{1}$, bounding variables $x_{-3}$ and
$x_{2}$, and a left-hand side inner variable $x_{-2}$.

Thus
\begin{equation*}
\begin{array}{l}
P(k=5) = P(C_{0}^{1})+P(C_{0}^{2})+P(C_{1}^{1})+P(C_{1}^{2})\equiv
p_{0}^{1}+p_{0}^{2}+p_{1}^{1}+p_{1}^{2}.
\end{array}
\eqno(S5)
\end{equation*}

\begin{center}
\includegraphics[width=26em, height=21em]{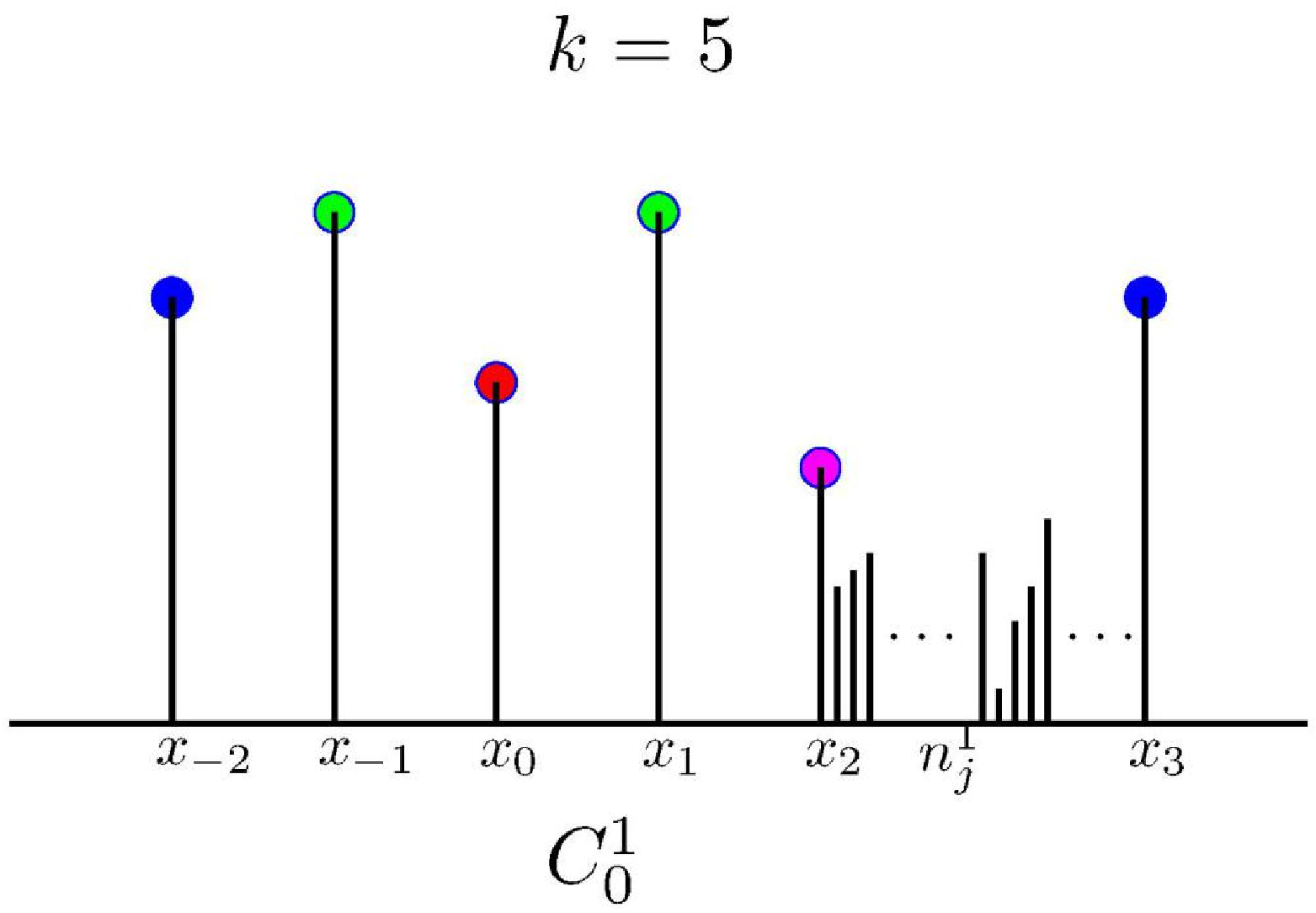}
\hspace{-1.5em}
\includegraphics[width=26em, height=21em]{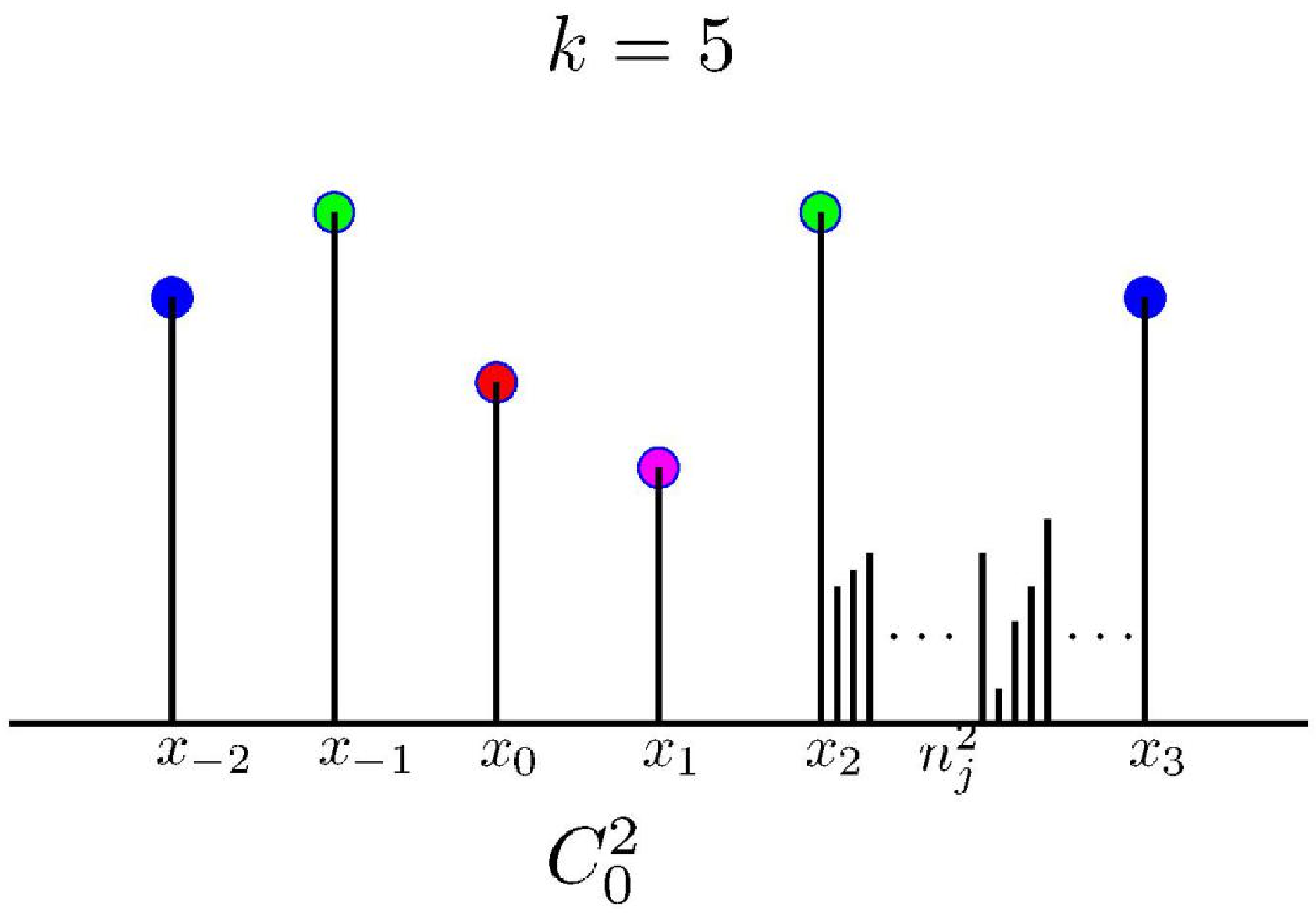}\\
\vspace{0em}
\includegraphics[width=26em, height=21em]{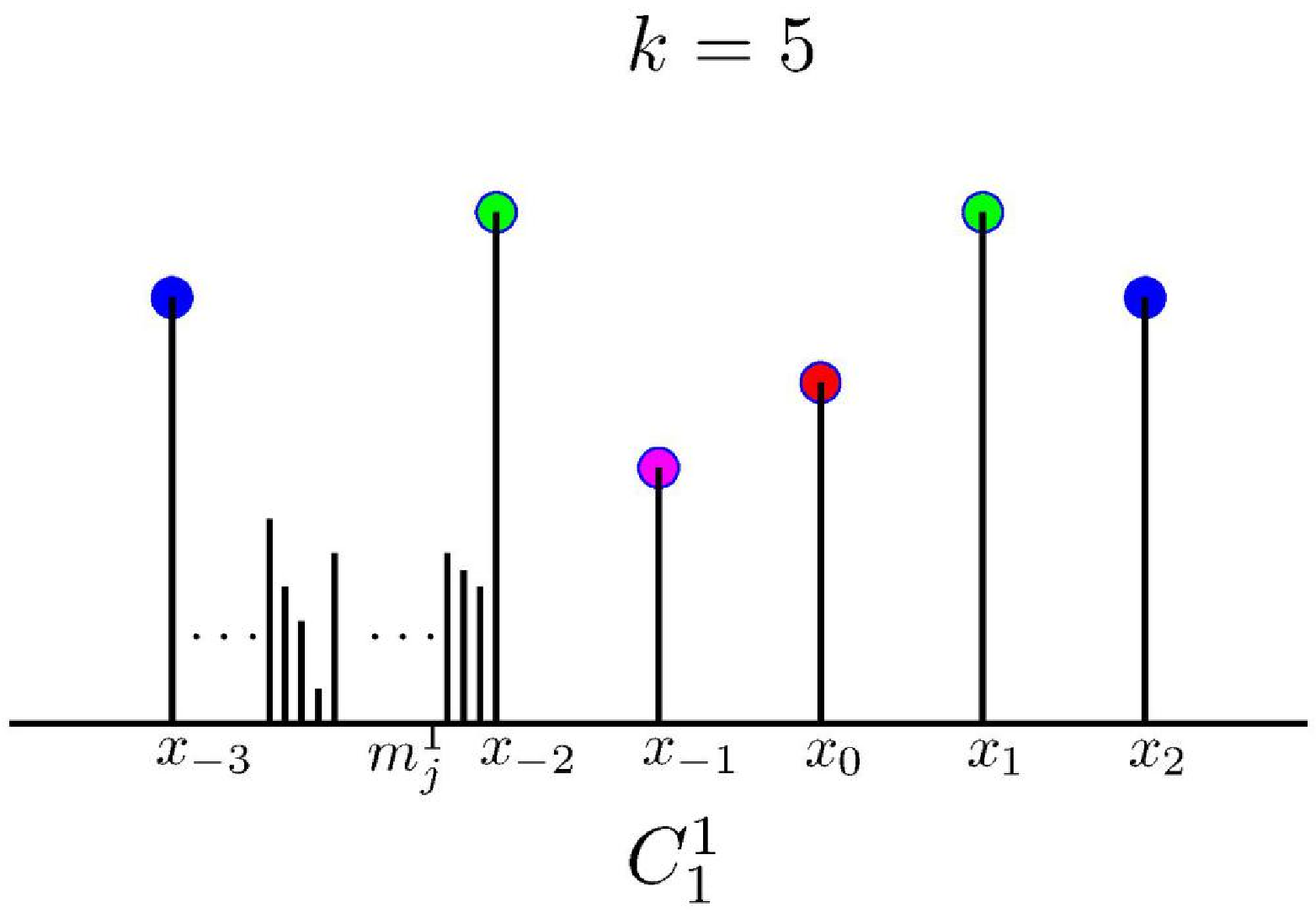}
\hspace{-1.5em}
\includegraphics[width=26em, height=21em]{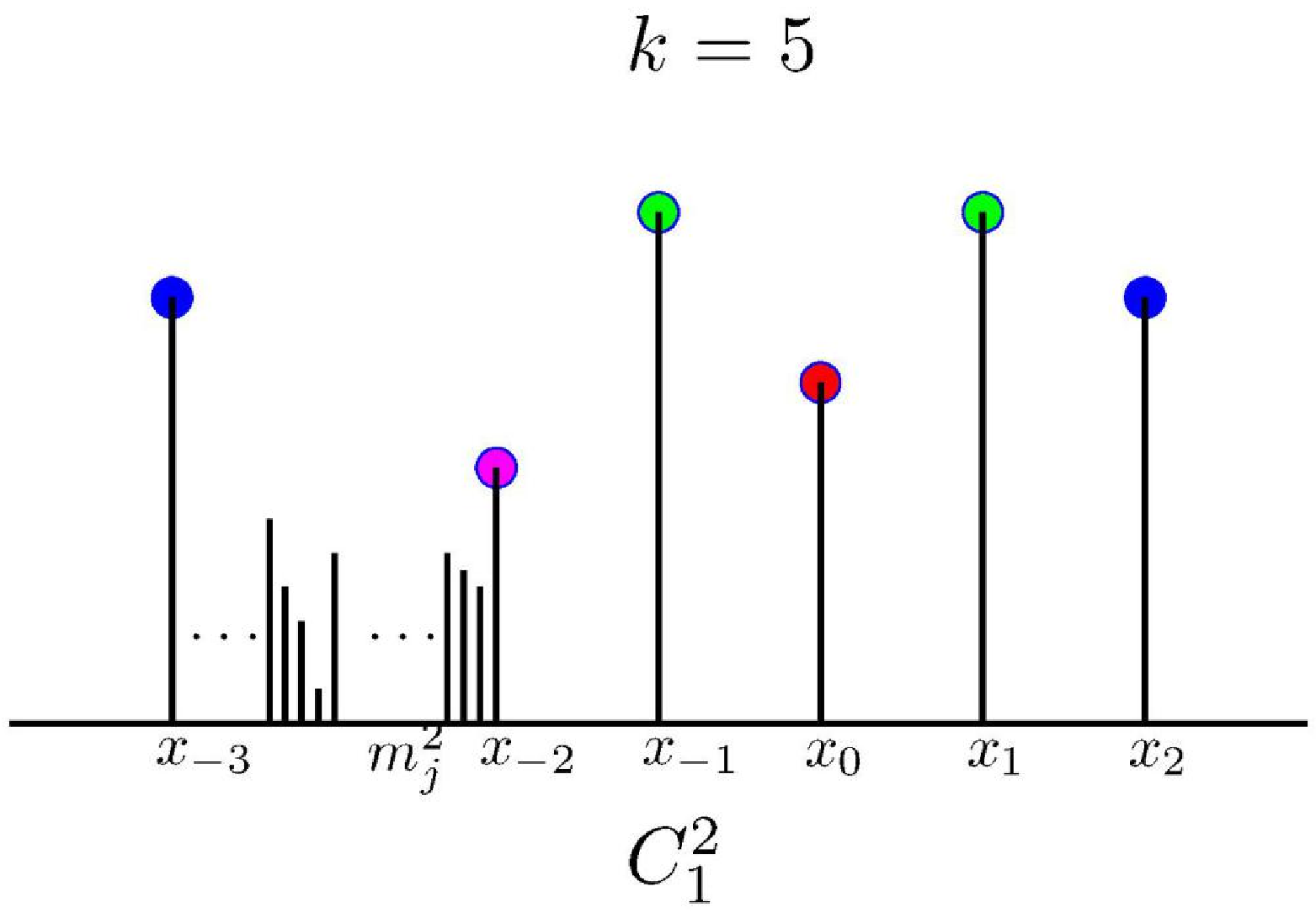}\\
\vspace{0em}
\end{center}
\begin{center}
\noindent {\small {\bf Fig.S2.} \emph{Set of possible configurations for a seed data $x_{0}$ with $k=5$. The sign of the subscript in $x_{i}$ indicates whether the data are located on the left-hand side of $x_{0}$ or on the right-hand side. The sigh of the subscript in $C_{i}^{j}$ indicates the number of inner data located on the left-hand side of $x_{0}$, the superscript in $C_{i}^{j}$ indicates the different cases. The signs $n_{j}^{1},n_{j}^{2},m_{j}^{1},m_{j}^{2}$ indicate the number of the hidden data.}}
\end{center}

Note that an arbitrary number of hidden variables
$n_{j}^{1},n_{j}^{2},m_{j}^{1},m_{j}^{2}$ eventually are located between
the inner data and the bounding variables (or the penetrable data and
the bounding data) and this must be taken into account in the
probability calculation. The geometrical restrictions for the hidden
variables are $n_{j}^{1}< x_{2},n_{j}^{2}< x_{1},j=1,2,...,r$ for
$C_{0}^{1},C_{0}^{2}$ and $m_{j}^{1}< x_{-1},m_{j}^{2}<
x_{-2},j=1,2,...,s$ for $C_{1}^{1},C_{1}^{2}$. Then
\begin{equation*}
\begin{array}{l}
p_{0}^{1} = Prob((x_{-2},x_{-1},x_{1},x_{3}\geq x_{0})\cap (x_{2} <
x_{0}) \cap (\{n_{j}^{1}< x_{2}\}_{j=1,2,...,r})),\\
p_{0}^{2} = Prob((x_{-2},x_{-1},x_{2},x_{3}\geq x_{0})\cap (x_{1} <
x_{0}) \cap (\{n_{j}^{2}< x_{1}\}_{j=1,2,...,r})),\\
p_{1}^{1} = Prob((x_{-3},x_{-2},x_{1},x_{2}\geq x_{0})\cap (x_{-1} <
x_{0}) \cap (\{m_{j}^{1}< x_{-1}\}_{j=1,2,...,s})),\\
p_{1}^{2} = Prob((x_{-3},x_{-1},x_{1},x_{2}\geq x_{0})\cap (x_{-2} <
x_{0}) \cap (\{m_{j}^{2}< x_{-2}\}_{j=1,2,...,s})).
\end{array}
\eqno(S6)
\end{equation*}
Because these are independent and identically distributed random
variables, $p_{0}^{1}$ can be calculated
\begin{equation*}
\begin{array}{l}
p_{0}^{1} =
\int_{0}^{1}f(x_{0})dx_{0}\int_{x_{0}}^{1}f(x_{-2})dx_{-2}
\int_{x_{0}}^{1}f(x_{-1})dx_{-1}\int_{x_{0}}^{1}f(x_{1})dx_{1}
\int_{x_{0}}^{1}f(x_{3})dx_{3}\int_{0}^{x_{0}}f(x_{2})dx_{2}\\
      +\sum\limits_{r=1}^{\infty}
\int_{0}^{1}f(x_{0})dx_{0}\int_{x_{0}}^{1}f(x_{-2})dx_{-2}
\int_{x_{0}}^{1}f(x_{-1})dx_{-1}\int_{x_{0}}^{1}f(x_{1})dx_{1}
\int_{x_{0}}^{1}f(x_{3})dx_{3}\int_{0}^{x_{0}}f(x_{2})dx_{2}\prod\limits_{j=1}^{r}
\int_{0}^{x_{2}}f(n_{j}^{1})dn_{j}^{1}.
\end{array}
\eqno(S7)
\end{equation*}
From Eq.~(S3) we now have
\begin{equation*}
\begin{array}{l}
p_{0}^{1} = \int_{0}^{1}f(x_{0})dx_{0}
\int_{x_{0}}^{1}f(x_{-2})dx_{-2}
\int_{x_{0}}^{1}f(x_{-1})dx_{-1}
\int_{x_{0}}^{1}f(x_{1})dx_{1}
\int_{x_{0}}^{1}f(x_{3})dx_{3}
\int_{0}^{x_{0}}\frac{f(x_{2})}{1-F(x_{2})}dx_{2}\\
= - \int_{0}^{1}f(x_{0})dx_{0}[1-F(x_{0})]^{4}ln[1-F(x_{0})]=\frac{1}{25}.
\end{array}
\eqno(S8)
\end{equation*}
Using the same method, we find the identical results for
$p_{0}^{2},p_{1}^{1}$ and $p_{1}^{2}$ and then we have
\begin{equation*}
\begin{array}{l}
P(k=5) = 4P_{0}^{1}= - 4\int_{0}^{1}f(x_{0})dx_{0}
[1-F(x_{0})]^{4}ln[1-F(x_{0})]=\frac{4}{25}.
\end{array}
\eqno(S9)
\end{equation*}

We thus conclude that a configuration $C_{i}^{j}$ contributes to $P(k)$
with a product of internals when (i) the seed variable $[S]$ provides a
contribution of $\int_{0}^{1}f(x_{0})dx_{0}$, (ii) each penetrable
variable $[P]$ provides a contribution of $\int_{x_{0}}^{1}f(x)dx$,
(iii) each boundary variable $[B]$ provides a contribution of
$\int_{x_{0}}^{1}f(x)dx$, and (iv) an inner variable $[I]$ provides a
contribution of $\int_{x_{j}}^{x_{0}}\frac{f(x)}{1-F(x)}$.

Using these four rules, we formally schematize the probability
associated with each configuration. For example, when $k=4$, $P(k)$ has
a single contribution $p_{0}$ shown in the formal diagram
$[B][P][S][P][B]$. When $k=5$, $P(k) =
p_{0}^{1}+p_{0}^{2}+p_{1}^{1}+p_{1}^{2}$ where $p_{0}^{1}$ is shown in
the diagram $[B][P][S][P][I][B]$, $p_{0}^{2}$ is shown in
$[B][P][S][I][P][B]$, $p_{1}^{1}$ is shown in $[B][P][I][S][P][B]$, and
$p_{1}^{2}$ is shown in $[B][I][P][S][P][B]$. Thus we derive a general
expression for $P(k)$ by applying the four rules for the contribution of
each $C_{i}^{j},i = 0,1,2,...,j = 1,2,...$.  When $k=6$, however, there
are 13 possible seed data $x_{0}$ configurations, and it is labeled
$C_{0}^{i},C_{1}^{j},C_{2}^{r}$.

Similar to $P(k=5)$, we derive
\begin{equation*}
\begin{array}{l}
P(k=6) = \sum\limits_{i=1}p_{0}^{i}+\sum\limits_{j=1}p_{1}^{j}+\sum\limits_{r=1}p_{2}^{r}.
\end{array}
\eqno(S10)
\end{equation*}
Here $C_{1}^{j}$ leads to the same expression as configurations in $k=5$
and thus we can derive $p_{1}^{j}$ by applying the four rules. Fig.~S3
shows that $C_{0}^{i}$ and $C_{2}^{r}$ are geometrically different and
are formed from a seed $x_{0}$ and two penetrable variables. In
configurations $C_{0}^{4}$ and $C_{2}^{4}$ there are three penetrable
variables, one of which ($x_{1}$ in $C_{0}^{4}$ and $x_{-1}$ in
$C_{2}^{4}$) is smaller than $x_{0}$. When calculating $P(k)$ the role
of this smaller penetrable variable is similar to the inner
variable. Thus without loss of generality we refer to this smaller
penetrable variable as the inner variable. There are two bounding and
two concatenated inner variables, and the concatenated variables produce
concatenated integrals.

\begin{center}
\includegraphics[width=13em, height=10em]{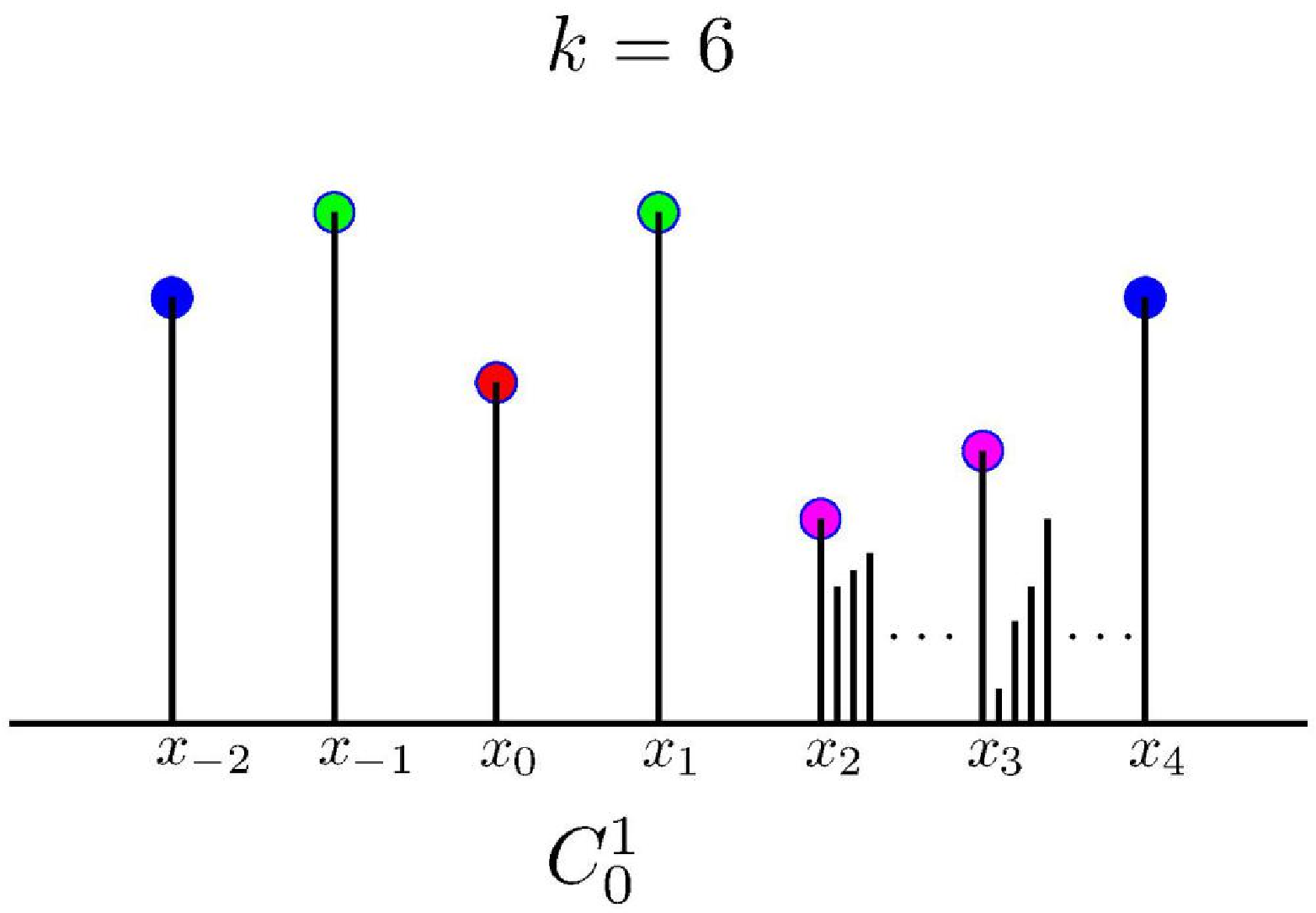}
\hspace{-1.5em}
\includegraphics[width=13em, height=10em]{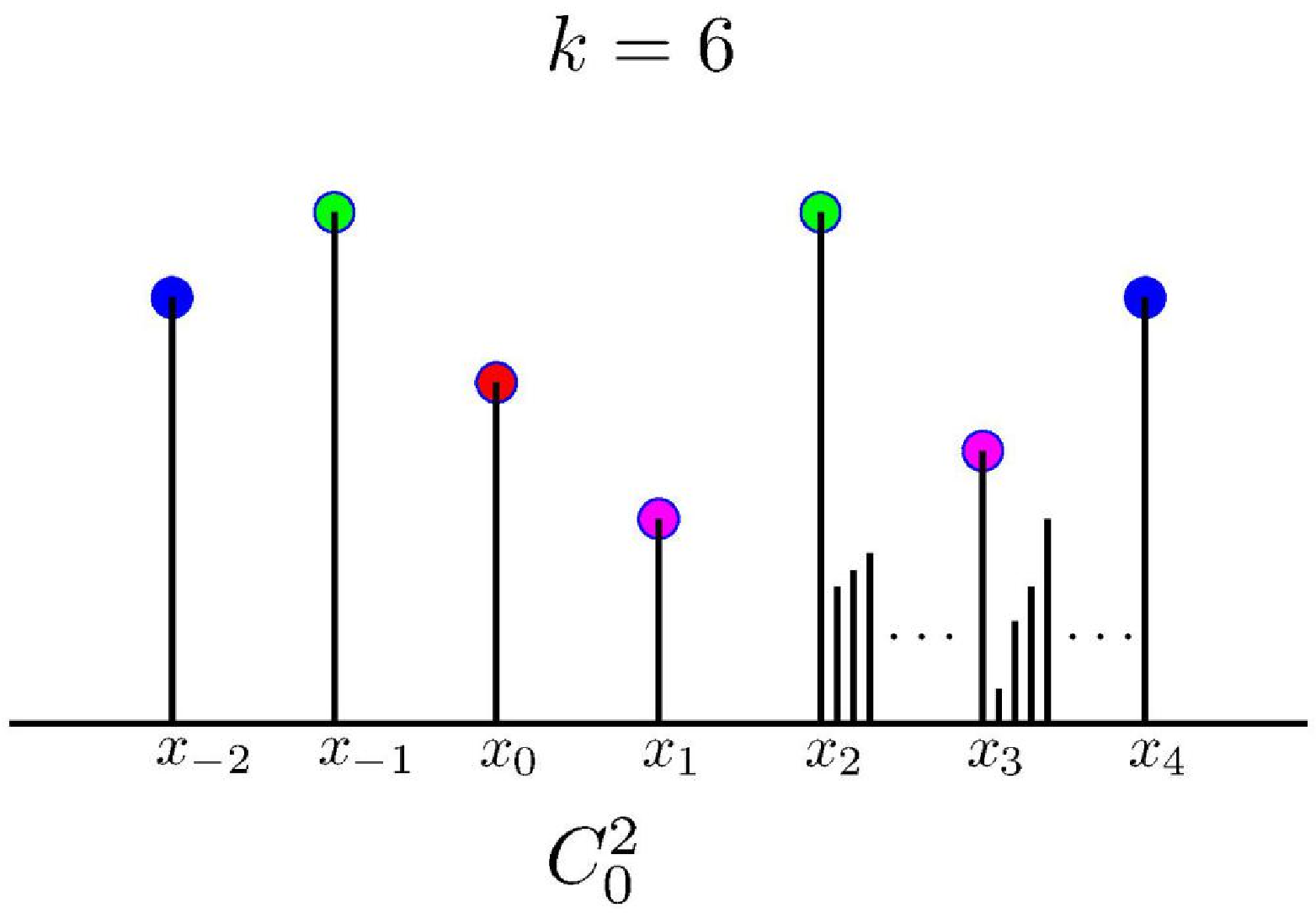}
\hspace{-1.5em}
\includegraphics[width=13em, height=10em]{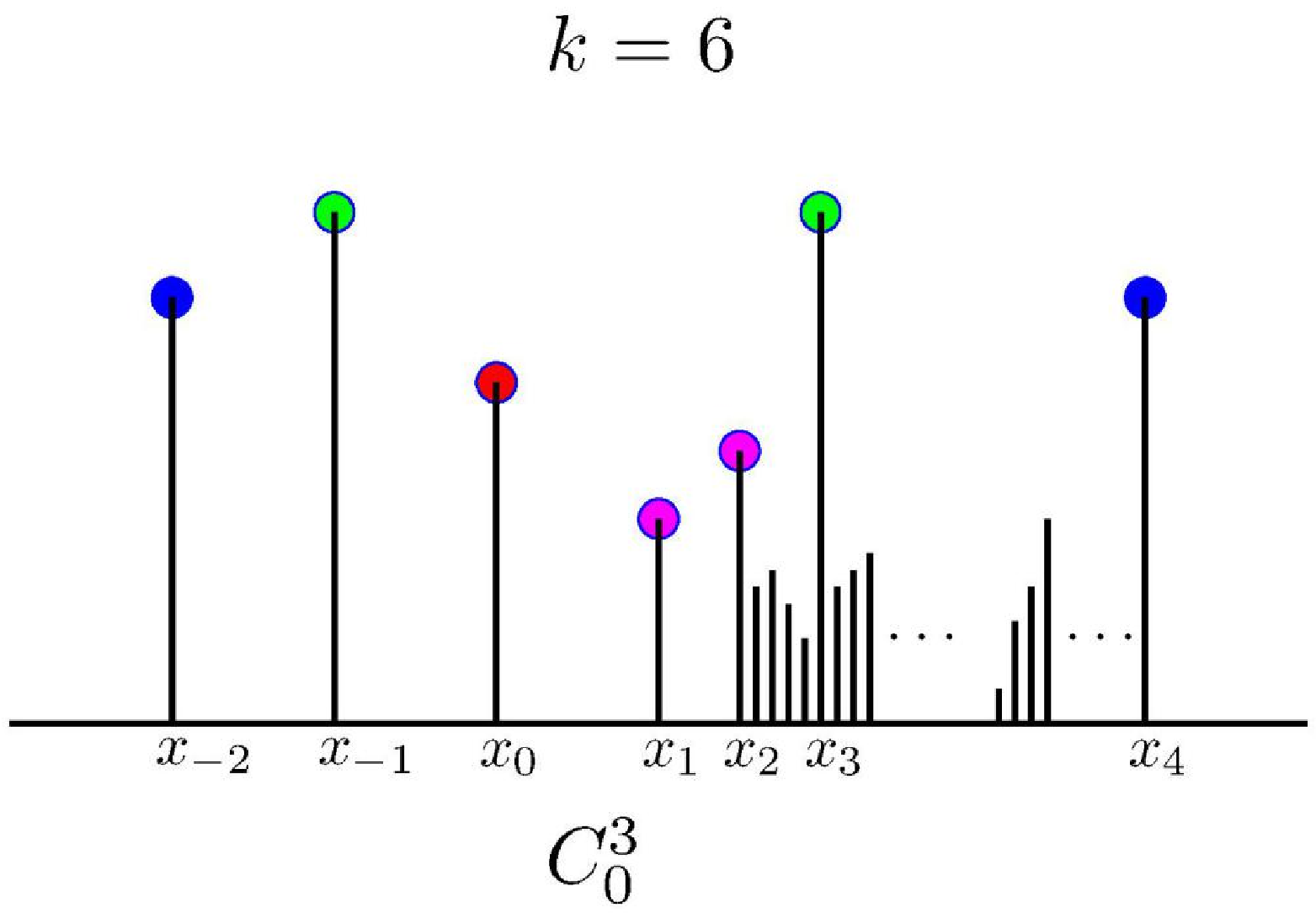}
\hspace{-1.5em}
\includegraphics[width=13em, height=10em]{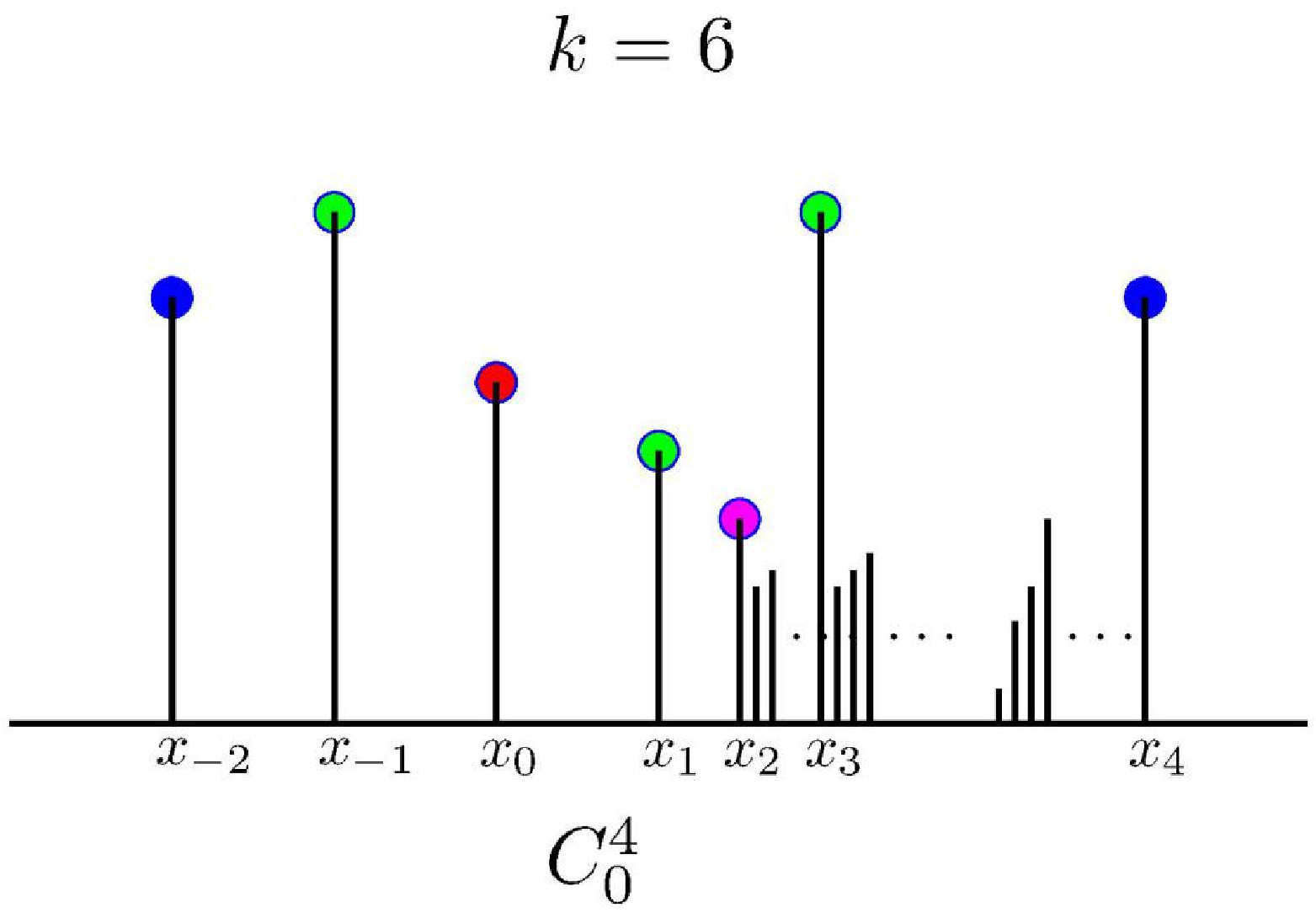}\\
\includegraphics[width=13em, height=10em]{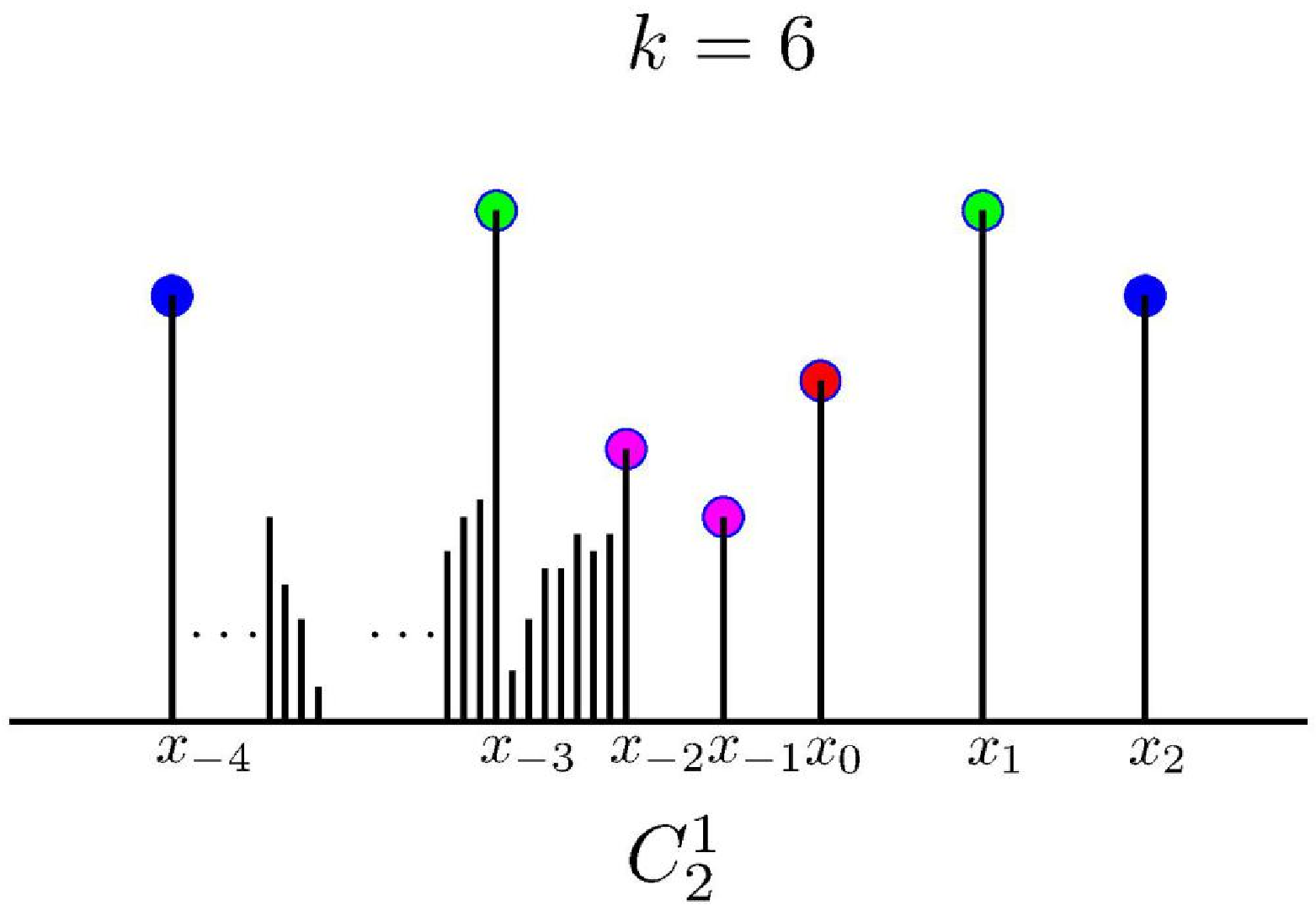}
\hspace{-1.5em}
\includegraphics[width=13em, height=10em]{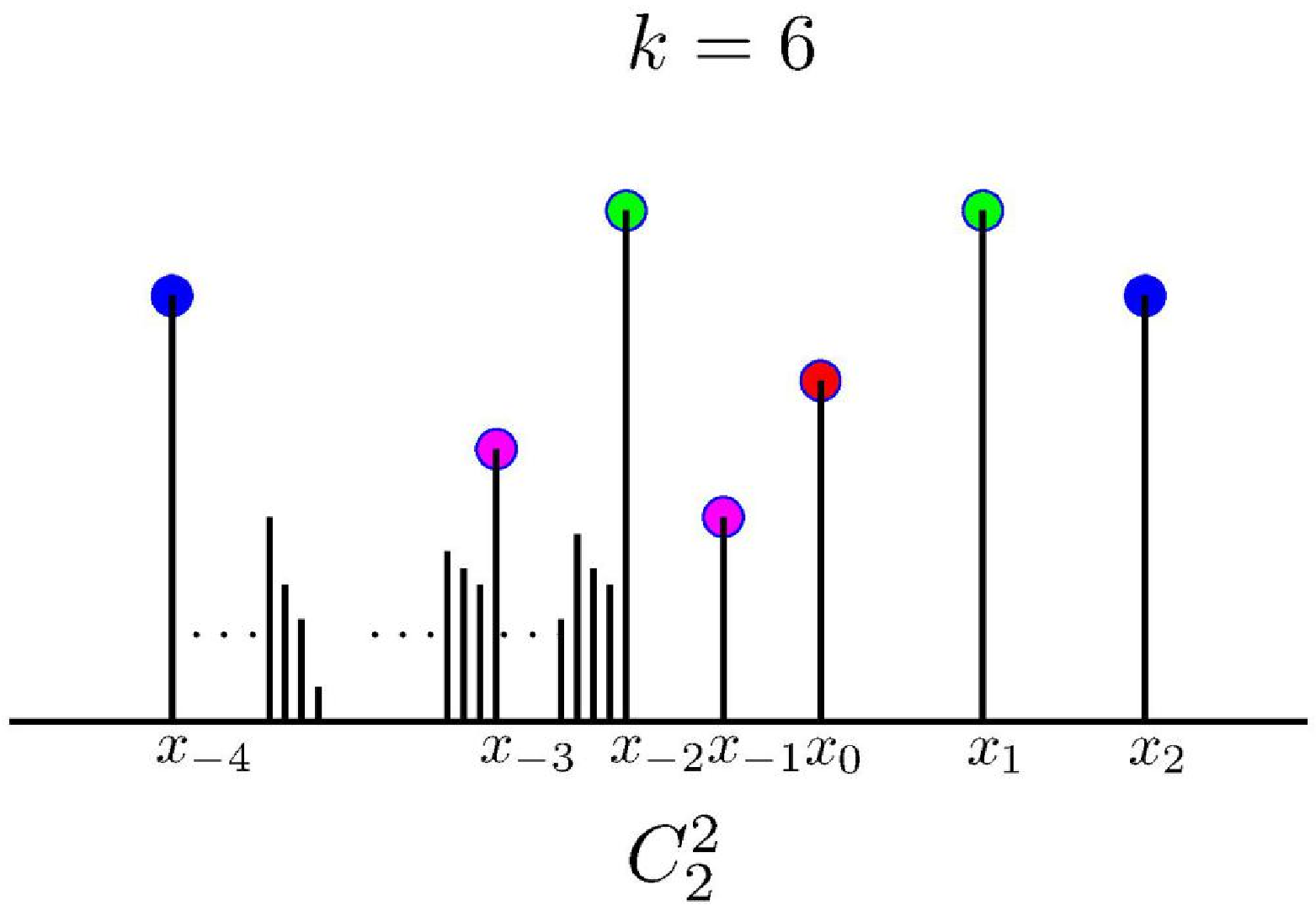}
\hspace{-1.5em}
\includegraphics[width=13em, height=10em]{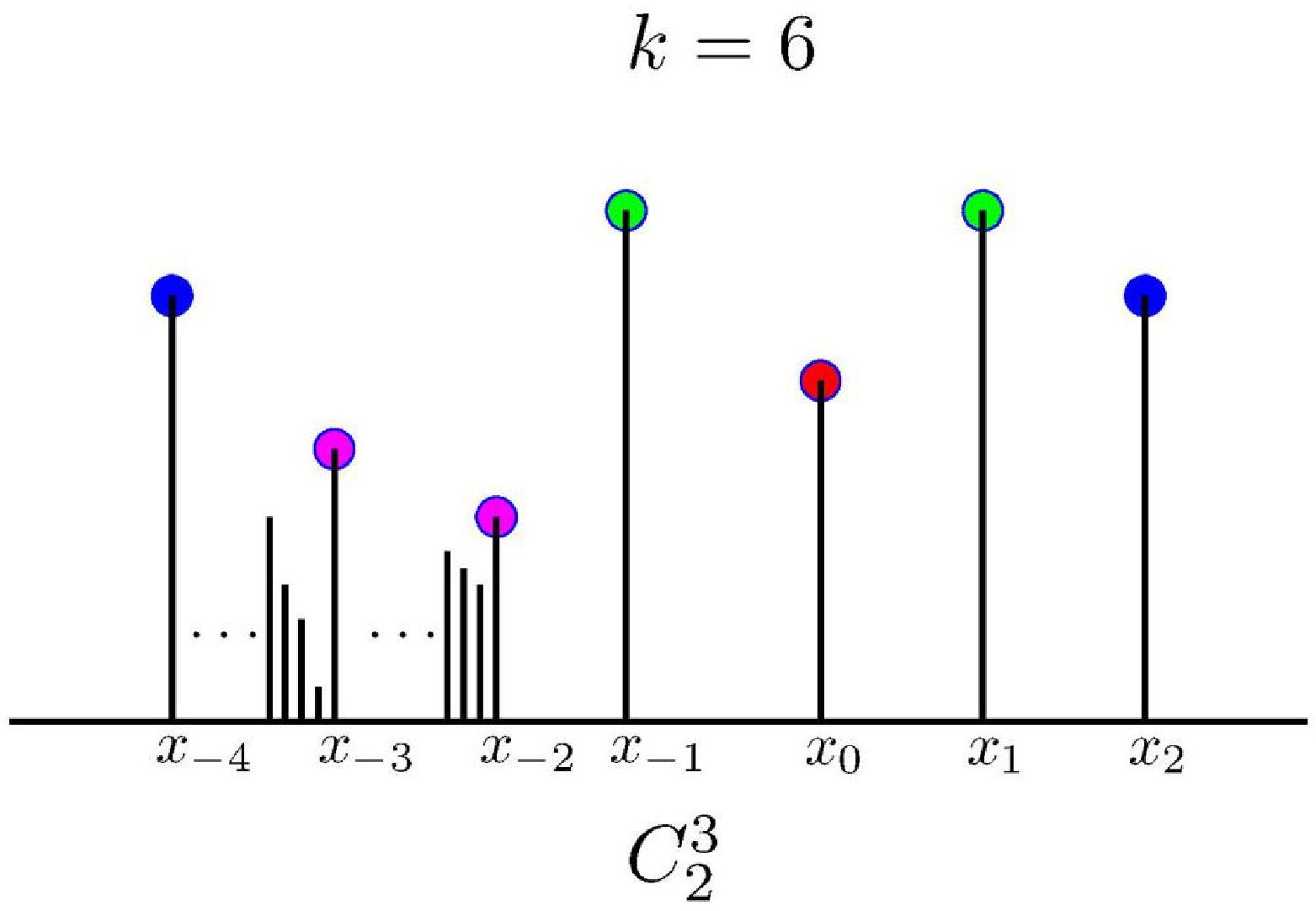}
\hspace{-1.5em}
\includegraphics[width=13em, height=10em]{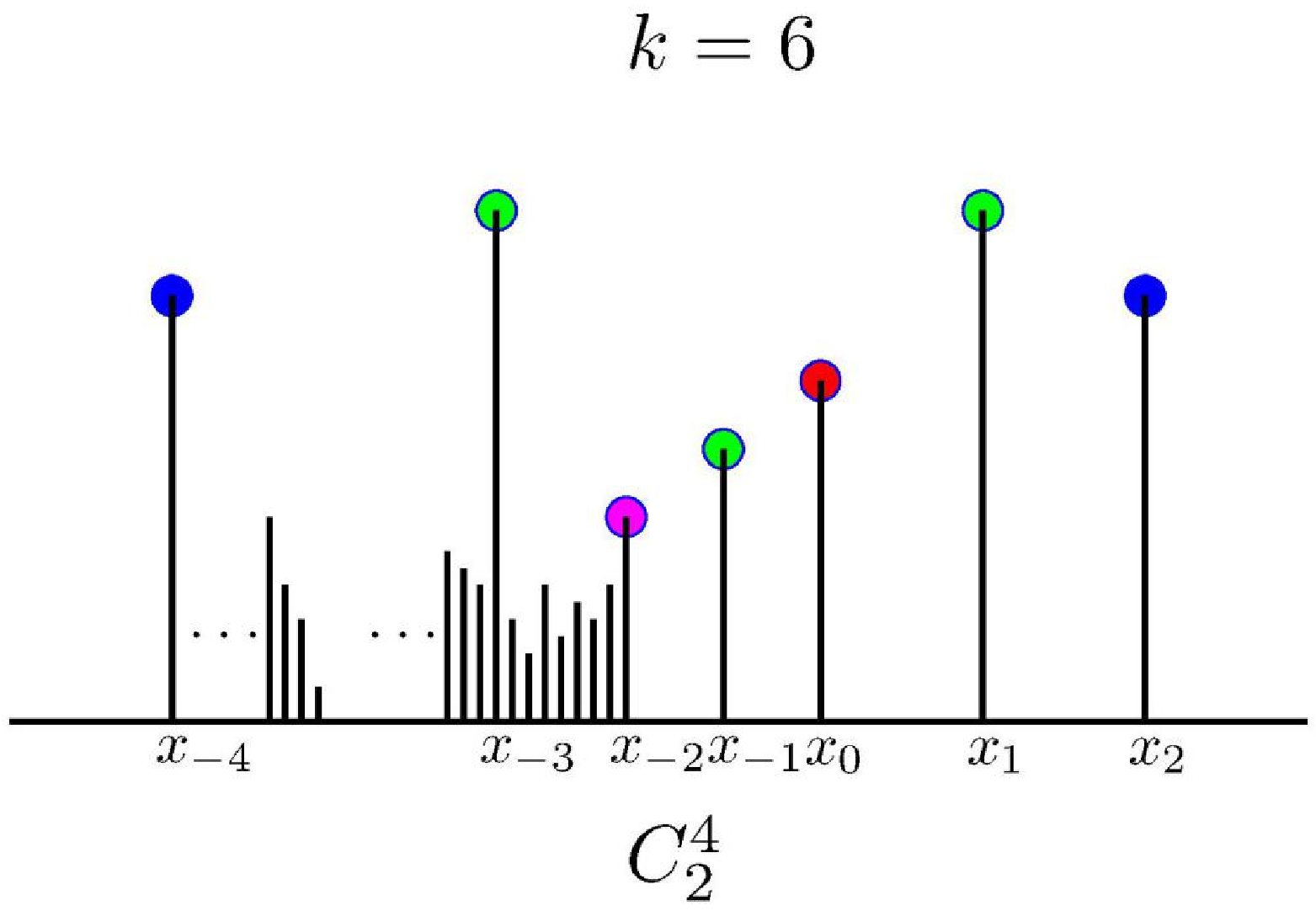}\\
\vspace{0em}
\end{center}
\begin{center}
\noindent {\small {\bf Fig.S3.} \emph{Set of possible configurations for
    $C_{0}^{i},C_{2}^{r}$ with $k=6$.}}
\end{center}

For example, when we apply the same formalism as for $k = 5$ we find
that when $k = 6$, in the case of $C_{0}^{i}$,
\begin{equation*}
\begin{array}{l}
p_{0}^{1} =\int_{0}^{1}f(x_{0})dx_{0}\int_{x_{0}}^{1}f(x_{1})dx_{1}\int_{0}^{x_{0}}\frac{f(x_{2})}{1-F(x_{2})}dx_{2}\int_{x_{2}}^{x_{0}}\frac{f(x_{3})}{1-F(x_{3})}dx_{3}\int_{x_{0}}^{1}f(x_{4})dx_{4}\int_{x_{0}}^{1}f(x_{-1})dx_{-1}\int_{x_{0}}^{1}f(x_{-2})dx_{-2}\\
p_{0}^{2} =\int_{0}^{1}f(x_{0})dx_{0}\int_{0}^{x_{0}}\frac{f(x_{1})}{1-F(x_{1})}dx_{1}\int_{x_{0}}^{1}f(x_{2})dx_{2}\int_{x_{1}}^{x_{0}}\frac{f(x_{3})}{1-F(x_{3})}dx_{3}\int_{x_{0}}^{1}f(x_{4})dx_{4}\int_{x_{0}}^{1}f(x_{-1})dx_{-1}\int_{x_{0}}^{1}f(x_{-2})dx_{-2}\\
p_{0}^{3} =\int_{0}^{1}f(x_{0})dx_{0}\int_{0}^{x_{0}}\frac{f(x_{1})}{1-F(x_{1})}dx_{1}\int_{x_{1}}^{x_{0}}\frac{f(x_{2})}{1-F(x_{2})}dx_{2}\int_{x_{0}}^{1}f(x_{3})dx_{3}\int_{x_{0}}^{1}f(x_{4})dx_{4}\int_{x_{0}}^{1}f(x_{-1})dx_{-1}\int_{x_{0}}^{1}f(x_{-2})dx_{-2}\\
p_{0}^{4} =\int_{0}^{1}f(x_{0})dx_{0}\int_{0}^{x_{0}}\frac{f(x_{1})}{1-F(x_{1})}dx_{1}\int_{0}^{x_{1}}f(x_{2})dx_{2}\int_{x_{0}}^{1}f(x_{3})dx_{3}\int_{x_{0}}^{1}f(x_{4})dx_{4}\int_{x_{0}}^{1}f(x_{-1})dx_{-1}\int_{x_{0}}^{1}f(x_{-2})dx_{-2}\\
\end{array}
\eqno(S11)
\end{equation*}

Using Eq.~(S8), when $k=5$, every integral depends on $x_{0}$, and thus
we integrate each term to find this dependence on. Here, however, there
are two concatenated inner variables, and two concatenated inner
variables generate the dependence on the integrals and hence on the
probabilities. Thus in the general case each configuration is not
equiprobable and does not provide the same contribution to the
probability $P(k)$. To weight the effect of these concatenated
contributions, we use the definition of $p_{i}$. Since $P(k)$ is formed
by $k-3$ contributions labeled $C_{0}^{i},C_{1}^{j},...,C_{k-4}^{r}$ in
which the subindex denotes the number of inner data present at the
left-hand side of seed $x_{0}$, we conclude that in general the $k-4$
inner variables make the following contributions to $P(k)$:

\begin{itemize}

\item[{(a)}] $p_{0}^{i}$ has $k-4$ concatenated internals (the
  right-hand side of seed $x_{0}$);

\item[{(b)}] $p_{1}^{j}$ has $k-5$ concatenated internals (the
  right-hand side of seed $x_{0}$) and an independent inner data
  contribution (the left-hand side of seed $x_{0}$);

\item[{(c)}] $p_{2}^{r}$ has $k-6$ concatenated internals (the
  right-hand side of the seed $x_{0}$) and another two independent inner
  data contributions (the left-hand side of seed $x_{0}$);

  $\vdots$

\item[{(d)}] $p_{k-5}^{j}$ has $k-5$ concatenated internals (the
  left-hand side of seed $x_{0}$) and an independent inner data
  contribution (the right-hand side of seed $x_{0}$); and

\item[{(e)}] $p_{k-4}^{i}$ has $k-4$ concatenated internals (the
  left-hand side of seed $x_{0}$).

\end{itemize}

\noindent
Note that $p_{m}^{n}$ is symmetric with respect to the seed and the
penetrable variables. Adding this modification to the four rules we
calculate a general expression for $P(k)$, i.e.,
\begin{equation*}
\begin{array}{l}
P(k) =
\sum\limits_{i}p_{0}^{i}+\sum\limits_{j}p_{1}^{j}+\sum\limits_{r}p_{2}^{r}+...+
\sum\limits_{j}p_{k-5}^{j}+\sum\limits_{i}p_{k-4}^{i}\\
=\sum\limits_{i}[S][P]^{2}[B]^{2}[I]_{0}^{i}[I]_{k-4}^{i}+
\sum\limits_{j}[S][P]^{2}[B]^{2}[I]_{1}^{j}[I]_{k-5}^{j}+
\sum\limits_{r}[S][P]^{2}[B]^{2}[I]_{2}^{r}[I]_{k-6}^{r}+...\\
...+\sum\limits_{j}[S][P]^{2}[B]^{2}[I]_{k-5}^{j}[I]_{1}^{j}+
\sum\limits_{i}[S][P]^{2}[B]^{2}[I]_{k-4}^{i}[I]_{0}^{i}.
\end{array}
\eqno(S12)
\end{equation*}
Using mathematical induction, we prove that
\begin{equation*}
\begin{array}{l}
P(k) = \sum\limits_{h=0}^{k-4}3^{h}[S][P]^{2}[B]^{2}[I]_{h}[I]_{k-4-h},
\end{array}
\eqno(S13)
\end{equation*}
where the concatenation of $h$ inner variable integrals $[I]_{h}$ is
\begin{equation*}
\begin{array}{l}
[I]_{h} = \int_{0}^{x_{0}}\frac{f(x_{1})}{1-F(x_{1})}dx_{1}
\prod\limits_{j=1}^{h-1}\int_{x_{j}}^{x_{0}}\frac{f(x_{j+1})}{1-F(x_j+1)}dx_{j+1}=
\frac{(-1)^{h}}{h!}[ln(1-F(x_{0}))]^{h}.
\end{array}
\eqno(S14)
\end{equation*}
Using Eq. (S13) and Eq.~(S14), we have
\begin{equation*}
\begin{array}{l}
P(k) = \sum\limits_{h=0}^{k-4}3^{h}\frac{(-1)^{k-4}}{h!(k-4-h)!}
\int_{0}^{1}f(x_{0})[1-F(x_{0})]^{4}[ln(1-F(x_{0}))]^{k-4}dx_{0}\\
=(\frac{1}{5})^{k-3}\sum\limits_{h=0}^{k-4}\frac{3^{h}(k-4)!}{h!(k-4-h)!}
=\frac{1}{5}(\frac{4}{5})^{k-4},\forall f(x).
\end{array}
\eqno(S15)
\end{equation*}
Note that $P(k)$ can be rewritten
\begin{equation*}
\begin{array}{l}
P(k)\sim exp[-(k-4)ln(5/4)],k=4,5,6,...,\forall f(x).
\end{array}
\eqno(S16)
\end{equation*}

\textbf{Theorem S2.} Let $X(t)$ be a real valued bi-infinite time series
of $i.i.d.$ random variables with
a probability density $f(x)$ and with $x\in [a,b]$, and consider its
associated LPHVG at a limited penetrable distance $\rho$. Then
$$P(k)\sim exp\{-(k-2\rho-2)ln[(2\rho+3)/(2\rho+2)]\},
\rho = 0,1,2,3,...,k=2\rho+2,2\rho+3,...,\forall f(x).$$

\textbf{Sketch of the proof.} The proof follows a similar path as for
LPHVG with the limited penetrable distance $\rho=1$ (see \emph{Theorem
  S1}). Instead of Eq.~(S13) we now have
\begin{equation*}
\begin{array}{l}
P(k) = \sum\limits_{h=0}^{k-2(\rho+1)}(2\rho+1)^{h}[S][P]^{2\rho}
[B]^{2}[I]_{h}[I]_{k-2(\rho+1)-h}.
\end{array}
\eqno(S17)
\end{equation*}
We prove by induction that
\begin{equation*}
\begin{array}{l}
P(k) = \sum\limits_{h=0}^{k-2(\rho+1)}(2\rho+1)^{h}\frac{(-1)^{k-2(\rho+1)}}{h![k-2(\rho+1)-h]!}\int_{0}^{1}f(x_{0})[1-F(x_{0})]^{2(\rho+1)}[ln(1-F(x_{0}))]^{k-2(\rho+1)}dx_{0}\\
=(\frac{1}{2\rho+3})^{k-2\rho-1}\sum\limits_{h=0}^{k-2(\rho+1)}\frac{(2\rho+1)^{h}(k-2(\rho+1))!}{h![k-2(\rho+1)-h]!}=\frac{1}{2\rho+3}(\frac{2\rho+2}{2\rho+3})^{k-2(\rho+1)},\forall f(x),
\end{array}
\eqno(S18)
\end{equation*}
i.e.,
\begin{equation*}
\begin{array}{l}
P(k)\sim exp\{-(k-2\rho-2)ln[(2\rho+3)/(2\rho+2)]\},\rho = 0,1,2,3,...,
k=2\rho+2,2\rho+3,...,\forall f(x).
\end{array}
\eqno(S19)
\end{equation*}

When $\rho = 0$ using Eq.~(S18) we find
$P(k)=\frac{1}{3}(\frac{2}{3})^{k-2}$, the result in Ref.~[11]. When $\rho
= 0$ the LPHVG becomes the HVG. When $\rho = 1$ the result is the same
as in \emph{Theorem S1}.

\textbf{Theorem S3.} Let $X(t)$ be a real valued bi-infinite time series
of $i.i.d.$ random variable with probability density $f(x)$ with $x\in
[a,b]$, and consider its associated LPHVG with the limited penetrable
distance $\rho$. Then the local clustering coefficient is
\begin{equation*}
\begin{array}{l}
C_{\rm min}(k) = \frac{2}{k}+\frac{2\rho(k-2)}{k(k-1)},\rho = 0,1,2,k\geq 2(\rho+1),\\
C_{\rm max}(k) = \frac{2}{k}+\frac{4\rho(k-3)}{k(k-1)},\rho = 0,1,2,k\geq 2(2\rho+1).
\end{array}
\end{equation*}

\textbf{Proof.} For a given node $x_{i}$, the local clustering
coefficient $C$ is the percentage of nodes connected to $x_{i}$ that are
connected to each other. Thus we calculate from a given node $x_{i}$ the
number of nodes from penetrable $\rho$ visible to $x_{i}$ have mutual
penetrable $\rho$ visibility (triangles), normalized with the set of
possible triangles $\binom{k}{2}$.

In the simplest $\rho=1$ case, Fig.~S1 shows that when a generic node
$x_{i}$ has a degree $k=4$ it has two penetrable data and two bounding
data, and there are thus five triangles and $C(k=4)=5/6$. Fig.~S2
shows that when a generic node $x_{i}$ has a degree $k=5$ it has two
penetrable data, two bounding data, and an inner datum. Here there are
two possible outcomes. For configurations $C_{0}^{1},C_{1}^{2}$ there
are seven triangles and $C(k=5) = 7/10$. For configurations
$C_{0}^{2},C_{1}^{1}$ there are eight triangles and $C(k=5)=
8/10$. Fig.~S3 shows that when a generic node $x_{i}$ has a degree
$k=6$ it has two penetrable data, two bounding data, and two inner
data. Here there are three possible outcomes. For configuration
$C_{1}^{3}$ there are nine triangles and $C(k=6) = 9/15$. For
configurations
$C_{0}^{1},C_{0}^{2},C_{1}^{1},C_{1}^{2},C_{1}^{4},C_{2}^{2},C_{2}^{3}$
there are 10 triangles and $C(k=6) = 10/15$. For configurations
$C_{0}^{3},C_{0}^{4},C_{2}^{1},C_{2}^{4}$ there are 11 triangles and
$C(k=6) = 11/15$. Thus nodes having the same degree can have different
clustering coefficients. Although the clustering coefficients of these
nodes are irregular, the minimum clustering coefficient and the maximum
clustering coefficient are regular.

The calculations of these minimum local clustering coefficients can be
rewritten
\begin{equation*}
\begin{array}{l}
C_{\rm min}(k=4) = [(k-1)+(k-2)\rho]/\binom{k}{2} = 5/6,\\
C_{\rm min}(k=5) = [(k-1)+(k-2)\rho]/\binom{k}{2} = 7/10,\\
C_{\rm min}(k=6) = [(k-1)+(k-2)\rho]/\binom{k}{2} = 3/5.
\end{array}
\eqno(S20)
\end{equation*}
In general, for a degree $k$ we can at a minimum form $(k-1)+(k-2)\rho =
(1+\rho)k-(2\rho+1)$ triangles out of $\binom{k}{2}$ possibilities,
and
\begin{equation*}
\begin{array}{l}
C_{\rm min}(k)=[(1+\rho)k-(2\rho+1)]/\binom{k}{2}=\frac{2}{k}
+\frac{2\rho(k-2)}{k(k-1)},\rho=0,1,2,k\geq 2(\rho+1).
\end{array}
\eqno(S21)
\end{equation*}
Similarly, the calculation of the maximum local clustering
coefficients can be rewritten
\begin{equation*}
\begin{array}{l}
C_{\rm max}(k=4) = [(k-1)+2\rho(k-3)]/\binom{k}{2} = 5/6,\\
C_{\rm min}(k=5) = [(k-1)+2\rho(k-3)]/\binom{k}{2} = 8/10,\\
C_{\rm min}(k=6) = [(k-1)+2\rho(k-3)]/\binom{k}{2} = 11/15.
\end{array}
\eqno(S22)
\end{equation*}
For a degree $k$ we can at a maximum form $(k-1)+2\rho(k-3) =
(1+2\rho)k-(6\rho+1)$ triangles out of $\binom{k}{2}$ possibilities, and
\begin{equation*}
\begin{array}{l}
C_{\rm max}(k)=[(1+2\rho)k-(6\rho+1)]/\binom{k}{2}=\frac{2}{k}
+\frac{4\rho(k-3)}{k(k-1)},\rho=0,1,2,k\geq 2(2\rho+1).
\end{array}
\eqno(S23)
\end{equation*}
This relation between $k$ and $C_{\rm min},C_{\rm max}$ allows us to
deduce the local clustering coefficient distribution $P(C_{\rm min})$
and $P(C_{\rm max})$,
\begin{equation*}
\begin{array}{l}
P(k) = \frac{1}{2\rho+3}(\frac{2\rho+2}{2\rho+3})^{k-2(\rho+1)},\\
k = \frac{\varphi+\sqrt{\varphi^2-8C_{\rm min}(2\rho+1)}}{2C_{\rm min}},
\varphi = C_{\rm min}+2\rho+2\\
k = \frac{\phi+\sqrt{\phi^2-8C_{\rm max}(6\rho+1)}}{2C_{\rm max}},
\phi = C_{\rm max}+4\rho+2.
\end{array}
\eqno(S24)
\end{equation*}
Then
\begin{equation*}
\begin{array}{l}
P(C_{\rm min}) = \frac{1}{2\rho+3}exp\{[\frac{\varphi+
\sqrt{\varphi^2-8C_{\rm min}(2\rho+1)}}{2C_{\rm min}}-2(\rho+1)]
ln(\frac{2\rho+2}{2\rho+3})\},
\end{array}
\eqno(S25)
\end{equation*}
\begin{equation*}
\begin{array}{l}
P(C_{\rm max}) = \frac{1}{2\rho+3}exp\{[\frac{\phi+
\sqrt{\phi^2-8C_{\rm max}(6\rho+1)}}{2C_{\rm max}}-2(\rho+1)]
ln(\frac{2\rho+2}{2\rho+3})\}.
\end{array}
\eqno(S26)
\end{equation*}

\textbf{Theorem S4.} Let $\{x_{t}\}_{t=0,1,...,n}$ be a bi-finite
sequence of $i.i.d.$ random variables
extracted from a continuous probability density $f(x)$. Then the
probability $P_{\rho}(n)$ that two data separated by $n$ intermediate
data are two connected nodes in the graph is
$$P_{\rho}(n) = \frac{2\rho(\rho+1)+2}{n(n+1)},\rho=0,1,2,...$$

\textbf{Proof.}  Without loss of generality, we restrict $x$ to $[0,
  1]$. When $\rho = 0$, Ref.~[11] derives
\begin{equation*}
\begin{array}{l}
P_{0}(n) = \int_{0}^{1}\int_{0}^{1}f(x_{0})f(x_{n})dx_{0}dx_{n}
\int_{0}^{min(x_{0},x_{n})}...\int_{0}^{min(x_{0},x_{n})}
f(x_{1})...f(x_{n-1})dx_{1}...dx_{n-1}=\frac{2}{n(n+1)}.
\end{array}
\eqno(S27)
\end{equation*}
When $\rho = 1$, because an arbitrary value $x_{0}$ from this series
will be connected to node $x_{n}$ if there is no more than one
$x_{i}\geq min(x_{0},x_{n})$ for all $x_{i},i=1,2,...,n-1$. Then
$P_{1}(n)$ is
\begin{equation*}
\begin{array}{l}
P_{1}(n) = \int_{0}^{1}\int_{0}^{1}f(x_{0})f(x_{n})dx_{0}dx_{n}\int_{0}^{min(x_{0},x_{n})}...\int_{0}^{min(x_{0},x_{n})}f(x_{1})...f(x_{n-1})dx_{1}...dx_{n-1}\\
+\int_{0}^{1}\int_{0}^{1}f(x_{0})f(x_{n})\int_{min(x_{0},x_{n})}^{1}f(x_{1})dx_{1}\int_{0}^{min(x_{0},x_{n})}...\int_{0}^{min(x_{0},x_{n})}f(x_{2})...f(x_{n-1})dx_{2}...dx_{n-1}\\
+...+\int_{0}^{1}\int_{0}^{1}f(x_{0})f(x_{n})\int_{min(x_{0},x_{n})}^{1}f(x_{n-1})dx_{n-1}\int_{0}^{min(x_{0},x_{n})}...\int_{0}^{min(x_{0},x_{n})}f(x_{1})...f(x_{n-2})dx_{1}...dx_{n-2}.\\
\end{array}
\eqno(S28)
\end{equation*}
Since the integration limits are independent, when we rewrite $x\equiv
min(x_{0},x_{n})$, we have
\begin{equation*}
\begin{array}{l}
P_{1}(n) = \int_{0}^{1}\int_{0}^{1}f(x_{0})f(x_{n})F^{n-1}(x)dx_{0}dx_{n}+\binom{n-1}{1}\int_{0}^{1}\int_{0}^{1}f(x_{0})f(x_{n})[1-F(x)]F^{n-2}(x)dx_{0}dx_{n}\\
=\binom{n-1}{1}\int_{0}^{1}\int_{0}^{1}f(x_{0})f(x_{n})F^{n-2}(x)dx_{0}dx_{n}-(n-2)\int_{0}^{1}\int_{0}^{1}f(x_{0})f(x_{n})F^{n-1}(x)dx_{0}dx_{n}.
\end{array}
\eqno(S29)
\end{equation*}
Without loss of generality we can fix $x_{0}$ and move $x_{n}$ such that
the latter equation becomes
\begin{equation*}
\begin{array}{l}
P_{1}(n) = (n-1)[\int_{0}^{1}\int_{0}^{x_{0}}f(x_{0})f(x_{n})F^{n-2}(x_{n})dx_{0}dx_{n}+\int_{0}^{1}\int_{x_{0}}^{1}f(x_{0})f(x_{n})F^{n-2}(x_{0})dx_{0}dx_{n}]\\
-(n-2)[\int_{0}^{1}\int_{0}^{x_{0}}f(x_{0})f(x_{n})F^{n-1}(x_{n})dx_{0}dx_{n}+\int_{0}^{1}\int_{x_{0}}^{1}f(x_{0})f(x_{n})F^{n-1}(x_{0})dx_{0}dx_{n}]\\
=\frac{2}{n}-\frac{2(n-2)}{n(n+1)}=\frac{6}{n(n+1)}.
\end{array}
\eqno(S30)
\end{equation*}
When $P_{\rho}(n),\rho>1$, the calculation follows a path similar to
that for $P_{1}(n)$ such that instead of Eq.~(S29) we have
\begin{equation*}
\begin{array}{l}
P_{\rho}(n) = \int_{0}^{1}\int_{0}^{1}f(x_{0})f(x_{n})F^{n-1}(x)dx_{0}dx_{n}+\binom{n-1}{\rho}\int_{0}^{1}\int_{0}^{1}f(x_{0})f(x_{n})[1-F(x)]^{\rho}F^{n-(\rho+1)}(x)dx_{0}dx_{n}.
\end{array}
\eqno(S31)
\end{equation*}
Then by induction we prove that
\begin{equation*}
\begin{array}{l}
P_{\rho}(n) = \frac{2\rho(\rho+1)+2}{n(n+1)},\rho=0,1,2,...
\end{array}
\eqno(S32)
\end{equation*}

\end{document}